\documentclass[acmsmall,authorversion]{acmart}

\usepackage{macros}
\usepackage{ulem}
\newcommand{\newtext}[1]{\authorcomment{black}{}{#1}}
\newcommand{\newtexttwo}[1]{\authorcomment{black}{}{#1}}

\AtBeginDocument{%
  \providecommand\BibTeX{{%
    \normalfont B\kern-0.5em{\scshape i\kern-0.25em b}\kern-0.8em\TeX}}}

\begin{document}

%%
%% The "title" command has an optional parameter,
%% allowing the author to define a "short title" to be used in page headers.
\title{Learning by Teaching: Key Challenges and Design Implications}

%%
%% The "author" command and its associated commands are used to define
%% the authors and their affiliations.
%% Of note is the shared affiliation of the first two authors, and the
%% "authornote" and "authornotemark" commands
%% used to denote shared contribution to the research.

% \renewcommand{shortauthors}{Amy Debbané}
\author{Amy Debbané}
\affiliation{%
  \institution{University of Waterloo}
  \city{Waterloo}
  \state{Ontario}
  \country{Canada}}
\email{agdebban@uwaterloo.ca}

\author{Ken Jen Lee}
\affiliation{%
  \institution{University of Waterloo}
  \city{Waterloo}
  \state{Ontario}
  \country{Canada}}
\email{kenjen.lee@uwaterloo.ca}

\author{Jarvis Tse}
\affiliation{%
  \institution{University of Waterloo}
  \city{Waterloo}
  \state{Ontario}
  \country{Canada}}
\email{jarvis.tse@uwaterloo.ca}

\author{Edith Law}
\affiliation{%
  \institution{University of Waterloo}
  \city{Waterloo}
  \state{Ontario}
  \country{Canada}}
\email{edith.law@uwaterloo.ca}

\renewcommand{\shortauthors}{Amy Debbané et al.}

%%
%% The abstract is a short summary of the work to be presented in the
%% article.
\begin{abstract}
  Benefits of learning by teaching (LbT) have been highlighted by previous studies from a pedagogical lens, as well as through computer-supported systems. However, the challenges that university students face in technology-mediated LbT---whether it be teaching oneself, teaching a peer, or teaching an agent---are not well understood. Furthermore, there is a gap in knowledge on the challenges that students encounter throughout the process of teaching (content selection, preparation, teaching, receiving and giving feedback, and reflection) despite its importance to the design of LbT platforms. Thus, we conducted a study with 24 university students where they taught content they had not fully grasped, without guidance, and participated in a semi-structured interview. Results demonstrate that participants encountered the following challenges: psychological barriers relating to self and others, and lack of know-how. Furthermore, we illuminate design implications required to overcome these challenges and benefit from LbT without requiring prior training in pedagogy.
%   In addition, we outline university students' perception on various tools and configurations a LbT platform could include.
\end{abstract}


%%
%% Keywords. The author(s) should pick words that accurately describe
%% the work being presented. Separate the keywords with commas.

\setcopyright{acmlicensed}
\acmJournal{PACMHCI}
\acmYear{2023} \acmVolume{7} \acmNumber{CSCW1} \acmArticle{68} \acmMonth{4} \acmPrice{15.00}\acmDOI{10.1145/3579501}

% \received{January 2022}
% \received[revised]{July 2022}
% \received[accepted]{November 2022}
%%
%% This command processes the author and affiliation and title
%% information and builds the first part of the formatted document.

\maketitle

%======================================================================
\section{Introduction}
%======================================================================

Learning by teaching (LbT) is a pedagogical method where a student is tasked to teach unfamiliar material typically to peers \cite{leikin2006learning,duran2017learning}.
% enriching
It produces more enriching experiences than simply being taught the material and learning for oneself \cite{duran2017learning}. 
% name
%LbT can be organized in a collaborative learning environment where students achieve their learning goals via a group-based approach. 
% peer benefit
Peer teaching materials to others has been observed to enhance the student's own learning of those materials (e.g., \cite{galbraith2011peer, hoogerheide2016gaining, roscoe2007understanding, roscoe2008tutor}). More specifically, the benefits of LbT in a group include learning concepts more effectively, encouraging more participation, improving learning satisfaction, developing teamwork skills, and promoting higher-order thinking \cite{chuang2015sscls}. 
% benefits
Furthermore, studies have shown that students who learned material with the expectation that they will be required to teach it performed better than students who learned to pass a test for themselves \cite{bargh1980cognitive, benware1984quality}. 
% reflective knowledge building
Students have also shown to be more motivated to grasp the material and when their partner misunderstands the material, the student reflects on their teaching, notices their
% review their
own misconceptions, constructs new explanations and knowledge on the material \cite{roscoe2007understanding}. 
% benefit for peer too
The partner who is listening to the student's explanation can also benefit from reciprocal peer tutoring scenarios when the explanations are correct, for the most part, conceptual, elaborate, and target the partner's own misconceptions \cite{walker2014adaptive,webb2003promoting}.

Currently, there is a lack of technology \newtexttwo{for supporting LbT} in a peer-to-peer context, both in-person and online. 
% \newtexttwo{\sout{, where students select unfamiliar materials to teach with the goal of learning primarily for oneself}}.
While prior work has focused on the effectiveness of LbT in various settings and forms (e.g., \cite{roscoe2007understanding}), the challenges students face throughout the process of LbT \newtexttwo{in the contexts of online teaching} 
% \newtexttwo{\sout{, from the preparation phase to the feedback phase,}} 
has not been explored.
% \newtexttwo{\sout{despite its importance to the design of LbT platforms}} 

\newtexttwo{Studying the online context is important. Online conferencing platforms, such as Zoom, have been widely adopted for instructional purposes around the world \cite{toney2021fighting, sayem2017effective}. 
 % By investigating LbT using conference call platforms, we aim to provide the basis to design online LbT platforms that not only can help university students more easily connect to their potential tutors/tutees, but also provide the tools necessary for LbT session conductors to effectively conduct their LbT sessions.  
Having an online LbT platform would provide more flexible alternatives to conventional in-person LbT approaches that might not be feasible in certain contexts (e.g., during a pandemic lockdown, for Massive Open Online Courses, to accommodate students with incompatible schedules). Research has also found challenges that are unique to online learning, such as technical difficulties \cite{ramadani2020teachers,garcia2020stem} and the lack of human touch \cite{dhawan2020online}; similarly, our research is aimed at discovering LbT challenges that may be unique to, or more prominent in, online learning contexts.} 

%For instance, based on our study, a future online LbT may help a pair of university students from different countries match with each other and schedule LbT sessions, such that one of the university students can improve their understanding of a course to prepare for their exam by teaching the other university student using their preferred tools (e.g., a real-time annotation interface), while the other student learns the same material for free.}

\newtexttwo{In this work,} we conducted an exploratory study---observing 24 university students teaching unfamiliar material \newtexttwo{(i.e., material they perceived to have learned but not fully grasped)} online and interviewing them about their experience---to gain a new understanding of how they approach the preparation and teaching of the materials, and the challenges that they encounter during this process. Our study does {\it not} focus on validating the effectiveness of LbT, as the validation of LbT's effectiveness has been achieved by a wide range of existing studies \cite{duran2017learning,galbraith2011peer, hoogerheide2016gaining, roscoe2007understanding, roscoe2008tutor,chuang2015sscls,bargh1980cognitive, benware1984quality,roscoe2007understanding}. Our focus on how the teaching happens among participants who have not been professionally trained in teaching enables us to uncover the unique challenges they face, so that new forms of technology can be later designed to enable students to get the most out of LbT.  \newtexttwo{Specifically, our study makes the following two contributions:}
\begin{itemize}[noitemsep,topsep=5pt]
    \item an analysis of university students' unguided LbT sessions and semi-structured interview responses aimed at uncovering their response to the LbT sessions and potential feature design solutions;
    \item design implications for creating online LbT platforms for university students that support and encourage the effectiveness of the LbT method. %and
    % \item An outline of potential features that could be incorporated into LbT platforms and university students' perceptions of them.
\end{itemize}

We will begin by outlining relevant research on pedagogical LbT practices, existing challenges with LbT in classroom settings, and systems that support this practice in Section \ref{sec:relwork}. Then, we detail the study design created to fill the gap in understanding university students' process of teaching something unfamiliar, when teaching someone or when teaching alone in Section \ref{sec:studydesign}. \newtexttwo{ In Section \ref{sec:analysis}, we present the method of analysis conducted on 24 semi-structured interviews and 48 LbT sessions.
The findings are grouped into the following three types of LbT challenges, namely psychological barriers relating to self, psychological barriers relating to others, and know-how barriers in Section \ref{sec:findings}. This section also includes how these challenges may be influenced by various peer matching and feedback configurations.  
In Section \ref{ref:discussion}, we discuss the design implications and a specific method for future exploration: a ``crash and learn'' approach. 
% \newtext{including designing online environments to make students more comfortable with the teaching material imperfectly, and supporting peer matching and collaborative feedback between peer teachers}. 
}
We conclude by outlining possible limitations, and discussing avenues for future work in Section \ref{sec:conclusion}.

\section{Related Work}\label{sec:relwork}
%======================================================================

One of our goals is to gain a better understanding of how university students LbT online and what kind of support would be required to best enable the process of teaching unfamiliar material in this context. Therefore, the related literature is detailed through both a pedagogical and computer-support lens.
% End of each paragraph explain how our work goes beyond what we already know about it
% e.g. this is well studied in in-person environment but not the online world and with adults 
% And for challenges most comp support is not for lbt

\newtexttwo{\subsection{\bf Phases of LbT}}
% LbT phases
LbT is a multi-step process, including phases like preparing teaching materials, explaining the materials, interacting with peers, and self-reflection.
% prep
Preparing materials with the purpose to teach them requires the student to learn enough to develop educational materials \cite{duran2017learning} and requires, at the minimum, a basic understanding of the materials with a plan to convey them \cite{carberry2008learning}.
% reformulation - prep
The reformulation of this gathered material also leads to gaining a better understanding of it since this encourages organization and finding the basic structures of the content \cite{gartner1971children}. This makes the learned material more accessible in memory through associations with known knowledge \cite{zajonc1960process}. 
% explanation
Then the explanation phase allows the student to test how their mind reviews and reformulates information into knowledge---similar to how we consolidate thoughts by explaining them to friends \cite{duran2017learning}---
% and 
even when the listener is passive \cite{zajonc1966social}. In addition, the student who is teaching (i.e., the student tutor) participates in {\it reflective knowledge-building}, which leads to recognizing their areas of improvement, reorganizing their own knowledge, deducing errors, repairing them, and producing better explanations \cite{roscoe2007understanding}. 
% self-explanation
The explanation phase could also be verbalized without peers, which is known as ``self-explanation'' \cite{ploetzner1999learning}. It consists of the creation of inferences that are instantiations of the principles and definitions introduced in the learned material, thus helping the student with problem-solving and grasping a better understanding of the material \cite{chi1989self}. During the construction of self-explanations, learning happens through the identification of knowledge gaps which are necessary to learn in order to complete the self-explanation. This leads to constructive cognitive activities that usually culminate in the acquisition of new knowledge \cite{ploetzner1999learning}. More specifically, self-generated questions \cite{chi1994eliciting} can be used to help identify knowledge gaps. 
% link to others and goals
These mechanisms that are used during self-explanations should also be involved when explaining to others \cite{duran2017learning}. 
This paper aims to deepen our understanding of the LbT challenges university students face \newtext{in the contexts of online teaching and technology usage}, both when LbT alone and to someone else without guidance or instructions.
% \newtexttwo{Note that guidance and training for in-person LbT and peer tutoring may not be completely applicable to the online context. Therefore, by providing minimum guidance for the LbT sessions (e.g., not limiting the teaching material's domains) and minimum training to the participants, we aim to explore a wide range of challenges that university students may face while conducting LbT sessions online, including ones that may not be recognized and addressed if guidance and training were provided.}

\subsection{\bf Challenges with LbT in the Classroom}

\newtext{
A bias students often face with carrying out LbT is the knowledge-telling bias, which is ``a reliance on knowledge-telling to the exclusion of knowledge-building'' \cite{roscoe2014self}. 
A plausible reason for this is the expertise hypothesis; in other words, ``tutors may not engage in much knowledge-building because such reasoning is largely beyond their capabilities'' \cite{roscoe2014self}.
Prior research found several remedies to overcome knowledge-telling bias, including drawing while explaining learning outcomes \cite{Fiorella_2020}, structured interactions with peers \cite{King_1998}, providing training \cite{McNamara_2004}, or self-explanation prompts \cite{Berthold_2008}. 
Besides that, Matsuda et al. found that students might exhibit the biased rehearsal effect in teachable agent contexts (i.e., where students teach a computer agent(s) as a way to LbT), where students might repeatedly use only similar and easier problems when teaching an agent, resulting in an increase in problem solving accuracy, but no improvements in pre-post test scores \cite{matsuda2011learning}.
To contribute to current research gaps, the challenges and biases during the various stages of LbT is the main focus of this paper, instead of an incidental research finding complementing the investigation of certain LbT interventions or tools.
% Instead of exploring the effectiveness of certain LbT interventions or incidentally finding evidence of certain challenges or biases, this work aims to focus on investigating challenges and biases that university students face in the context of online teaching and technology usage through the various stages of LbT.
% Instead of incidentally finding evidence of certain challenges or biases when exploring the effectiveness of certain LbT interventions or tools, this work aims to focus on the exploration of the challenges and biases that university students university students face in the contexts of online teaching and technology usage through the various stages of LbT.
}

As previously outlined, LbT can be done by teaching peers, and is hence closely related to peer tutoring, which is a building block of LbT that has been studied in the following configurations: teaching to one peer, teaching to a group of peers, and teaching with a group of peers (e.g., each member learns to become an expert at one specific topic \cite{aronson2011cooperation, slavin1980cooperative}, this is also known as cooperative learning). In general, students who have mastered the material previously assist the development of understanding for struggling students \cite{carberry2008learning}. Peer tutoring typically occurs in an environment where one student plays the role of someone who has the knowledge or skills of the materials to be taught, and others play the role of the learner or tutee \cite{damon1989critical}.
% cooperative learning or taking turns teaching/ teaching with a group 
% The best known and most researched cooperative learning method is undoubtedly the Jigsaw technique (Aronson & Patnoe, 2011; Slavin, 1995). In Jigsaw, each member of the team learns to become an expert on a specific topic that is only a part of the whole that they need to learn in order to achieve the didactic goal; and then share it with their teammates, thus learning-by-teaching. https://journals-scholarsportal-info.proxy.lib.uwaterloo.ca/pdf/14703297/v54i0005/476_leaiaapm.xml
% more on peer tutoring
% Additional studies examine the benefits of explaining materials in the context of understanding its effects on the teacher in peer tutoring and cooperative learning. Peer tutoring is the most prevalent form of peer-assisted learning, where students who have mastered the material previously assist the development of understanding for struggling students \cite{carberry2008learning}. Peer tutoring typically occurs in an environment where one student plays the role of someone who has the knowledge or skills of the materials to be taught, and others play the role of the learner \cite{damon1989critical}. Benefits of peer tutoring include providing academic and motivational help for the learners, developing relationships between students, and engaging the teacher. \cite{topping1988peer}.
The configuration of student pairing could impact their learning outcomes.
% These configurations are important considerations since they can have a significant impact on the learning outcomes of the students. 
For example, if a particular setup fosters peer question asking, this can impact the tutor's understanding of the content. Roscoe's study showed how peers' questions are a significant predictor of the tutor's knowledge building and suggested further investigations of how each member's level of understanding in a LbT scenario affects the amount of knowledge building \cite{roscoe2014self}. 
% It was also suggested to further investigate how the peers' relevant background knowledge impacts the knowledge building that can occur in a LbT session.
Some factors that can contribute to limited question asking include an inability to identify their own knowledge gaps \cite{graesser1994question}, and various social influences \cite{Engel2011Need} such as the fear of peers' negative judgment \cite{post2019development}.
The quality of the LbT sessions is not only impacted by peers' question-asking abilities but also by students' personalities and cognitive styles \cite{miller1994group}, motivation levels \cite{rienties2009role}, and group dynamics in terms of contribution equity \cite{shah2014analyzing}.
% The potential to improve the quality of the LbT sessions is not limited to matching students with the correct expertise level but peers' personality \cite{}, cognitive style \cite{}, motivation levels \cite{}, and group dynamics \cite{} can also play a role in the effectiveness of this learning method.
For example, when analyzing equity in collaborative learning discussions among students, those who perceived themselves as less competent in the subject at hand were dominated by their peers who perceived themselves as more competent, resulting in reduced opportunities to learn \cite{shah2014analyzing}.
% variability with tutor and tutee 57 73
%  why pairing are important to consider: ``Roscoe’s study found that human tutee’s questions were significant predictors of human tutors’ knowledge building and deep understanding [71]" ... `` Roscoe suggested future studies to investigate how the absolute and relative levels of tutor and tutee expertise affect the amount of knowledge building" --> understand what factors play a role in generating the best pairing
% group dynamics: ``, it is not well understood how a group’s dynamics affect each tutor’s learning. Within collaborative learning literature, researchers found that its effectiveness is affected by students’ personality (e.g., extroverted vs. introverted), cognitive  style [58], motivation levels [69], and the group’s contribution equity [77]. For instance, Shah etal. [77] found that students who perceived themselves as being less competent in the subject were dominated in discussions within dyads, and as a result, were negatively affected in their learning process. These results raise questions of how teachable agent systems should be designed to manage group dynamics, e.g., encourage or enforce equal engagement and contribution among tutors"
Even though there is some understanding of how peer groupings can impact LbT in various classroom settings, studies observing university-level participants have been done only without computer-mediated support. There is also a lack of understanding of how technology can play a role in supporting these challenges to generate better knowledge-building during LbT among this age group. While there have been computer-aided studies done with children students who were in the same class (e.g., \cite{biswas2004incorporating, leelawong2008designing}), learning processes and collaboration evolve with age, and could impact the design of LbT tools. More specifically, among adults, the frequency of executively controlled learning increases \cite{kuhn2006children}, metacognitive skills such as knowledge retrieval skills followed by failed attempts improve \cite{flavell1975metamemory}, and frequency of applying knowledge acquisition strategies is higher \cite{kuhn2000metacognitive}.
% find more about how group dynamics can change between these age groups. 
%  these studies were done with children who were in the same classes. how do these results change with adults in different programs or universities to make lbt more accessible... since ``the learning process itself evolves with age" p6... existing relationship between participants.. rapport background knowledge... -> understanding the implications can help design the appropriate configurations for adults (recall: impression management, bandwagon effect, ambiguity effect) 
By gaining a better understanding of the design implications for adult collaborative LbT, we aim to support the potential benefits of peer tutoring such as providing academic and motivational help for the learners, developing relationships between students, and engaging the teacher \cite{topping1988peer}.
% form prev section:dThe explanation phase allows the teacher to test how their mind reviews and reformulates information into knowledge, similar to how we consolidate thoughts by explaining them to friends \cite{duran2017learning}, and even when the listener is passive \cite{zajonc1966social}. Furthermore, the teacher participates in \emph{reflective knowledge-building} which leads to recognising their areas of improvement, reorganising their own knowledge, deducing errors, repairing them and producing better explanations \cite{roscoe2007understanding}. Occasionally, the explanation phase has also been done without peers which is known as ``self-explanation" which can be verbalized \cite{ploetzner1999learning}. It consists of the creation of inferences that are instantiations of the principles and definitions introduced in the learned material, thus helping the student with problem-solving and grasping a better understanding of the material \cite{chi1989self}. During the construction of self-explanations, learning happens during the identification of knowledge gaps which are necessary to learn in order to complete the self-explanation. This leads to constructive cognitive activities that usually lead to the acquisition of new knowledge \cite{ploetzner1999learning}. More specifically, self-generated questions \cite{chi1994eliciting} can be used to help identify knowledge gaps. These mechanisms that are used during self-explanations should also be involved in explaining to others \cite{duran2017learning}.

\subsection{\bf Peer Tutor Training}\label{sec:training}
% TLDR: since we wanna know about challenges in the teaching process of peer LbT, one proxy is to see what peer tutor training aims to impart in inexperienced tutors.

Related to LbT is the concept of peer tutoring, defined as ``people from similar social groupings who are not professional teachers helping each other to learn and learning themselves by teaching'' \cite{topping1996effectiveness}.
Specifically, as part of this paper, we aim to investigate challenges faced by university students in their teaching process in the context of LbT. Although no existing works answer this question directly, an indirect way to understand possible challenges, beyond those discussed above, is through the motivations and structures of peer tutor training programs. Since peer tutor training programs aim to equip peer tutors with skills relevant to teaching their peers, it is possible that students who did not go through such training would face similar issues while teaching in a slightly different context---LbT; while peer tutoring aims to educate both tutor and tutee \cite{topping1996effectiveness}, LbT focuses on the student's (i.e. tutor) learning \cite{duran2017learning}.

% what happens when there is no training
When peer tutors are not trained, their tutoring might involve questioning behaviours that are limited in frequency, infrequent error corrections, and providing unsuitable feedback (i.e., uninformed and unconstructive \cite{huff1987contemporary}) \cite{graesser2014evolution}. In the context of writing, untrained peers were found to spend much time discussing the essay's subject without relation to the actual writing \cite{george1984working}.
% what to train about
Given the importance of peer tutor training, researchers have expressed diverse opinions on what peer tutors should be trained on, including: the content intended to be taught by the tutors \cite{mckellar1986behaviors,fremouw1978peer}, the
knowledge to apply tutoring practices rooted in theories \cite{Murphy2006}, interpersonal skills (e.g., friendliness, rapport) \cite{shamoon1995critique,reigstad1984training}, the ability to deal with high-order concerns before lower-order concerns \cite{reigstad1984training}, their metacognitive skills, explanatory potential and awareness of behaviours related to the tutor's role (e.g., to observe the tutee attentively) \cite{ensergueix2010reciprocal}, proper peer criticism practices, and awareness of potential learning difficulties tutees might face \cite{bruffee1980two}. Research also found the importance of having qualified trainers lead these training programs \cite{choi2011peer}.
% training outcome
Training has been found to lead to many benefits, including greater learning gains and cognitive benefits for both tutors and tutees \cite{sharpley1981,choi2011peer}, increased task engagement and commitment \cite{stanley1992coaching}, the provision of higher quality feedback (e.g., addresses both low and high-level concerns) \cite{stanley1992coaching,choi2011peer}, a deeper understanding of the material \cite{berg1999}, and more accurate self-efficacy beliefs \cite{ensergueix2010reciprocal}. As such, peer tutor training has been recognized as being a core part of organizing peer tutoring programs \cite{Topping2005,ensergueix2010reciprocal,bruffee1980two}. Interestingly, Robin and Heselton found interactive tutor training to produce higher quality tutoring behaviour than training tutors using only a written handbook, albeit having no differences in tutee learning outcomes \cite{robin1977proctor}.
% , especially since the magnitude of these benefits is affected by how structured the programs are \cite{Cohen, Kulik and Kulik (1982) }.
Since these studies have been conducted outside the context of online LbT, we aim to fill the gap by first understanding if and where there are specific areas that university students may need training or support with their teaching approach, which can provide design implications for supporting such training.

\subsection{\bf Computer Support for LbT }

%  ``Existing learning-by-teaching research involving teachable agents (i.e., agent tutees) are mostly for younger children (usually in elementary school) [9, 24, 52, 53, 60, 87]. Research on learning by teaching for older students (e.g., university students), on the other hand, mostly investigated human tutor/tutee contexts without the use of teachable agents [2, 19, 55, 71, 72]." ->> or web based technology? 

Existing research on educational technology has investigated multiple types of software that are aimed at supporting the processes in LbT. 
% discussed in Section \ref{sec:pedagogical}.
More than two decades ago, Kumar et al. introduced the term Virtual Learning Environment (VLE), which allows students to ``access a complete course, take tests, and interact with the professors as well as classmates'' \cite{kumar1998virtual}. Another type of support technology is computer-supported collaborative learning (CSCL), which is a paradigm of educational technology that focuses on using technology to support peer interaction, and sharing of knowledge and expertise in a collaborative learning environment \cite{kumar1996computer}.
% The third relevant type of technology is groupware, which is any computer-based system that supports group collaboration on a common task \cite{zhang2012human}. % TODO check this citation
Some educational technologies also incorporate the use of virtual agents, often conversational, in various educational settings \cite{biswas2005learning, biswas2004incorporating, anderberg2013exploring, matsuda2013studying,mehdi20,ravarieffects,cehaHumour,Lee2021}. 
% Moreover, there exists frameworks and studies on peer tutoring support tools that are relevant. 
We discuss how these technologies support different parts of the LbT process below.

% \noindent {\bf Support for Preparing Materials. }
\subsubsection{Support for Preparing Materials. }
Cheng and Yen introduced a VLE that supports course instructors in producing course material using a ``Preparation Room'' \cite{cheng1998virtual}. In a comparative study of ten VLEs, Al-Ajlan observed that VLEs support the process of designing curricula with various features \cite{al2012comparative}. These features include course templates, instructional design tools, and ways for sharing and reusing content \cite{al2012comparative,akobe2019web}. However, these systems are meant to be used by instructors, and not peer tutors.
In designing a peer-tutoring orchestration tool, Phiri et al. suggested three specific tasks, including activity management (for specifying metadata associated with the teaching activity), resource management (for uploading and organizing materials) and activity sequencing (for specifying the order of materials according to the teaching activity) \cite{phiri2017peer}. Moreover, multiple types of materials were supported, be it PDF documents, videos, or audio files for increased flexibility. Although this tool was evaluated using student tutors, participants did not prepare their own materials; official course materials were used instead \cite{phiri2017peer}. 
Within the peer-tutoring literature, Walker et al. extended the Cognitive Tutor Algebra (CTA) system to support peer tutoring and help prepare tutors by first asking both tutor and tutee to individually solve an equation and then providing questions to the tutor to help them prepare tutoring questions (e.g., “A good question asks why something is done, or what would happen if the problem was solved a certain way.” \cite{walker2008tutor}).
Our work aims to contribute to this by investigating how to design tools supporting material preparation, specifically in peer-to-peer LbT contexts. This is especially important since i) peer tutors usually have inadequate tutoring expertise without proper support or training \cite{ensergueix2010reciprocal}, and ii) in-the-wild peer-to-peer LbT might involve materials and topics prepared by peers, instead of predefined alternatives (e.g., existing course materials). 

\subsubsection{ Support for Explaining Materials. }
% Beyond teachable agents, TALK ABOUT SCAFFOLDING IN PEER TUTORING
Many learning environments with teachable agents also have features aimed at supporting the process of explaining educational materials. For instance, Betty's Brain helps students ``develop structured networks of knowledge that have explanatory value'' through the activity of building concept maps that relate causal effects between concepts (e.g., for concepts relevant to river ecosystems) \cite{biswas2005learning}. Curiosity Notebook, on the other hand, provides conversations of different types, including one aimed at asking the student to generate explanations for relationships between entities and characteristics (e.g., if an animal lays eggs) \cite{Lee2021}.
In peer tutoring contexts, Walker et al. designed an adaptive system that supports tutors when ``explaining a problem-solving step using a domain support'' to tutees \cite{walker2014adaptive}. It prompts the tutor to provide extra conceptual explanations when detecting tutor chat messages that lack conceptual reasoning. It also sends encouraging messages to tutors when they successfully explain concepts while tutoring. 

\subsubsection{Support for Interacting with Peers. }
The VLE introduced by Cheng and Yen supports interactions between students, and also between students and teachers so that students can get organized feedback \cite{cheng1998virtual}. Discussion forums are a common feature used by VLEs for supporting student interactions \cite{al2012comparative}. 
Similarly, CSCL enables a diverse group of collaborators to contribute their individual opinions and knowledge in a project \cite{zhang2012human}, hence allowing interactions between learners, teachers and peers. 

Peer tutoring tools could support peer interactions by allowing tutees to request tutoring sessions from specific tutors for a particular course according to tutors' self-reported expertise \cite{akobe2019web}. 
Westera et al., on the other hand, built an online self-organized peer allocation mechanism to pair peers for peer tutoring purposes \cite{Westera2009}. Instead of manual requests by students, the system automatically selects the most appropriate peer, based on tutor competency and workload fairness (i.e., workload distribution should be fair over all competent tutors). Based on a pilot study, they made several suggestions; specifically, such a system is more suitable for online learning contexts where there are a large number of students (100 or more) who do not know each other. Moreover, the system should ``foster group awareness and community feeling'' \cite{Westera2009}.
To investigate methods of supporting peer tutors in the context of learning algebra on CTA, Walker et al. built two methods of support---fixed and adaptive domain support \cite{walker2008tutor}. Specifically, peer tutors could see tutees' attempt at solving an algebraic equation live, and interact with tutees by either marking the tutees' steps (in solving the problem) as correct or wrong, adjusting the value of the tutees' skill bars, or via the chat tool. The fixed domain support provides tutors with the problems' answers, while the adaptive version has additional features, including i) providing tutors with hints if tutees ask for them, and ii) highlighting the correct answer if tutors mark their tutees incorrectly (e.g., marking a wrong step as correct). While no significant differences were found between the two types of support, they found that tutors benefited from tutees' impasses, which were negatively correlated with the tutees' learning gains \cite{walker2008tutor}.
% TODO look into walker2014adaptive
Support systems could also support tutors in their ability to provide help that is timely (i.e., providing help when tutees make errors or request for help) and appropriate (i.e., prompt tutees to self-explain and provide appropriate feedback for errors) \cite{walker2014adaptive}.

Interestingly, benefits can be gained from interacting with human students and with teachable agents. Systems like Betty's Brain and SimStudent contain agents that provide feedback to users either via conversations or examples (e.g., similar mathematical algorithms) \cite{biswas2005learning,matsuda2013studying}. Processes like administering quizzes for the purpose of gauging learning progress have also been simulated using agents \cite{biswas2005learning,Lee2021}. Ravari et al., on the other hand, used an adaptive teachable agent to encourage equal participation within student-teacher dyads and found it to be effective \cite{ravarieffects}.

% \subsection{Support for Self-explanation \& Self-reflection.}
\subsubsection{Support for Self-Reflection.}
Self-reflection can be supported with reflexivity tools in VLEs that allow for reflections on students' own learning processes \cite{cheng1998virtual}.
Moreover, teachable agent systems, like Betty's Brain, can be used to support ``the development of reflection or meta-cognitive skills'' by using a mentor agent that directs users to reflect on specific parts of the materials based on their concept maps \cite{biswas2005learning}.

Another aspect of self-reflection for peer tutors is to reflect on not just the content, but also their teaching approach. Walker et al. added support for tutor reflection in CTA by asking three questions after their tutoring session (e.g., ``What was the best question asked by the tutee? If the tutee didn't ask any questions, what was a good question he/she could have asked?'') \cite{walker2007student}. They found that the reflection questions made tutors skip less difficult questions and spend more time solving each question, even though no additional learning gains were observed. 
Moreover, researchers have found the use of sentence classifiers for labeling the content of peer interactions to be beneficial in encouraging peers to reflect on what collaborative activities are most suitable \cite{weinberger2005epistemic}, and have implemented them in peer tutoring support systems \cite{walker2014adaptive}.

\bigskip

In summary, prior work has shown that the various stages of LbT unfamiliar material can be supported through computer systems. 
% \newtext{Moreover, such tools have been shown to support a wide range of age groups, including elementary school \cite{Lee2021}, middle school \cite{biswas2005learning}, high school \cite{matsuda2011learning,walker2008tutor} and university \cite{cheng1998virtual} students. }
However, there exists a lack of support tools built specifically for LbT in a peer-to-peer context without any required instructor involvement; our work aims to build the foundation for such tools that are specifically designed for university students.
\section{Study Design}\label{sec:studydesign}
%======================================================================

We conducted a study to gain a new understanding of how university students approach teaching unfamiliar material, what the individual differences are in their processes and their struggles, and how to support this process and mitigate these challenges using web-based tools.
% removed for Aug 22: a) how university-level students naturally teach content when asked to learn by teaching and what challenges do they face, b) what kind of tools and scaffolding would be required to support this process on a learning-by-teaching platform, and c) what kind of implications would there be when forming a virtual learning-by-teaching community. 
The study involved two LbT sessions, where the participants taught someone \newtexttwo{(i.e., an investigator)} and themselves, followed by a semi-structured interview. \newtexttwo{The use of an investigator in the first scenario is to ensure consistent tutor-tutee interactions across participants \cite{roscoe2007understanding}.} \newtexttwo{Whereas, the design choice to investigate the latter scenario is informed by past research that found LbT to be effective through self-explanations~\cite{duran2017learning}, in addition, this could provide more flexibility for students.}
\newtext{Below, we present the research questions, and} explain the tasks that participants were asked to perform during the study.
% ``\section{Building a Sense of Community in Online Learning-by-Teaching} During the interview, we asked participants the questions ``If there was a feature that enabled you to upload your teaching video with other students who are learning the same content, what do you think the advantages and disadvantages are of having this kind of community of students sharing videos?" --> create goal from this section

% and specific hypotheses driving 

% {|rll|} old format
\begin{table}[htbp!]
\small
\caption[Interview Question Guide]{The sequence of questions used to guide the conversation during the interview.}
\label{tab:interview-script}
\begin{tabular}{|r p{0.92\linewidth}|}
\hline
{\textbf{\#}} & \textbf{Question}    \\ \hline
\textbf{1}    & How satisfied were you with the teaching?    \\
\textbf{2}    & How difficult was the teaching task?    \\
\textbf{3}    & How would you compare your experience teaching someone vs teaching alone?    \\
\textbf{4}    & Why did you choose those particular two pieces of materials to teach?      \\
\textbf{5}    & Did you do any preparations, why or why not?  What kind of preparations did you do? How much time did you spend preparing?\\
\textbf{6}    & Have you heard of “learning by teaching” and what does it mean to you? Have you tried to teach someone in order to better learn something for yourself?    \\
\textbf{7}    & Do you feel more comfortable teaching while interacting with a partner or alone? Why? \\
\textbf{8}    & Do you feel you learn more during the process of teaching someone, or do you feel you learn more while teaching alone, why?   \\
\textbf{9}  &  If we were to design a tool to support learning by teaching, would you prefer that this tool would let you teach alone and reflect on the materials, or connect you with a partner so that you can teach/interact with them? \\
\textbf{10}   & If you want to have a partner when teaching, who would you want that partner to be, who are you most comfortable with when you are teaching?  Would you choose to teach a student who already understands the content or not? Why?    \\
\textbf{11}   & What support would you like to have when you are learning by teaching? If there was a tool that supported learning by teaching, would you use it?    \\
\textbf{12}   & Imagine now you have a tablet, phone or web app for learning by teaching, would you feel comfortable with using [feature in list of features], why or why not? Would you use [feature in list of features], why or why not?    \\
\textbf{13}   & If there is a feature that allows different configuration for learning by teaching, which version would you prefer?  Rank in order of preference: a) Synchronous and asynchronous b) teaching to a group, an individual, with a group (students in a group taking turns teaching others), alone (teaching to yourself), and to a virtual agent (e.g., chatbot).    \\
\textbf{14}   & If there was a feature that enabled you to upload your teaching video with other students who are learning the same content, would you be comfortable with that, why or why not? What do you think the advantages and disadvantages are of having this kind of community of students sharing videos? How would you feel about uploading a PDF file, text or images?  \\
\textbf{15}   & If you do another learning by teaching session again (like repeat what we have done here), would you do something differently? What would it be? How would you explain it more effectively?    \\ \hline
\end{tabular}
\end{table}

\subsection{Research Questions}
This work investigates the following research questions: 
\begin{itemize}
  \item \textbf{RQ1: } Which challenges do students face during the LbT process and where do they happen?
  \item \textbf{RQ2: } Which tools can support the challenges students face when LbT?
\end{itemize}

\subsection{Procedure}

% \newtext{I. Teaching Preparation.}  
\indent\textbf{I. Teaching Preparation.}
Prior to the study session, each participant was asked to prepare to teach two university assignments or quiz questions that should each take about 5 minutes to teach. The only requirement was that each question should not come from the same course. \newtext{In order to determine if the material was indeed unfamiliar or not fully familiar to them, we asked participants to only select questions that they did not receive full marks on.} Participants were told that they could teach in whichever way they want, and optionally use whichever tools they would like (e.g., whiteboard, Miro, paper and pencil, images, diagrams, and extra devices or webcams are allowed in the video call). They were also informed that their teaching does not need to be perfect, as we are only interested in what they naturally do. \newtext{If any students followed up with questions about the expectations we had for this study, we reiterated that the aim of the study is to observe how students teach unfamiliar material. They were reminded of this at the beginning of their sessions as well.} \newtexttwo{We decided against having participants teach completely unfamiliar materials (e.g., materials they have never seen before) as the current setup allows us to investigate the more realistic scenarios of university students using an online LbT platform for courses they are taking at the time \cite{gurung2010focusing}.}

\newtext{We chose not to provide any resources or guidance for the participants' teaching preparation; this is because the goal of the study is not to demonstrate the merit/success of LbT as a method. Instead, it is to observe the type of preparation students are inclined or willing to take before teaching, and the variety of challenges they may encounter in the entirety of the LbT process (i.e., in the absence of explicit support or instructions on how to teach, what would students struggle with the most). 
% \newtexttwo{This type of unstructured teaching within LbT is commonly researched \cite{Ehly_1987,Cohen_1982} and is more realistic and natural. Particularly, most university students are not trained in teaching, and although tutor training is beneficial for LbT, making it a requirement causes LbT to be less accessible in real-world contexts.}
Similarly, the use of material that was unfamiliar to students allows the study to explore ``the haziness of the situations in which [students] learned ... when teaching'' \cite{leikin2006learning}.} 
\newtexttwo{Using participants' previous assignment mistakes as the topics to teach also points towards a possible real-world context where LbT could be beneficial, since university students often fail to learn from their assignment and examination mistakes \cite{yerushalmi2006guiding,Mason_2010,Henderson_2009}. }

% ``If you decide to participate, please bring two homework/quiz/exam questions from TWO of your courses, that you got wrong or want to have a better understanding, and teach it to us. The only requirement is the two questions should NOT come from the same course. You can teach in whatever way you want and use whatever tools you want (e.g., whiteboard, Miro, paper and pencil, images, diagrams, and extra devices or webcams are allowed in the video call), the use of tools is not required and are optional. Your teaching does not need to be perfect, we are only interested in what you naturally do. Before teaching, there will be a pre-study survey asking you for your demographic information (e.g., age, gender, major, etc.) as well as prior experience with teaching. Teaching each question should take about five minutes. After that, you will fill in a survey and the researchers will ask you a few follow-up questions. The whole study takes about 1 hour - 1 hour and 30 min."

\textbf{II. Pre-Study Questionnaire. }
First, a pre-study questionnaire was digitally administered to assess the students' demographic information, their prior experience with teaching, their perceived confidence and comfort level to teach each piece of material they prepared for the study, and a brief overview of the topic or subject they plan to teach for each session. The section on prior experience in teaching captured data about their self-perceived experience level in teaching and the areas of teaching that they have experience in (e.g., marking, holding tutorials, creating course material, giving lectures, holding offices hours, tutoring, teaching outside of school, etc.) to gain a better understanding of their experience, and their confidence level in teaching in general.

\textbf{ III. Explanations. }
At the time of the study session, the student joined a video call and was given up to fifteen minutes to teach each piece of material. For one explanation, they were allowed to interact with the investigator by asking questions, assessing visual and auditory feedback (or in whichever form they chose). For the other explanation, the student also taught their prepared material, except this time they could not interact with the investigator. For this part of the study, the investigator's left their camera on, left the study room, and returned after fifteen minutes, to show that deception was not used. 
\newtexttwo{These two LbT settings (teaching while having a live tutee and having no live tutee) have been shown to be effective \cite{Duran_2016,roscoe2007understanding,Hoogerheide_2016,Hoogerheide_2014,Fiorella_2013,Fiorella_2014}.}
The first 12 participants taught with the investigator present and then taught alone, while the rest were instructed to teach following the reverse order to reduce bias and carryover effect. The explanations were done over Zoom where they had the ability to draw, import documents, screen share, or use other features if they chose to.
\newtexttwo{Although different types of teaching (e.g., drawing vs. not drawing) have varying effectiveness in LbT \cite{Fiorella_2020}, not restricting the tools used during LbT allows us to gain insights into participants' preferences and the barriers associated with them.}
% \newtexttwo{Not restricting what tools could be used while teaching follows similar procedures in existing works \cite{Ehly_1987} and allows us to gain insights into participants' preferences and barriers associated with them.}
\newtexttwo{Note that guidance and training for in-person LbT and peer tutoring may not be completely applicable to the online context. Therefore, by providing minimum guidance for the LbT sessions (e.g., not limiting the teaching material's domains) and minimum training to the participants, we aim to explore a wide range of challenges that university students may face while conducting LbT sessions online, including ones that may not be recognized and addressed if guidance and training were provided.}

% At the beginning of the study the participants were told that they would have up to fifteen minutes to teach each prepared material and that they did not need to use up all the given time. For the one piece of material they were allowed to interact with the investigator, and for the other piece of material they would be teaching out loud, to themselves.

\newtext{
\textbf{ IV. Approach of the Investigator. }
During the sessions where the investigator was present, the investigator remained as consistent as possible with the way they interacted verbally and non-verbally with participants. If a participant asked the investigator a question about how they should approach the task, the investigator reiterated the instructions they received and told them to use any method of their choice. Throughout the explanations, the investigator would occasionally nod to indicate that they were following along. If a student chose to engage with the investigator by asking them skill-testing questions, the investigator would attempt to answer the skill-testing question. If the investigator was asked about their academic background or expertise, they informed them that they had a basic to foundational level understanding of their material for the majority of the sessions since they have taken a wide range of courses. Work from Roscoe et al. demonstrates that tutees play a significant role in peer teaching outcomes for the tutor \cite{roscoe2004influence}, however, since our aim is not to understand the effectiveness of LbT, the behaviour of the investigator can remain more consistent.}

% I am also concerned about the teaching to the study investigator option without providing the students info about their academic background, which might have affected their performance especially if they feel the superiority of the investigator. 
% Did the investigator remain passive in all the sessions. 
% It seemed in some sessions, the students asked questions but there is not enough details about the investigator responses in all the sessions, and whether it has affected the students. (R1)
% One participant mentioned that receiving more validation made her feel more confident.

% The fact that this is not a peer teaching experiment makes it a bit easier to analyze because the behaviour of the investigator is somewhat consistent (e.g., \cite{roscoe2004influence} demonstrates tutee plays significant role)
% https://escholarship.org/content/qt3g30r749/qt3g30r749.pdf

\textbf{ V. Interview and Post-Study Questionnaire. }
%\vspace{-1em}
% Interview questions:
% a guide or template to help you prepare content or lesson plans (try to get more on scaffolding)?
% a matching tool to connect you with students to teach or to volunteer to learn
% a whiteboard tool or a concept map building tool
% screen sharing functionality 
% presentation pointers
% lesson ratings only visible to you, as a teacher 
% public ratings
% sending out a feedback form to the student(s) you taught
% history tool to view previously taught content
% session recordings
% student’s questions
% student’s questions that were not answered
% Annotation tool
Once the participant finished teaching both explanations, the student participated in a semi-structured interview (Table \ref{tab:interview-script}) to assess their experience in teaching to someone versus teaching alone, their previous understanding and experiences of LbT, and various feature design questions about what they would like to see in a LbT platform to gain a better understanding of the requirements the platform would have to meet. Since the study was done over video calls, the post-study questionnaire was integrated into the interview and screen sharing was used to answer feature configuration ranking questions. \newtext{While it is well known that people are generally poor at predicting tool usage or preferences these questions were included in the questionnaire to help the participants imagine various LbT scenarios and configurations to ask further probing questions since these concepts may be unfamiliar to them. The direct responses to these questions do not derive the key findings of the study. Instead, the data was thematically analyzed for other themes describing their challenges.}

\subsection{Participants}
In total, 24 University of Waterloo undergraduate and graduate students participated (Table \ref{tab:interview-demographics}). The participants who had some experience with teaching (72\%) indicated gaining this experience from marking (n=5), holding tutorials (n=2), creating course material (n=4), giving lectures (n=1), holding office hours (n=3), answering questions on Piazza (n=7), and tutoring (n=14).
The participants were recruited using snowball sampling through emails, the HCI Lab SONA recruitment system (Appendix \ref{AppendixA} displays the study information published on the system), and social media networks (e.g., Facebook). In exchange for their participation, they were remunerated with a \$25 CAD Amazon gift card.

\begin{table}[htbp!]
\small
\caption[Participant Demographic Information]{Participant demographic information (i.e., gender, age, undergraduate (U) or graduate (G) level of studies, faculty, and major) and their teaching experience level (on a scale from no experience to some experience to lots of experience) and teaching confidence level (on a scale of 1 to 5, where 1 is low).}
\label{tab:interview-demographics}
% Please add the following required packages to your document preamble:
% \usepackage{multirow}
\begin{tabular}{l|l|l|l|l|l|l|}
\cline{2-7}
&\multicolumn{4}{c|}{\textbf{Demographics}}  & \multicolumn{2}{c|}{\textbf{Teaching}} \\ \hline
\multicolumn{1}{|l|}{\textbf{P\#}} & \textbf{Gen.} & \textbf{Age} & \textbf{U or G} & \textbf{Faculty} & \textbf{Experience} & \textbf{Confidence} \\ \hline
\multicolumn{1}{|l|}{{P1}} & F & \color{black}{20-21}\color{black}{} & U & Science & Some Experience & 4  \\ \hline
\multicolumn{1}{|l|}{{P2}} & F & \color{black}{20-21}\color{black}{} & U & Engineering & Some Experience & 2  \\ \hline
\multicolumn{1}{|l|}{{P3}} & M & \color{black}{20-21}\color{black}{} & U & Engineering & Some Experience & 4  \\ \hline 
\multicolumn{1}{|l|}{{P4}} & M & \color{black}{24-25}\color{black}{} & G & Engineering & Some Experience & 2  \\ \hline
\multicolumn{1}{|l|}{{P5}} & M & \color{black}{24-25}\color{black}{} & U & Engineering & Some Experience & 3  \\ \hline
\multicolumn{1}{|l|}{{P6}} & F & \color{black}{18-19}\color{black}{} & U & Mathematics & Some Experience & 3  \\ \hline
\multicolumn{1}{|l|}{{P7}} & F & \color{black}{18-19}\color{black}{} & U & Arts & Some Experience & 3  \\ \hline
\multicolumn{1}{|l|}{{P8}} & F & \color{black}{22-23}\color{black}{} & U & Science & No Experience & 2  \\ \hline
\multicolumn{1}{|l|}{{P9}} & F & \color{black}{22-23}\color{black}{} & G & Applied Health Sciences & Some Experience & 3  \\ \hline
\multicolumn{1}{|l|}{{P10}} & F & \color{black}{18-19}\color{black}{} & U & Arts & Some Experience & 4  \\ \hline
\multicolumn{1}{|l|}{{P11}} & F & \color{black}{22-23}\color{black}{} & U & Arts & No Experience & 2  \\ \hline
\multicolumn{1}{|l|}{{P12}} & F & \color{black}{18-19}\color{black}{} & U & Mathematics & Some Experience & 2  \\ \hline
\multicolumn{1}{|l|}{{P13}} & F & \color{black}{20-21}\color{black}{} & U & Science & Some Experience & 2  \\ \hline
\multicolumn{1}{|l|}{{P14}} & M & \color{black}{20-21}\color{black}{} & U & Arts & No Experience & 3  \\ \hline
\multicolumn{1}{|l|}{{P15}} & M & \color{black}{20-21}\color{black}{} & U & Mathematics & Some Experience & 3  \\ \hline
\multicolumn{1}{|l|}{{P16}} & F & \color{black}{22-23}\color{black}{} & G & Arts & No Experience & 1  \\ \hline
\multicolumn{1}{|l|}{{P17}} & M & \color{black}{22-23}\color{black}{} & U & Mathematics & Some Experience & 3  \\ \hline
\multicolumn{1}{|l|}{{P18}} & F & \color{black}{20-21}\color{black}{} & U & Mathematics and Arts & No Experience & 2  \\ \hline
\multicolumn{1}{|l|}{{P19}} & F & \color{black}{22-23}\color{black}{} & U & Science & Some Experience & 2  \\ \hline
\multicolumn{1}{|l|}{{P20}} & F & \color{black}{22-23}\color{black}{} & U & Science & No Experience & 3  \\ \hline
\multicolumn{1}{|l|}{{P21}} & F & \color{black}{22-23}\color{black}{} & U & Engineering & Some Experience & 3  \\ \hline
\multicolumn{1}{|l|}{{P22}} & M & \color{black}{18-19}\color{black}{} & U & Engineering & Some Experience & 3  \\ \hline
\multicolumn{1}{|l|}{{P23}} & M & \color{black}{18-19}\color{black}{} & U & Mathematics & No Experience & 2  \\ \hline
\multicolumn{1}{|l|}{{P24}} & M & \color{black}{18-19}\color{black}{} & U & Arts and Mathematics & Some Experience & 5  \\ \hline
\end{tabular}
\end{table}

%\begin{table}
%  \caption{Characteristics of participating students}
%  \label{tab:freq}
%  \begin{tabular}{ccl}
%    \toprule
%    Characteristic&n = X\\
%     \toprule
%    Gender& \\
%    \midrule
%    \\Female&X\\
%     \midrule
%    \\Male&X\\
%     \toprule
%    Age\\
%    \midrule
%    \\range 1&X\\
%     \midrule
%    \\range 2&X\\
%    \midrule
%    \\range 3&X\\
%     \toprule
%    Program Department\\
%    \midrule
 %   \\program a&X\\
%     \midrule
%    \\program b&X\\
%    \midrule
%    \\program c&X\\
%  \bottomrule
%\end{tabular}
%\end{table}

% data: https://docs.google.com/forms/d/1YpIfmvKso3IZ79273mYYi8DCpJuzACb0IBGUI_l93uI/edit#responses
% \begin{figure*} [h]
%   \centering
%   \includegraphics[width=.5\columnwidth]{ct0.png}
%   \caption{General Confidence in Teaching Abilities}
%   \label{ct0}
% \end{figure*}

\begin{figure*} [ht]
  \centering
  \includegraphics[width=1\columnwidth]{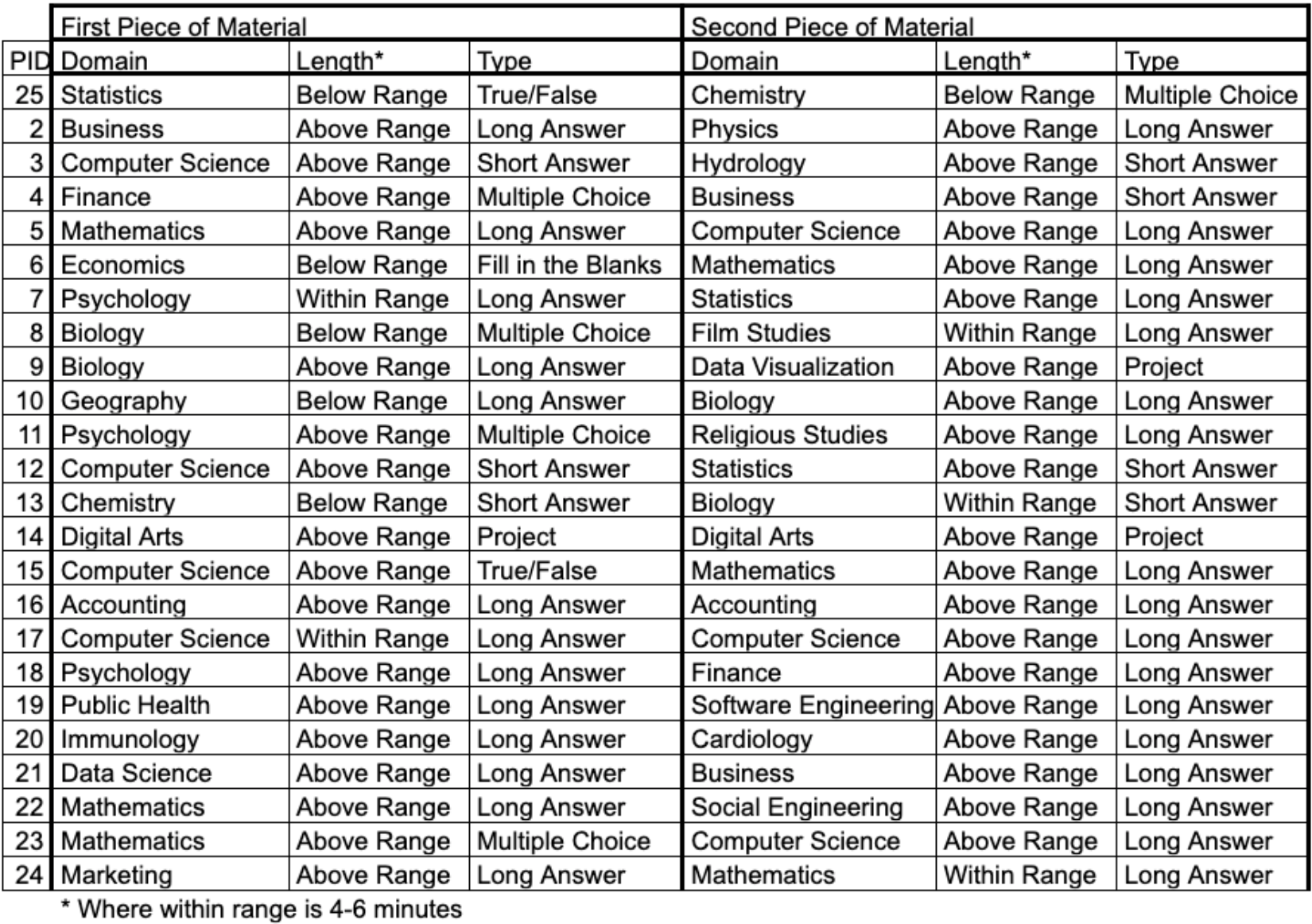}
  \caption{Domains and assignment format.}
  \label{tpcs}
\end{figure*}

\newtext{
As previously outlined, participants did not receive any restrictions on the topic they were going to teach since they come from diverse backgrounds and experience levels. Figure \ref{tpcs} illustrates the range of concepts participants chose to teach (e.g., statistics, finance, immunology, etc.), the assignment format (e.g., short answer, multiple choice, fill in the blank, etc.), and if they were able to complete their LbT session within the asked time range of about four to five minutes. Regardless of the topic that was taught and the assignment format, most students' sessions were above the given time range, and only five sessions were completed within the time range. Since this is an exploratory study and we are not trying to draw conclusions about the effectiveness of LbT, we aim to see a diverse set of participants and concepts for teaching, which gives us insight into what kinds of challenges may occur across a range of domains and tools used to teach. While this is not enough data to drive conclusions about how a participant’s background influences their choices, there has already been work done on the relationship between the material chosen and the LbT effectiveness. For example, tutoring has been observed in reading, mathematics, science, and other domains where the tutor is able to learn regardless of the subject, but mathematics and science programs may have higher gains than reading ones \cite{roscoe2007understanding}. }

% https://www.researchgate.net/profile/Rod-Roscoe/publication/252722912_Understanding_Tutor_Learning_Knowledge-Building_and_Knowledge-Telling_in_Peer_Tutors'_Explanations_and_Questions/links/55bfa9d308aed621de139d71/Understanding-Tutor-Learning-Knowledge-Building-and-Knowledge-Telling-in-Peer-Tutors-Explanations-and-Questions.pdf

\section{Analysis}\label{sec:analysis}
% notes: this could be changed to inductive content analysis 
% notes: might be missing: more explanation on why we observed teaching someone vs teaching alone

The LbT and interview sessions were screen-captured and transcribed to facilitate the analysis. The data was thematically analyzed \cite{braun2006using} by three researchers. Two researchers created an initial coding scheme to categorize the teaching sessions and interview responses. Once a consensus was reached, all data was independently coded by one researcher. A third researcher reviewed 25\% of the participants' interviews and LbT sessions by random selection. With the result, we aim to provide a breakdown of what happened during the LbT sessions when university students taught unfamiliar material without guidance. The key primary and secondary codes included:
\begin{itemize}
    % \item \textbf{Participant} characteristics of the participant: language fluency, humour, nervousness, hesitation
    \item \textbf{Teaching Task} (i.e., properties of the LbT task): domain, length and type (of question)
    \item \textbf{Teaching Approach} (i.e., properties of the teaching method): introduction, conclusion, past errors, current hesitations, tips, key points, reiteration, definition, importance, examples, analogies and similes 
    \item \textbf{Tools} (i.e., tools chosen by participants to aid the teaching): screen share, notes, references and visuals 
    \item \textbf{Interaction} (i.e., types of exchanges between the participants and the investigator): check-ins and assumptions
\end{itemize}
More detailed charts with all the codes can be found in Appendix \hyperref[AppendixA]{A}. Codes that ultimately did not provide further insight to our research questions were omitted from the results.

\newtexttwo{The codes that are included describe what occurred at each stage of LbT while also describing which tools and ways participants chose to use. We noticed that there were various barriers found across the Teaching Task, Teaching Approach, Tools, and Interaction codes that shared similar themes on a higher level. These themes could then be used to appropriately address the research questions in such a way that is digestible to understand when applying to future LbT scenarios and support tools. The codes} were synthesized into the following themes: psychological challenges relating to self, psychological challenges relating to others, and know-how challenges.

\section{Findings}\label{sec:findings}
%  psychological, social, know-how and technical
% note the highlight have not been updated but they are more or less similar
% Cone - social
% Ctwo - psychological 
% Cthree - technological
% Cfour - Know-How

% Start with what we think the psychological challenge is

% 1. ``the first psychological challenge is this...''  
% 2. ``here are some evidence, e.g., when participants are asked question X they mentioned Y'' ... 
% 3. ``more about this type of psychological challenge''

% Alt: open with really good quotes then explain. 

% End: What we think the psychological challenge is

% The proportion of ``in-our-own-words summary'' to ``quotes'' should something like 60-40 or maybe even 70-30.  basically, avoid a dump of quotes (especially when they are not very powerful or illustrative

% prestudy survey: comfort/confidence
% poststudy: difficulty and satisfaction

% prep: choosing the correct level of difficulty: fear of teaching it wrong, over confidence, motivation (disliking a topic but knowing it is important to learn) -> too hard when teaching or not prepared, over preparing due to lack of confidence
% teaching: nervousness (speaking too fast, quietly, mumbling, stuttering, laughing at their explanation) using humour
% reflection: 

\subsection{Psychological Barriers Relating to Self}
% https://www.sciencedirect.com/science/article/pii/S0065260108604074
% https://journals-scholarsportal-info.proxy.lib.uwaterloo.ca/details/03090590/v35i0005/420_litpzsfmla.xml
%  http://www.behavsci.ir/article_81286_397a2c0a04404711c7e997e7f54e1bd4.pdf
%  https://thedecisionlab.com/biases/ 

Below, we explore how psychological barriers relating to self can impact the way university students select what they will be teaching, their teaching approach, as well as the amount of time and effort spent in teaching preparations.

\subsubsection{\bf Selection of Teaching Materials \& Approach.}

We observed two psychological barriers relating to self, cognitive dissonance and zero risk bias, that can both impact the kind of materials and teaching approach university students may select when using LbT. Below we outline what barriers these are, how they come into play while LbT, and the way they can negatively impact potential learning gains when using LbT. 

% Jack Brehm was the first to investigate the relationship between dissonance and decision making in 1956, psychologist Leon Festinger was the first to formulate it into a theory of social psychology. In his seminal book published in 1957, A Theory of Cognitive Dissonance, Festinger details his theory and points to its influence in the psychology of learning.
When LbT, the goal of the student is to learn by preparing and teaching material. During this process, they have to select materials to teach---either materials they already were mostly comfortable with (which could be less intimidating but present fewer opportunities for learning) or materials they are unfamiliar with (which could be more intimidating but present more opportunities for learning). Many of our participants seem to be attached to the idea that the lessons need to be accurate. There is an inherent and perhaps deeply ingrained concept of need to be correct and perform well in comparison with others that makes LbT (i.e., the idea of being wrong, then using feedback to improve) foreign and counter-intuitive to students. In particular, when we asked participants if they have heard of or used LbT before, several participants mentioned that they tended to teach things that they are either familiar with or confident about (P4-5, P11-13), and ask for help on the parts that they do not know well (P2, P4-5). P13 explained, ``I feel like when you learn by teaching, it tends to be content that you're more comfortable with, which kind of defeats the purpose.'' P4 also mentioned having used this method for courses they know well because they believed that they would otherwise ``confuse'' themselves when teaching someone else without ``really [yielding] anything'' and would rather ask for help. In other words, there is a cognitive dissonance (the psychological bias that occurs when people aim to preserve cognitive consistency with their actions, values, feelings, or beliefs \cite{festinger1957theory}) between what the participants think they should do and what they are asked to do, and this bias becomes a barrier to LbT.

\begin{figure}[!tbp]
  \centering
  \begin{minipage}[b]{0.45\textwidth}
    \includegraphics[width=\textwidth]{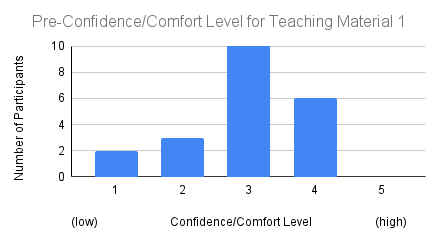}
    \caption{Material 1 Pre-Confidence/Comfort Levels}
    \label{ct1}
  \end{minipage}
  \hfill
  \begin{minipage}[b]{0.45\textwidth}
    \includegraphics[width=\textwidth]{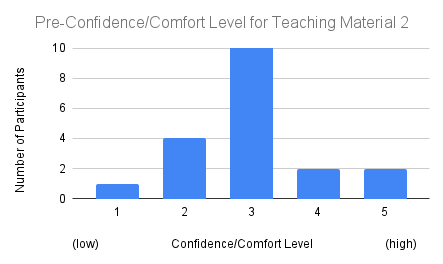}
    \caption{Material 2 Pre-Confidence/Comfort Levels}
    \label{ct2}
  \end{minipage}
\end{figure}

To investigate further whether students avoided selecting more unfamiliar material, we analyzed participants' choices of what to teach during the LbT session. For this study, participants were required to select two problems from any of their university coursework; however, they were encouraged to choose material which they received either partial marks or none at all. Prior to the study, participants were asked to rank their confidence/comfort level in teaching each material, to better understand if there were any hesitations to choose unfamiliar materials, which may be harder to teach but where more learning gains could be expected. Figures \ref{ct1} and \ref{ct2} display the survey results. 

In general, most participants selected material that they were neither not confident/comfortable with nor very confident/comfortable with, as indicated by a score of 3 (10 participants for their first material, 10 participants for their second material). To gain a better understanding of these results, we asked participants why they selected each piece of material in the interviews. Of the participants who selected material with higher confidence/comfort levels of 4 or 5 (total = 10), 
% \Cone{Sentence summarizing the confidence graph results to be added (data could be grouped instead of separating the confidence for both materials).}
% combined graphs: In general, participants were not very confident in teaching \begin{sparkline}{5} \sparkspike .15 3/24 \sparkspike .317 27/24 \sparkspike .483 30/24 \sparkspike .65 9/24 \sparkspike .817 3/24 \end{sparkline} and had moderate confidence in the materials chosen \begin{sparkline}{5} \sparkspike .15 4.5/24 \sparkspike .317 10.5/24 \sparkspike .483 42/24 \sparkspike .65 12/24 \sparkspike .817 3/24 \end{sparkline}.
% In general, participants were not very confident in teaching
% \begin{sparkline}{5} \sparkspike .15 3/24 \sparkspike .317 27/24 \sparkspike .483 30/24 \sparkspike .65 9/24 \sparkspike .817 3/24 \end{sparkline}; participants were also slightly more confident in Teaching Material 1 \begin{sparkline}{5} \sparkspike .15 6/24 \sparkspike .317 9/24 \sparkspike .483 39/24 \sparkspike .65 18/24 \sparkspike .817 0/24 \end{sparkline} than in Teaching Material 2 \begin{sparkline}{5} \sparkspike .15 3/24 \sparkspike .317 12/24 \sparkspike .483 45/24 \sparkspike .65 6/24 \sparkspike .817 6/24 \end{sparkline}.
P11 explained they picked topics that they became more confident about after realizing their mistakes and did not want ``to teach these things that [they] know nothing about''.
Additionally, they thought it was suitable for teaching since teaching is ``the translation of knowledge'' and it was ``relatively easy for [them] to explain''.
% and found the solution to be ``pretty easy to understand'' which 
% after being ``able to go back deconstruct and analyze [themselves]'' and found the solution to be ``pretty easy to understand'' which they thought was suitable for teaching since there were no longer new 
%  another part of the quote but relating to others: since they didn't want to appear ``dumb'' while explaining something they didn't understand. 
% it was a question i got wrong on a test and then it was something that i was able to go back deconstruct and analyze myself and be like okay yes this is why i'm not going to email my prof and be like this is a dumb question because. i understand why i got it wrong so that was good and then the second one uh i just found it really interesting i it was content that i hadn't uh encountered before the course and it was also pretty easy to understand and i figured it'd be pretty easy to explain and to teach somebody else like is i i know that teaching is basically just like uh not transition but the translation of knowledge and i basically could just like tell you i didn't have to i don't know i feel like there wasn't necessarily there was stuff to understand but yeah both topics were just interesting and relatively easy for me to explain even though i did so in a very robust way
P18 also felt that they needed to ``[know] the material well enough to explain it to someone else''.
% so they ``took a while...to finally understand [the material] after going through all of that time learning it [themselves]''. 
P22 even mentioned that they would ``have to be confident in 1.5 times the material of what [they'll] be teaching ... so that they will be able to answer 1.5 times the questions and understand where [the material] will expand to'' as well as be able to ``articulate the material in different ways in case someone does not understand''. 
% Conversely, P19 said that ``[they're] not confident in both of their [teaching materials]'' but ``[they] do still have a general idea of them ...rather than choosing something...[they] just didn't know how to explain at all''.
% 
% tables for pre-teaching confidence/comfort with each piece of material
% 

Participants were also asked why they chose to teach each piece of material during their interview. A few participants admitted that they chose topics they already partially understood (P1, P22) and considered the complexity of the problem as a factor of their choice (P4, P8, P10, P17, P23). P8 thought that ``both of [their materials] had really straightforward answers'' and were not ``high level difficultly questions''. Interestingly, P3 and P15 said that they would have chosen ``more difficult'' (P15) questions with ``more depth'' (P3) instead. P18 explained that ``after going through all of that time learning [the material] myself, I felt I knew the material well enough to explain it to someone else''. This is in alignment with the preconception that when teaching, the listener should be able to also understand the material being taught, which is not always necessary or desired for LbT.

If the student should arrive to teach something they have not fully grasped, a grand challenge for LbT as a pedagogical method would be to get people to be comfortable teaching something that they are uncomfortable or not confident about. This requires a drastic shift in mindset to counteract the existing cognitive dissonance.
% This aligns with past findings showing how teachers do not want to be embarrassed by being incorrect \cite{lambiotte1987manipulating}.

% \subsubsection{\bf Zero Risk Bias. }
% Like many cognitive biases, the zero risk bias is a mental shortcut. The common narrative behind these shortcuts is that they reduce cognitive strain. Instead of having to calculate the optimal solution, something that would require a lot of time and energy, we opt for the choice with less effort and uncertainty. - thedecisionlab.com

Another reason why we may have observed the results above could be due to zero risk bias, which occurs when people select options where they can have (near) absolute certainty over options that may have fewer drawbacks \cite{Allais1953,Kahneman1979,Viscusi1987}. During LbT, we believe that participants exhibited this by opting for choices (e.g., teaching materials, teaching strategies) that require less effort and uncertainty. 
% Although there are earlier findings showing a preference for certainty over overall risk reduction such as Prospect Theory, the zero risk bias as a unique cognitive phenomenon is often attributed to a 1987 paper published by Kip Viscusi, Wesley Magat, and Joel Hubert.4 The researcher’s found evidence for “certainty premiums” in eliminating risk by asking participants how much they would pay to reduce the possible risk of side effects from cleaning products (insecticide and toilet bowl cleaner). 
% https://journals-scholarsportal-info.proxy.lib.uwaterloo.ca/pdf/21908370/v225i0001/31_mtzb.xml mearusing zrb: Methodological Artefact or Decision-Making Strategy?
%  people are willing to take into account considerable drawbacks in order to gain certainty (e.g., Allais, 1953; Kahneman, 2012; Kahneman & Tversky, 1979; Viscusi, Magat, & Huber, 1987)
For example, for some participants, even when they realize the benefit of selecting material that is more challenging, they may be inclined to select material they know they can teach well. 
% P16 explained that they occasionally forgot how to approach tax cases, so they chose to teach this as an ``opportunity to help further ...ingrain the process into [their] memory'' as opposed to teaching something they were less familiar with.
P3, who only spent 10 minutes preparing for each teaching session by selecting questions that are easier to teach, specifically those that ``wouldn't take too long to explain'' and ``doesn't require prerequisite knowledge''. However, when interviewed, P3 admitted that they would have chosen questions with ``more depth'' if they were to do another LbT session. Similarly, P15 would have chosen questions that are ``more difficult'' for any future sessions, instead of the questions they had chosen, which, despite being confusing, have ``solutions [that are] really easy''. Another participant reported selecting questions that they ``fully understood'' to teach (P19).
% P15
% P3 also mentioned that they would want to choose questions that have ``more depth to it'', with multiple levels or parts to explain and P15 said that they would want to choose ``a more difficult question and also maybe ... choose something from a different like not cs or math''.
% P3: the entire preparation took only around 11-20 minutes
% Another participant, P3 said that he spent 10 minutes preparing for each teaching session by ``selecting a suitable question that that wouldn't take too long to explain, or something that doesn't require like a previous prerequisite knowledge that I can explain in one sitting''.
% add examples of wanting to choose something less familiar after the teaching session
% Furthermore, their confidence score (too low) could be indicative of the participant knowing they would complete the teaching session with high success rates and no risk as to avoid ingraining incorrect explanations. 

% (related note:  “the fear of learning something new”, or “the fear of changing, based on a fear of the unknown” (Schein, 1994) ...)
% https://journals-scholarsportal-info.proxy.lib.uwaterloo.ca/pdf/03090590/v35i0005/420_litpzsfmla.xml
Another example of this bias is observed when participants explained why they opted for audio-only explanations without using visual representations, even if having drawings may add value to their explanation. 
This was due to their lack of ability to draw legibly (P9) or success with using the drawing software (P17). P9 said ``I've tried to draw and it has thus far never worked out for me because I just can't do it'' and ``my writing is very bad''. 
This risk that the potential use of visuals might be ineffective becomes, in this case, a perceived risk. This risk and the anxiety due to ``a fear of the unknown'' \cite{schein1995organizational} that comes with making risky decisions \cite{loewenstein2001risk} are completely mitigated by not attempting to create and use them during the teaching sessions, resulting in a zero-risk situation.
Similar to cognitive dissonance, when zero risk bias is present, the students who select material or teaching methods that present lower risks of producing incorrectly, are subject to not make the best use of LbT.

\subsubsection{\bf Time \& Effort Spent in Teaching Preparations}

During the semi-structured interviews, participants expressed having spent minimal time and effort preparing to teach. While time constraints can be an obvious reason for not spending enough time as they may have liked in preparation, another reason could be due to overconfidence in their ability to articulate the concept at hand. As a result, the teaching session may not go as planned, and they may not get as much out of the feedback. A psychological obstacle that could explain this phenomenon is optimism bias, which refers to our tendency to underestimate the likelihood of experiencing negative events, making us overoptimistic \cite{weinstein1980unrealistic,seaward2000optimism}.
% Therefore, we compare participant's pre-confidence/comfort level with teaching their selected material against their post-satisfaction scores. 
For example, P19 ranked their pre-confidence/comfort level with teaching both their selected materials as $4$ (where $1 =$ no confidence/comfort and $5 =$ high confidence/comfort) explaining that they ``have a clear idea and structure of how to explain ... the topic''. However, after their teaching sessions, they ranked their post-satisfaction score in their ability to teach the material a $2$ (where $1 =$ not satisfied and $5 =$ very satisfied), since they ``felt like it didn't go as planned'' because they were ``stumbling on [their] words'' and were ``not sure if [they] got [their] definitions across properly''. When asked if they would do anything differently, P19 said that they would have liked to have prepared an ``outline of how [they were] going to go about teaching.''
% P19 was fairly confident but not satisfied after
Similarly, over half of the participants mentioned wanting to have been more prepared by spending more time explaining background knowledge (P1, P6, P8, P14), having a better understanding of the content (P3, P10), or ``[painting] a clear concept for the topic'' (P24), by doing more research of relevant terms (P3), writing down keywords (P10), and by adding some analogies (P3) or examples (P7) to make the lesson easier to understand. 
% P4 
% P12
% 5-10 min prep
While at first, optimism bias can seem to have a positive impact on learning gains in LbT, because stumbling during teaching would reveal gaps in knowledge, it can also have a negative one. This can occur when the gaps in the knowledge are gaps such as not understanding a question at all or gaps that are less complex than they could have been due to a lack of understanding in fundamental knowledge. For example, in P24's LbT session, they began to read the question they expected that they would teach then said ``I don't know how to... I'm gonna move on to the next question'' and did not try to approach the initial question at all. 

\begin{figure}[!tbp]
  \centering
  \begin{minipage}[b]{0.45\textwidth}
    \includegraphics[width=\textwidth]{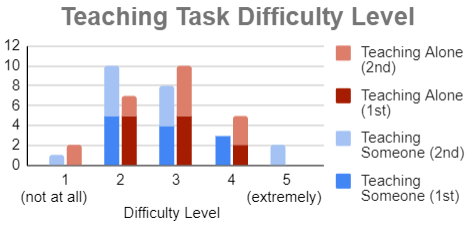}
    \caption{Difficulty Levels}
    \label{difficulty}
  \end{minipage}
  \hfill
  \begin{minipage}[b]{0.45\textwidth}
    \includegraphics[width=\textwidth]{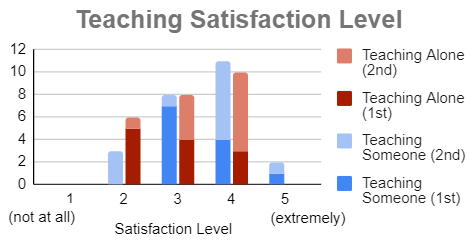}
    \caption{Satisfaction Levels}
    \label{satisfaction}
  \end{minipage}
\end{figure}

\subsection{Psychological Barriers Relating to Others}\label{psychbarrierothers}
% the way the teaching is organized (group, individual, preparation around who the individual is, reactive devaluation
% Since LbT can be organized in a way that students teach peers, psychological barrier relating to others such as impression management, bandwagon effect, and ambiguity effect were observed.

\subsubsection{\bf Impression Management. }
Impression management in LbT can take up unnecessary cognitive effort when students devote their attention to think about how their peers will perceive potential mistakes, rather than to reflect on their teaching. For example, another reason why P11 picked topics that they became more confident in was because they did not want to appear ``dumb'' while explaining something they did not understand or stumbled through. Here, we can observe that the participant aimed to ``maintain a desired situational identity'' \cite{modigliani1971embarrassment} (i.e., a desired flow of conduct in a given context) by explaining material clearly. Hence, failure (i.e., a loss of situational self-esteem) is perceived to result in embarrassment reflecting an undesired self-presentation to others \cite{modigliani1971embarrassment}.

% https://psycnet-apa-org.proxy.lib.uwaterloo.ca/fulltext/1971-09853-001.pdf
% Embarrassment, the author argues, is aspecial, short-lived, but often acute, loss ofself-esteem. More specifically, because thisloss of esteem is always related to someongoing social situation, we shall say thatembarrassment reflects a loss of situationalself-esteem.
% Goffman (19SS), who has written most extensively and creatively about the structure of face-to-face encounters, argues that man's behavior in social gatherings is marked by a pervasive desire to maintain face. ``Face is an image of self delineated in terms of approved social attributes [p. 214].'' Since different social situations call for the manifestation of somewhat different attributes, a certain open-ness and flexibility is required in order to establish an appropriate face or situational identity. 
%  Particularly, since most participants were not familiar with LbT and teaching in general, we observed that during the sessions, participants dedicated much effort to deal with teaching-related challenges discussed above.
% impression management:
Similarly, other participants 
% Students might also devote attention to impression management while teaching; participants 
reported being ``afraid of judgement'' (P2). An example is P19, who did not want to ``seem like [they're] reading every single line'' while teaching someone. This is especially the case if their sessions were recorded and accessible by others at a later time, since it could feel like having ``a lot more eyes on [them]'' (P3) and an ``awful [experience due to having] multiple people judging [them]'' (P2), or if they were teaching someone they were unfamiliar with, due to the fear of making the session ``awkward'' (P16). These factors might lead to a lost opportunity to learn and self-reflect during teaching, as the fear of failure drains the student's cognitive and emotional resources \cite{duley2005fear}.

% alone vs someone

We also observed this phenomenon when asking participants to rank the difficulty of the teaching task (where 1 = not at all difficult and 5 = extremely difficult, Fig. \ref{difficulty}) and their satisfaction level with their teaching (where 1 = not at all satisfied and 5 = extremely satisfied, Fig. \ref{satisfaction}) with each of their teaching materials as shown in Figures \ref{difficulty} and \ref{satisfaction}. When explaining their reasons for their ratings regarding the level of teaching task difficulty and satisfaction for both sessions, participants made comparisons between teaching someone (the investigator) and teaching alone (to themselves or to a pretend audience). For example, P19 explained that not wanting to ``seem like [they are] reading every single line'' while teaching someone resulted in lower satisfaction. Interestingly, some participants struggled due to not being used to, or comfortable with, using a virtual teaching setup. Even when teaching alone, P13 felt uncomfortable due to a feeling like ``someone's watching'', which may be due to the whole study being recorded (P3). However, P13 did rank their satisfaction score higher in their teaching alone session (satisfaction = 4) than the first, explaining that there was the removed ``pressure of having someone there'' but they might feel differently about this in a non-virtual context.

\subsubsection{\bf Bandwagon Effect. }
One of the ways that LbT can be organized is in a small group session, where a student is tasked with teaching unfamiliar material to several peers. This configuration could present other challenges, such as the bandwagon effect.
The bandwagon effect refers to when an opinion gains more support for superficial reasons, like if an opinion is more popular than other alternatives \cite{rikkers2002bandwagon, moy2012attitudinal}. It can result in people's tendency to ``remain silent if they feel---rightly or otherwise---that they are in the minority'' \cite{rikkers2002bandwagon}. When participants were asked about their perceptions of LbT in group settings with peers, they raised concerns about the bandwagon effect, either when LbT a group, or LbT with a group (where group members take turns teaching). Particularly, in these settings, if it seems like ``everyone understands what one person does not'', then that person would be more afraid and less likely to ask questions or provide feedback that could be useful for everyone's learning (P1).
% Previously but not right meaning: Generally, when someone else is present in a teaching session, the student teaching might experience ``a little bit more pressure to get [the material] right'' (P7), as opposed to being more ``honest or say whatever'' (P13), indicating the need for some way of encouraging honest and open communication to prevent the resulting self-censoring of the bandwagon effect, which could hamper learning.
% P13 - line 254 6-0-rq2.tex prev version
% On the flip side, teaching alone may be more authentic, as P13 said, ``you take away the pressure of having someone there... it's really just you there, so you can be honest or just say whatever.'' about the teaching not the feedback giving
Similarly, P3 felt that when teaching in a group setting, they could be ``the only one who's technically going to be speaking and relaying information'' if the peers choose to listen passively. 
% prev listed as P6 but was P3 ^
% P3 `` to a group i'm i'm the only one who's technically going to be speaking and kind of relaying information''
To avoid the potential bandwagon effect of group passive listening, some participants pointed out that they'd expect to have more meaningful and engaging one-on-one interactions when they can focus their teaching to one peer. P11 explained they ``would appreciate the intimacy... [of] one-to-one interactions'' because they may be ``more focused'' and P10 said they would be more comfortable interacting in those sessions.
Other students noted that active participation in group-based LbT sessions can work---students would be able ``to get more out of [LbT]'' through ``better conversation'' and by receiving ``all types of questions'' (P4). P7 added that you can learn from ``different [peers'] perspectives''.  

\subsubsection{\bf Ambiguity Effect. } 
% % introduce example first and then definition
% The ambiguity effect is a cognitive bias that compromises decision making when ambiguous options are presented amongst less ambiguous ones. The options where people feel well-informed about are generally preferred \cite{frisch1988ambiguity}.
% % To an extent, the ambiguity effect is an adaptive response. People prefer options that they feel well-informed about to options that they feel leave too much to the imagination. This can be useful for avoiding options for which we genuinely have too little information to go on. Even better, the ambiguity effect can lead us to seek out more information about the ambiguous option, so as to make a more informed decision.
% In LbT scenarios where the student can select who they want their peer to be, this bias can potentially impact negatively their experience when only selecting peers they know instead of unfamiliar peers. This can result in having less availability to practice LbT with peers and reduce the variety in the  perspectives and feedback that a student may receive. To understand if this may occur in LbT, 

We asked participants whom they would prefer to be matched with.
% wanting to match with people that you already know -> downside is goofing around, not having multiple perspectives etc. or perspectives from peers in upper years
About two-fifths of the participants said they would like to be paired with a peer who is in the same program 
% (P2, P6, P8, P11, P13, P18, P20-21) 
and would prefer that they were a friend.
% (P2, P5, P17-P19).
P19 explained ``I would definitely prefer a friend just because we would be used to how we explain questions or talk to each other. With the other [people] ...I'm kind of shy and I feel intimidated. There's also the possibility of the other person being terrible.''
% nvivo q10 for ``friend'':
% P2 ``yeah like I definitely wanted to be like one of my roommates who's like also in my program. I'm just comfortable with them because like we live together and we study together anyway so we kind of already do this sometimes''
% P2 ``i would prefer it to be like a friend''
% P5 ``most comfortable teaching like a friend''
% P17 ``I have some kind of preference but not not necessarily okay friends uh yes some someone who I know better or closer with that's definitely a plus''
% P18 most comfortable with a friend
% ---> P19 ``I would definitely prefer a friend just because we're just we would be used to how we explain questions or talk to each other with the other options again I'm kind of shy and I feel intimidated and there's also the possibility of the person other person being terrible so yeah I definitely would prefer a friend'' 

In LbT scenarios where the student can select who they want their peer to be and only select peers they know or are friends with can be a result of the ambiguity effect. This effect is a cognitive bias that compromises decision making when ambiguous options are presented amongst less ambiguous ones. The options where people feel well-informed about are generally preferred \cite{frisch1988ambiguity}. It can potentially impact negatively students' experience when only selecting peers they know instead of unfamiliar peers for various reasons, including and not limited to having less availability to practice LbT with peers, which could reduce their opportunities to use the LbT method and reduce the variety in the perspectives and feedback that a student may receive. By overcoming the ambiguity effect in the LbT peer matching context, students can gain access to people they would not have otherwise been able to teach or have a more authentic audience with a wider variety of knowledge and opinions. In line with this, participants noted that 
%  has the same level of understanding as them (P3, P5-7, P9, P11-12, P17, P21, P24).
% or the required prerequisites (P2, P6, P14-16, P18, P20-22, P24) to avoid teaching redundant material (P18). 
it could be beneficial to be paired with a classmate and not a friend, so that while there is some ``familiarity'', there is not enough to lose focus or “goof around” (P11, P13). 
Being matched with strangers could also reduce the consequences of “mess[ing] up” (e.g. saying “something embarrassing”) (P16). A few participants preferred to be paired with people who have the same background knowledge (P4, P12-13) or people who already understand the content since they might be able to point out the teacher’s mistakes (P6, P13), share better insights (P3, P7), and reduce the amount of time spent on explanation (P6). Sessions with these people might provide a larger return on investment (P4), potentially result in better questions being asked (P12), and increase the teacher’s motivation to help them (P13). However, a quarter of the participants 
% (P7, P9, P13, P19, P21-22)
did mention that they would have to know about their partner's background and expertise before agreeing to pair with them. 

\subsection{Know-How Barriers}
% purple
% place some of visual challenges here that is related to the teaching itself if any

% Dunning-Kruger effect: ``The Dunning-Kruger effect effect occurs when a person’s lack of knowledge and skills in a certain area cause them to overestimate their own competence. By contrast, this effect also causes those who excel in a given area to think the task is simple for everyone, and underestimate their relative abilities as well.'' why we cannot perceive our own abilities 
% could be related to not preparing enough and making content comprehension mistakes during the teaching session or why they might not have provided a thorough enough explanation 

\subsubsection{\bf Missed Opportunities to Learn from Content Incomprehension.}
During the LbT sessions, various participants encountered the problem of content incomprehension, which could in fact deepen their understanding of the material if they had attempted to understand the content. However, participants who moved on without trying to correct or improve their understanding missed these knowledge-building opportunities. Participants' struggles with content comprehension included skipping steps (P24), hesitations about correct terminology (P1-2, P5, P7, P16, P19-20, P22), wording/phrasing (P9), spelling and pronunciation (P11, P20), and their overall approach to answering the question (P11, P19) \footnote{Appendix \ref{sec:app} describes the LbT session codes for these findings}. These incomprehensions were not resolved during their LbT sessions. On the other hand, a few students did pause to review notes to correct a potential error or forgotten content (P6, P9, P12, P18, P24) without being prompted to do so. To further illustrate the contrast between both situations in which knowledge-building opportunities were missed versus taken, we outline P2's approach in both their teaching sessions and how they responded to these hesitations.

\bigskip

\textit{LbT Session 1: Learning from Content Incomprehension. }
% indent and give different font. make case same format. describe sequence of the session and around text have explanation of what happened (like a long quote)
% make quote smaller: https://tex.stackexchange.com/questions/401930/modifying-font-size-in-quotation-and-chapter-headings
In P2's LbT Session 1, they taught the investigator how to create a go-to-market launch strategy plan.
\begin{quote}
    P2 started with a brief introduction explaining that they would screen share and opened a blank document to write down key points on the topic they chose to teach. They proceeded to outline key steps by naming them and providing various definitions and real-world examples that most people would understand (e.g., the distribution of COVID-19 vaccines). Once they finished explaining how distribution needed to be included in their plan, they forgot what the next step of the plan was and told the investigator that they were going to take some time to check their notes. Once they found the information, they provided a brief reiteration of the previous steps and proceeded to the next one. Further into their go-to-market launch strategy plan, they realized that they missed another step in creating the timeline, so they took time to explain why this is important to consider early on and what should be included. At the end, they provided extra resources of information they skimmed over, a brief conclusion, and asked if the investigator had any questions.   
\end{quote}
In P2's LbT Session 1, we can see how they took the time to pause, question their knowledge gaps in the steps they wanted to provide in their launch strategy plan and filled them in by checking their notes. They also went a step further to explain why they fixed their errors and the importance of each step. By pausing after making an error, correcting it, and reiterating the steps, P2 was able to practice \textit{reflective knowledge-building}. 

\bigskip

\textit{LbT Session 2: Missing the Opportunity to Learn from Content Incomprehension. }
In P2's second LbT session, they taught how to take a mechanical model and turn it into an electrical one. 
\begin{quote}
    P2 started with an introduction on how to take a mechanical model with springs and masses and turn it into an electrical model with circuits, capacitors and batteries. They also explained how they were told that this is an important skill to learn for future applications. P2 screen shared various pre-made hand-drawn diagrams and notes and walked through them verbally as if they were solving the problem with an imaginary peer (e.g., ``now we understand...''). Halfway through the explanation, P2 said ``capacitances are created from springs'' paused and asked, ``am I saying that right?'' Then they begin to try to correct their explanation of the rules for converting specific elements from mechanical systems to electrical but said ``I'm going into this assuming that these are things that you know''. Then they paused and decided it was important to explain the following rule: ``dampers turn into resistors with the resistance value being one over the damper, like the damping coefficient'' but did not explain why this was the case and moved on. Later on, they concluded their explanation and provided information about additional diagrams that could be created to analyze the voltages at each node but forgot the name of these diagrams and ended their explanation.
    
\end{quote}
Here, we can observe how P2 skipped over key information, such as the reason behind certain rules, which could have helped them remember these rules for future applications, which they said would be important in their introduction. P2 also did not pause to check the name of the diagram that could be used for future analysis of the system. A few reasons why P2 did stop and fix an explanation but not in another could be due to thinking it was not a crucial part of the explanation, due to time limits, or because they were teaching alone as opposed to someone who may have not understood. Regardless of their exact reason for skipping material, if P2 had taken the time to deepen their understanding of various terminology and rules, they would be better prepared for solving future problems related to converting mechanical models to electrical ones.  In order to increase the amount of knowledge-building that occurs in LbT sessions among university students, they could be encouraged to be open to and learn how to overcome their content incomprehension. 
% Concluding sentence: In order to increase the amount of knowledge-building that occurs in LbT students 

\subsubsection{\bf Lack of Experience in Teaching. }

Other hesitations with the LbT task were about how to approach the teaching itself. More specifically, a few participants were not sure if they should read the question they were given (P7, P9, P19), if background information should be provided (P9), and whether a practice question should be provided to apply the knowledge that was taught (P11). They were also not sure about the order of the material they taught throughout the explanation (P11-12, P18, P22), how they should refer to potential listeners in the teaching-alone sessions (P14), the clarity of their explanation overall from the listeners' perspective (P1), the amount of content to include (P1, P15), and how to conclude their session (P23). 

Some participants were also unsure about how to teach using visual representations of their content. For example, P2 had screen shared a static mechanical diagram but was unsure of how to explain the direction of displacement and eventually used hand motions to aid the explanation. P20 was not sure how to set up their laptop in such a way that the camera showed their face and they could write on it at the same time, which caused their drawings to not be legible, resulting in the use of additional verbal explanations of their visual content. Similarly, P11, P12, and P17 opted to use verbal explanations since the drawings they made on paper when shown to the camera were too difficult to see. Ultimately, P12 was left unsatisfied with their explanation, since they found it hard to visualize multiplexers in their head without visuals. Two participants, P6 and P13 were successful with showing their paper drawings to the camera but encountered difficulties while trying to point at various locations; specifically, it was difficult to locate where exactly they were pointing so they would turn back the paper every so often to check or try to see how it was displaying on the screen by moving the paper to the side. 

The various hesitations with selecting the most appropriate teaching approach or visual tools could negatively impact the amount of learning that a student can get out of a LbT session, since certain methods are more effective than others. Moreover, since several participants were unfamiliar with the idea of LbT, they were unsure about the range of possible techniques they could use for LbT sessions or what may have been expected of them.

\begin{figure*} [ht]
  \centering
  \includegraphics[width=1\columnwidth]{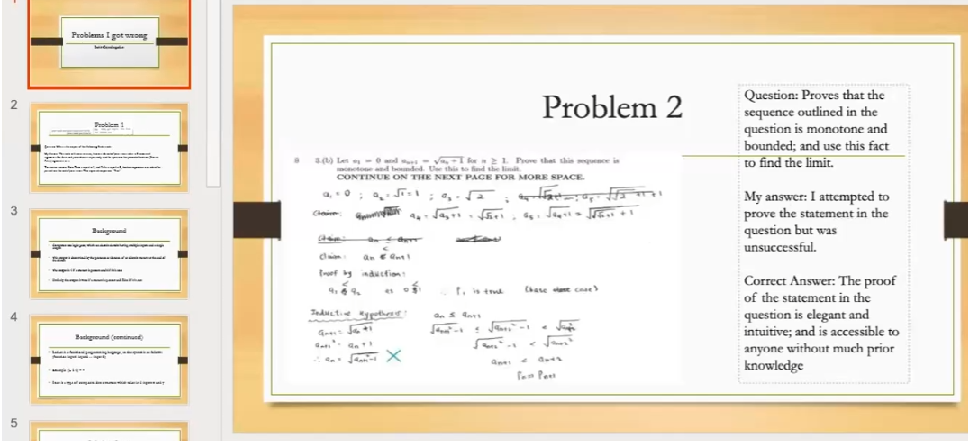}
  \caption{Presentation Deck for LbT by P15.}
  \label{ppt}
\end{figure*}

To illustrate how the use of the effective visual tools and teaching approach in a LbT session can increase the learning outcomes for the teacher, we outline P15's teaching approach. P15 explained that they spent over 50 minutes preparing for their teaching session, this included reviewing notes, making a slide deck, and preparing various examples. They added that they spent extra time due to a lack of confidence in ``teach[ing] on the fly''. Figure \ref{ppt} displays how P15 started by sharing the question they planned to teach, followed by their initial answer including their errors, and then shared the correct answer. By doing this, they were able to clearly understand their knowledge gaps and confidently walk through a solution while still identifying more complex knowledge gaps and walking through additional examples shared later in their session. 

% from July paper: Interestingly, participants who took a long time (50 min. or more) reported reasons like needing time to prepare their teaching material thoroughly (e.g., Figure \ref{ppt} by P15) and practice (P20) due to a lack of confident in ``teach[ing] on the fly'' (P15) or ``winging it'' (P20), and to make the teaching session stay on track and more engaging (P18). Teaching a difficult material (P20) and difficulties in choosing materials that ``would be better ... to teach'' (P19) are other reasons for longer preparation times.

% someone vs alone
Another barrier in LbT among university students is the lack of experience specifically when teaching alone. This was observed when participants explained their rankings in Figures \ref{difficulty} and \ref{satisfaction}. Some participants preferred having someone present while teaching, since teaching alone could cause nervousness because they ``weren't sure if [they were] rambling'' (P1). A recurring theme was the absence of feedback when teaching alone (P1, P7, P16-17, P21-22). While teaching alone, P1 could not ``make sure ... if anything made sense or ... was confusing'', a sentiment also expressed by P17 and P22. P21 ``second-guess[ed]'' themselves more, and felt less ``satisfied'' and ``fulfillment'' due to the inability to interact with the investigator.
Moreover, P22 lacked ``someone ... to bounce ideas off of'', and P7 faced more difficulty when determining when to move on while teaching. This is reflected in Figure \ref{difficulty} and \ref{satisfaction}, where regardless of the order of teaching sessions (teaching alone vs. someone), participants rated teaching alone as being slightly more difficult and less satisfying than teaching someone (i.e., the investigator).

On the flip side, participants also were unsure about how to approach LbT with someone without having context on their background and expertise. Some participants ranked the satisfaction level in the middle as 3. P12 explained that they did not know what the investigator's background with the material being taught was, making it ``hard to judge ... where to start'' but had they known a bit more they would have ``been able to structure [the session] more'' appropriately. P15 provided another reason that ``it [was] hard to say without knowing if someone understood'' and said that ``I'm not really sure, like I put it in a way that I thought it made sense, but as for someone who is seeing it for the first time I don't know if it would.'' While it may be relevant to have context about peers, students can learn to tailor their LbT sessions on the fly and still identify their own knowledge gaps. For example, P21, who had some previous tutoring experience, engaged the investigator in their LbT session by asking them multiple questions to ensure they were following along. P21 also explained that they identified new knowledge gaps based on the unexpected answers the investigator followed up on within their session. Students who have less experience with tutoring may not be familiar with techniques that can be used to engage their peers in their LbT sessions to not only identify the peers' backgrounds and knowledge gaps, but also their own. 

\newtexttwo{
\subsection{Online Peer Collaboration for LbT}
}
%---- 
% brainstorm:
% Look into challenges of peer learning
% Or focus on online and  grownup problems of running peer lbt sessions
% scheduling, partner matching, interest, relate back to rw on the classroom setup
% Psychological Barriers Relating to Others
% \bf Impression Management.
% \bf Bandwagon Effect.
% \bf Ambiguity Effect.
% ----
As previously discussed in Section \ref{psychbarrierothers} Psychological Barriers Relating to Others, university students i) could incur unnecessary cognitive effort for impression management when teaching peers, ii) may be hesitant to pair with unfamiliar students who could have acted as beneficial LbT peers, and iii) can fall into the bandwagon effect trap, making LbT feedback potentially less useful. These challenges can be overcome by designing appropriate peer matching and feedback configurations that enable positive collaboration among university-level peers during LbT, as discussed below.

% ================================= I. Matching
{\bf I. Supporting Peer Matching. }
It is important to begin by understanding how the ambiguity effect can influence students' decision making when deciding to pair with a particular peer for LbT. It can actually lead to students seeking out more information about the ambiguous pairing (i.e., working with a stranger) through the web-based tool in order to be informed about a decision \cite{frisch1988ambiguity}. Therefore, it is important to understand which information students would like to know about their peers in order to feel familiar enough with them to comfortably engage in LbT. 
% https://onlinelibrary-wiley-com.proxy.lib.uwaterloo.ca/doi/epdf/10.1002/bdm.3960010303

Many participants said they would like to be paired with a peer who is in the same course or program,
% (P2, P6, P8, P11, P13, P18, P20-21)
has the same level of understanding as them,  
% (P3, P5-7, P9, P11-12, P17, P21, P24), 
or the required prerequisites 
% (P2, P6, P14-16, P18, P20-22, P24)
to avoid teaching redundant material. 
% (P18). 
Interestingly, participants also cautioned against pairing with friends since that might lead to a loss in focus and ``goofing around'' (P11, P13) instead of providing feedback.
Moreover, students should be given the option to be paired with those who are familiar with the specific content being taught, since they would be better able to point out mistakes by the student teachers (P6, P13) and share better insights (P3, P7).

In summary, information that would help students make an informed decision about which peer they would most benefit from pairing with in a LbT session would include the peer's expertise, educational background, and disclosing their name (so as to know if they already know them). Sessions where students can select peers based on this information might provide larger return-on-investment (P4), result in better questions being asked (P12), and increase the teacher's motivation to help peers (P13). All 23 participants who were asked whether they would use a matching tool reported intentions of using such a tool. Additionally, designing for this type of peer matching gives students access to a larger pool of potential peers they would want to work with due to feeling more comfortable with their options, thus increasing the number of perspectives and types of feedback they could receive in a LbT session.

{\bf II. Supporting Collaborative Feedback}
% brainstorm:
% encouraging comp sup reflection...
% and comp sup engagement sessions similar to Phase 2 of PF
% group size can also play a role...
% and building a community
% 
It is also important to consider how collaboration in the LbT sessions can be supported through web-based tools to get the most out of feedback as well as the question and answer phases in LbT to avoid negative impacts of the bandwagon effect and impression management.

% Once students have the ability to familiarize themselves with potential peers desinging the LbT sessions around the appropriate group size can 
Given the benefits of feedback for knowledge-building, tools can be built to support teaching \textit{to} a group (an audience larger than one), since it might lead to ``better conversations'' (P4), and allow the teacher to get ``more on-the-spot feedback'' (P16) because having more people involved would culminate in ``different amounts of knowledge'' and might encourage ``different types of questions'', which could lead to the discovery of ``different gaps in [the teacher's] knowledge'' (P22). Support for teaching \textit{with} a group (where students in a group take turns LbT each other) could also yield better feedback, since everyone is also ``expecting good feedback'' for themselves (P9). 

Regarding group sizes, allowing students to be in smaller groups can allow them to ``get to know each person's deficits'' and be more comfortable to provide critical ``follow-up[s]'' and ``feedback'' (P8).
This could also mitigate students' fear of asking questions in groups, as mentioned by P1.
% On the other hand, when an audience is not available, virtual agents could be used to provide feedback for LbT sessions, although participants commented that based on their experience with agents (e.g., chatbots), they do ``not [provide] actual feedback'' (P2), since they must be ``really well-developed ... [to] give you proper feedback'' (P4).
Although LbT in a larger group setting might reduce the amount of attention given to self-reflection due to a sense of ``insecurity'' (P11), these settings also have their own benefits, as reported above.
As such, a possible middle ground is to allow the organization of group-based LbT since with ``small groups [of] people [whom the teacher knows] will reciprocate the same level of attention to what [the teacher is] doing'' (P9).

\begin{figure*} [h]
  \centering
  \includegraphics[width=.8\columnwidth]{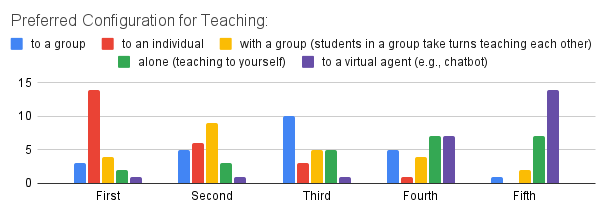}
%   \caption{Preferred Configuration Types for Learning-by-Teaching Sessions}
\caption{Preferred Configuration Types for LbT Sessions}
  \label{lbtconfig}
\end{figure*}

{\bf Teaching to an Individual, with a Group, or to a Group. }
To gain a better understanding of the participants preferences and perceptions of collaborative LbT in various configurations, they were asked about their thoughts on various ways of setting up a web-based LbT platform including the following: teaching synchronously, teaching asynchronously, teaching to a group of students, teaching to an individual student, teaching with a group of students (where students in a group take turns teaching each other), teaching alone (or to oneself), and teaching to a virtual agent (e.g., chatbot). Note that including the option of teaching alone can help us understand why participants may avoid or opt for peer collaboration.
% Comment from Jan 12 2022: new to fit in section II (highlighted previous sentence)
Figure \ref{lbtconfig} summarizes the participant's order of preference where the label ``First'' indicates the participant's most preferred configuration and ``Fifth'' indicates the participant's least preferred configuration.  
% Comment from Jan 12 2022: ''added explanation instead of percentages. I am open to changing  the graph. If it is summarized in percentages then the first and fifth choices could be shown in percentages in two different graphs'' (highlighted previous sentence)
Teaching to an individual is most frequently ranked as the most preferred configuration by participants. 
Familiarity with this configuration due to prior experience with one-to-one tutoring might be a reason behind a high rank (P10).
Aligned with analyses presented previously, the presence of feedback is a common reason for participants' favourable opinions of teaching a person. Feedback could take the form of visual cues (P2) like the peer's facial expressions (P6, P12), and peer's questions (P13). 
% Feedback can improve the teaching experience by letting the teacher know what the peer understands (P9-10) and the peer's level of engagement (P10), and by allowing adjustments (e.g., adjusting pace, teaching style, content, structure) while teaching (P3, P12). 
However, participants noted that feedback ``is not necessarily a confidence booster'' (P7); for instance, it could be rude to keep asking the peer if they understand the taught content (P9). 
Teaching with a group is ranked as the second most preferred configuration (Fig. \ref{lbtconfig}).
Teaching with a group is beneficial since everyone could ``teach and learn in the same session'' (P13). This removes ``the pressure of ...  being the sole teacher'' (P13) and the sole focus of the group (P6). Moreover, teaching with a group might encourage more social interaction and better feedback, since everyone is also ``expecting good feedback'' for themselves (P9).
However, teaching with a group might be ``really confusing'' because ``everyone has a different teaching style'' and there is a need to ``keep adapting'' to each person's teaching (P15).

% \noindent{\bf Teaching to a Group.}
Teaching to a group is ranked as the third most preferred configuration (Fig. \ref{lbtconfig}).
Teaching an audience of more than one might lead to ``better conversations'' (P4), and would allow the teacher to ``teach more people at the same time'' (P4).
% Moreover, the teacher can get ``more ... on-the-spot feedback'' (P16) because having more people involved would culminate ``different amounts of knowledge'' and might encourage ``different types of questions'', which could lead to the discovery of ``different gaps in [the teacher's] knowledge'' (P22).
That said, this configuration might need the teacher to be an ``expert'' who is capable of ``handling all types of questions'' (P4).
On the flip side, having ``multiple people judging you'' could be an ``awful'' experience (P2).

{\bf Teaching Alone or to a Virtual Agent. }
Opinions on teaching alone were not unified, but participants generally ranked them in the bottom three (Fig. \ref{lbtconfig}).
Participants reported reasons for not preferring to teach alone including that teaching alone might evoke ``a greater sense of apathy'' and a lack of purpose (P11), which might lead to a feeling that the teacher ``doesn't feel like [they are] really doing anything'' (P24).
Moreover, teaching alone only requires the teacher to ``regurgitate information because you're not actually trying to make someone learn the content'' (P13). 
However, in the case that the teach-alone session is recorded and viewable by others later, teaching alone might encourage more concentration (P5) and thorough preparation work, hence a better understanding since the teacher ``can't rely on someone else to fill in the gaps of where [the teacher's] knowledge is lacking'' (P8). Not teaching anyone also gives the teacher more control, as having someone else might result in comments and questions that ``throw [the teacher] off'' (P16).
In contrast, other participants felt that if they were just teaching themselves and no one else, their thought process could be less refined and as such, the process is ``less difficult'' due to a lack of confidence to teach others (P19) or a feeling of incompetency as a teacher (P19). In this sense, ``having a person listening or not actually doesn't change very much'' (P5).
For a few participants, preferring to not teach anyone else is a matter of personal preference, as they might be ``shy'' (P19), ``don't want to have to meet with somebody'' (P11) and ``more comfortable'' with teaching alone (P18). Moreover, not having to teach someone else might be ``a lot more efficient'' logistically, since there is no need to ``set up times or things with other people''.
Interestingly, P21 suggested using teach-alone sessions as a form of ``full-dress rehearsal'' for teaching others. This might be useful for participants who are not confident about their understanding of the material (P11, P19, P21).

% \noindent{\bf Teaching a Virtual Agent.}
Teaching to a virtual agent is by far rated as the least preferred configuration (Fig. \ref{lbtconfig}). Reasons for this rating include the perception that virtual agents are ``dumb'' (P2), ``feel fake'' (P2) and ``not very interactive'' (P5). Moreover, virtual agents do ``not [provide] actual feedback'' (P2), since they must be ``really well-developed ... [to] give you proper feedback'' (P4). Teaching to a virtual agent might also be seen as distracting and pointless (P4), and participants would rather teach alone (P2).
However, a few participants thought teaching virtual agents would be ``cool'' (P4), ``interesting and different'' (P12), which make participants ``genuinely curious'' about LbT virtual agents (P18). Virtual agents would act as ``another entity'' during teaching sessions, which might be analogous to ``talk[ing] your ideas out to a rubber duck'', and as such could be superior to teaching alone (P12). Teaching virtual agents could also ``help the algorithm'', and could be beneficial due to the agents' suggestions (P5). That said, a lack of understanding of virtual agents might result in hesitations towards using them (P14).

{\bf Synchronously vs Asynchronous LbT. }
While many responses by the participants for teaching an individual, with and to a group assume a synchronous nature, participants were also asked about their preference for synchronous vs. asynchronous LbT. Three participants preferred asynchronous (P14, P16, P19), with the rest preferring synchronous (P2-13, P15-18, P20-15). 
Most participants preferred to participate synchronously because teachers could receive immediate feedback or questions (P1-2, P8-10, P15, P18, P20-22, P24), see facial cues (P22, P24), adapt quickly (P2, P8, P10, P12, P15, P20, P22), and have direct interactions (P3, P13). 
Moreover, participating asynchronously would require repeated checks for feedback, instead of just focusing on a single block of synchronous activity (P13, P23). As such, participants might not have the motivation to perform the repeated follow-ups required from asynchronous sessions (P1).
Further, feedback received not during teaching but after the teaching is completed might be less helpful for learning (P8). The advantage of increased accountability when teaching others might also be dampened (P23).
On the other hand, LbT asynchronously allows for more flexible scheduling (P3, P14, P19) and the teacher can record their teaching at any time (P6). Participants watching recorded asynchronous sessions could also focus on only things they need help with, and ``skip through'' other parts, potentially saving time (P16).

% Having presented participants' perceptions and preferences for various LbT configurations, we now present implications for designing tools for these configurations that are either directly suggested by participants, or derived from previously presented comments.

% \noindent{\bf Encouraging and allowing informed decisions.} 

Each configuration has its own strengths, weaknesses, and varying degrees of suitability depending on many factors (e.g., content (P11), level of expertise (P4), relationship between users (P8), and amount of free time available (P4)), as discussed above. As such, for a platform supporting multiple configurations, it is essential to educate users of the differences between various configurations and guide them towards choosing the best one for their unique context. Specifically, if teaching a virtual agent is an option, designers should make clear the capabilities of the agent, since most users might be especially unfamiliar with this configuration. On the other hand, for users who might lack the confidence to use configurations that involve other people, additional tools for boosting users' confidence or guiding users through the LbT process that involve other people might be useful. Otherwise, they might be inclined to only teach alone or to a virtual agent, and miss out on the unique benefits of teaching someone, a group, or with a group. Furthermore, other features could be added to provide more flexible support for users, for example, a scheduling workflow that mitigates scheduling difficulties between users is commonly reported by participants as helpful. Additionally, tools could be designed to encourage collaborative feedback even when teaching alone, by collecting peers' comments on a session asynchronously.
Additionally, tools could be designed to encourage self-reflection regardless of configuration. These tools could ``reintroduce concepts
% Comment from Jan 12 2022: doesn't fit with section II. 
throughout the day and re-solidify them ... similar to Duolingo'' (P12).
To mitigate communication difficulties within larger groups (when teaching to or with a group), tools could be designed to facilitate teaching a larger group, and make it easier to direct questions to a specific person and speak without interrupting others.
Furthermore, the ideal group size in teaching to or with a group configurations might change depending on various factors (e.g., material complexity).

% \noindent{\bf Encouraging self-reflection.} A few participants reported preferring to teach alone or a virtual agent due to having more opportunities to perform self-reflection. However, tools could be designed to encourage self-reflection regardless of configuration. These tools could ``reintroduce concepts throughout the day and re-solidify them ... similar to Duolingo'' (P12).

% \noindent{\bf Encouraging productive group interactions.} 
% Participants suggested a need for some way to practice teaching in group settings as a way of increasing their confidence and preparedness (P19).
% Moreover, to mitigate the increase in communication difficulty within larger groups (when teaching to or with a group), tools could be designed to facilitate teaching a larger group, and make it easier to direct questions to a specific person and speak without interrupting others.
% Furthermore, the ideal group size in teaching to or with a group configurations might change depending on various factors (e.g., material complexity). As such, these configurations should support varying group sizes, and allow users the flexibility of changing the size. Even though most participants preferred synchronous sessions, tools should support both synchronous and asynchronous sessions to accommodate more users.

% ======== on reflection
% \noindent {\bf Building a community} 

Another possible feature is to encourage self-reflection, even when collaborating, by allowing a community of students who are learning the same content to upload their own teaching videos and share amongst the community. By having a community, participants mentioned that they can become both better teachers and better learners. As P12 said, ``Having people practice teaching other people and making tutorial videos ... will help them ... instill the concepts better for themselves.'' P10 said, ``You can learn something from somebody else’s lesson, whether you’re ... learning how to be a better teacher or whether you’re just like another learner... trying to pick up the concepts that were talked about.'' 
The key reasons for these learning benefits, as the participants see it, are the exposure to different forms of teaching approaches, questions, explanations and perspectives of the same topic (P2, P6, P19, P24), as well as the ability to reinforce knowledge, identify knowledge gaps, and get feedback (P18-19, P7).

{\bf LbT in a Community. }
To gain a better understanding of what a community feature may entail, we asked participants the question ``If there was a feature that enabled you to upload your teaching video with other students who are learning the same content, what do you think the advantages and disadvantages are of having this kind of community of students sharing videos?'' Advantages that have been previously outlined such as having exposure to different approaches (P2, P6-7, P12, P18-19, P24) and having familiarity with their peers (P2, P8-9, P22-24) were mentioned as well as having an opportunity to contribute to other peers' learning (P7, P10-11, P16), which could help to promote a healthier mindset amongst students ``to be helping each other versus competing against each other'' (P12). Participants also saw the benefit of using this space to be able to learn at their own time and at their own pace (P14, P20). Since content shared on such a platform ``is not a set class time'', students could use them ``whenever'' and ``revisit'' them, which could help ``solidify'' their understanding (P14, P20).

Disadvantages of this feature were also raised, such as having scrutiny and fear of judgment when using the platform (P2-3, P16). A participant mentioned feeling uncomfortable having their teaching being recorded, because of potential mistakes they might make (P10). P16 described that they may feel like an ``imposter'' since they are not used to the idea of making mistakes when teaching others. Another concern was that participants felt that they might need to take a significant amount of time to upload content, as there is some additional overhead when working asynchronously and recording (P19, P21, P23). Unequal contributions for peers could also occur (P2, P6, P8-9, P11). P9 explained the issue of ``group dynamic''---you could be ``shouting into a group chat void'' where group members do not deliver what they promise. Issues of academic dishonesty were also raised by P12 and P17. Another point participants raised were privacy concerns about the video being permanent, public, and shared beyond the group without the teacher's permission (P10, P11, P13, P18), and having personal identifiable information (P2, P7). Lastly, information overload could also occur (P4-5, P21). Participants mentioned that such a LbT community can ``get really cluttered when there’s just too many sources of information'' (P4), ``too many files to keep track of'' (P5), and that it would be nice to have ``less amount of information but better quality'' (P4).

Design considerations can be made for these advantages and disadvantages of LbT in a virtual community. For example, P10 suggested mitigating incorrect information by creating a rating system to signal the quality of the recorded lessons, or having users with specific privileges that can validate information. Another suggestion was to mitigate insecurities of teaching something unfamiliar by keeping the community small (P9, P11). To manage information overload, P21 suggested designing a community ranking system to sort through all the file uploads. 

\section{Discussion}\label{ref:discussion}

In this section, we discuss the design implications of our study's findings; in addition, we also outline future exploration of the Crash and Learn method in LbT.

\subsection{Design Implications}

\newtexttwo{

While LbT has been shown to enhance students' learning of the material \cite{duran2017learning, galbraith2011peer, hoogerheide2016gaining, roscoe2007understanding, roscoe2008tutor}, the current pitfalls in realistic settings range from challenges with LbT in the classroom (e.g., peer groupings impacting learning outcomes \cite{damon1989critical, Engel2011Need, roscoe2014self, shah2014analyzing}) to a lack of support tools built specifically for LbT without instructor involvement at the university level. 
This exploratory study is designed to uncover the challenges university students face while LbT without guidance and in the online context, as a first step towards building the appropriate LbT support. We uncovered the various psychological barriers relating to self and others, and the know-how barriers in LbT. Furthermore, we began to uncover the various strengths and weaknesses of multiple online configurations for supporting peer collaboration (e.g., peer grouping, collaborative feedback, synchronous versus asynchronous sessions, etc.), which can serve as future technological design recommendations.  

As previously motivated, one of the goals of the study is to understand if university students required unique types of training in the context of online LbT, in addition to prior works discussed in Section \ref{sec:training}. 
Aligning with prior work, we observed the need for some tutor training on teaching methods and processes (e.g., question asking and answering \cite{roscoe2007understanding}). However, university students could also be trained to be familiar with common features of online conferencing tools (e.g., online whiteboards, screen sharing), to learn about ways of dealing with the observed psychological barriers, and to carry out LbT as a learning method, which we discuss further in Section \ref{sec:futureexploration}.

Future CSCW (Computer Supported Cooperative Work) and LbT research can extend our work in several methodological directions. In general, our findings revealed psychological and know-how barriers that might not be unique to LbT in online contexts. However, we also observed LbT barriers unique to the online context, such as the preference for audio-only explanations. %These barriers were made more obvious since this study investigates online LbT through a HCI lens, and imply that there is much value to further research on LbT barriers from HCI angles.
Moreover, we hypothesize that the magnitude of psychological barriers during online LbT may depend on how the online LbT tools are designed. For instance, whether or not a student-tutor could view the live webcam footage of their student-tutee(s) might affect how much impression management manifests as a barrier while LbT; and how to properly design technology that balances the need for face-to-face interaction with privacy concerns or social anxiety remains an open question. This hypothesis is further complicated by the fact that the design of online LbT technologies may need to be domain-specific (e.g., some material might be more suited for visual representations). Other factors affecting LbT technological designs include cultural contexts  (e.g., impression management is done through different psychological mechanisms among individualists vs. collectivistic people \cite{riemer2011impression}) and age group (e.g., mature university students have unique learning motivations and methods \cite{richardson1994mature}). As such, how much each barrier observed in this study manifests itself in the LbT process would very likely depend on these factors and associated technological design choices.

We propose two ways of designing technologies based on the barriers discussed in this study. The first is to explore how technology could be designed to mitigate one or more barriers, with the focus being the minimization of situations that might give rise to the barrier(s). The second is to explore how technology could be designed to help students maximize their learning through LbT for a particular domain, while preventing them from suffering from the barriers. Researchers or designers working with educators to test possible online LbT technology in real-world classes might prefer the latter approach. Our study did not examine the exact technological features that can support LbT (e.g., virtual agents, specific visual aids tools, peer matching, etc.) since gaining a better understanding of the challenges would be initially required. However, the barriers that we discovered in our findings generated some insights into the {\it type} of technologies that might help the online LbT process; these envisioned future technologies are discussed in more detail in the next section.

}

\newtexttwo{
\subsection{Future Exploration: Crash and Learn}\label{sec:futureexploration}
% \subsection{Design Implications: Crash and Learn}
% future exploration: based on the findings, if a LbT platform were to be build, what are the useful functionality
% new intro
%  add sentence about future exploration?
To support the previously outlined challenges that university students face---psychological barriers relating to self, psychological barriers relating to others, and know-how barriers---we present potential design solutions drawing from our own analysis and existing work related to educational support methods and tools. More specifically, this section outlines a ``crash and learn'' approach, aimed at helping students become comfortable with learning from their mistakes.
}

The participants in this study were unfamiliar with the idea of teaching material to peers without having mastered it first. This cognitive dissonance could in part stem from students' education system since a ``classroom climate is likely to convey a performance-goal structure and may prompt fears of failure and avoidance motives among many students'' \cite{jackson2017fear}.
However, in order for LbT to be useful, students must overcome the fear of being wrong and the idea that they must do things correctly as previously noted. Below we outline two possible design solutions that could encourage students to become comfortable with making mistakes (``crashing'') and learning from them in their LbT sessions.

% overview of CSCL:
% In CSCL research, generally speaking, research focuses more on what is gained from structuring but not as much on what is lost. Structure, again, has been conceived broadly in CSCL research as well. It takes on a variety of forms. I give a few examples (also see Koschmann, Suthers, & Chan, 2005): task or problem structuring (e.g., Jonassen & Kwon, 2001; Kapur & Kinzer, 2007); meta-cognitive support through reflection prompts (e.g., Lin, Hmelo, Kinzer, & Secules, 1999); content support (e.g., Fischer & Mandl, 2005); interactional support through question prompts (e.g., Ge & Land, 2003); supporting group discourse through argumentation tools (e.g., Cho & Jonassen, 2002) and representational guidance (Suthers & Hundhausen, 2003); scripting inter-dependencies through a division of labor (e.g., Schellens, Van Keer, Valcke, & De Wever, 2005); supporting the problem-solving process through process scaffolds (e.g., Schwartz, Lin, Brophy, & Bransford, 1999), and so on

{\bf I. Incorporating Productive Failure. }
% productive failure 2008: https://journals-scholarsportal-info.proxy.lib.uwaterloo.ca/pdf/07370008/v26i0003/379_pf.xml
% designing for productive failure 2012: https://journals-scholarsportal-info.proxy.lib.uwaterloo.ca/pdf/10508406/v21i0001/45_dfpf.xml
% website w/des principles: https://www.manukapur.com/productive-failure/
% when PF fails: https://www.researchgate.net/profile/Tanmay-Sinha-2/publication/333005127_When_Productive_Failure_Fails/links/5d1b725c299bf1547c9259e2/When-Productive-Failure-Fails.pdf
Productive Failure (PF) outlines the design of conditions for learners to persist in generating and exploring representations and solution methods (RSMs) for solving complex, novel problems \cite{kapur2008productive, kapur2012designing}. Contrary to traditional ordering of instruction, PF starts with a problem-solving phase where learners explore and generate RSMs to complex problems based on material they have not learned yet (similar to LbT), followed by an instruction phase by an expert or teacher who builds on the learners' solutions \cite{sinha2019productive}. While students typically fail to produce correct solutions, they benefit more from subsequent instruction (i.e. guidance) \cite{kapur2016examining, schwartz1998time} since PF primarily impacts conceptual understanding and transfer and maximizes learning in the longer term \cite{kapur2008productive}. This sequence follows a similar approach to LbT where a student may fail to teach material correctly to a peer, but the peer may be able to provide guidance by asking questions or providing feedback. In order to promote PF in LbT the core mechanisms of PF (``(a) activation and differentiation of prior knowledge, (b) attention to critical features, (c) explanation and elaboration of these features, and (d) organization and assembly into canonical RSM'' \cite{kapur2012designing}) can be integrated using a modification of Kupar's two phases of PF. The first phase (Phase 1) encourages the generation and exploration of multiple RSMs (where failure typically occurs) during the preparation and explanation phases of LbT. The second phase (Phase 2) provides opportunities for consolidation (e.g., reviewing notes, asking peers questions, etc.) and knowledge assembly (e.g., comparing and contrasting failed or suboptimal RSMs) into canonical RSMs. In addition to incorporating these two phases, there are three design layers that can be applied to embody the PF principles including a) the task the students take part in, which should be adequately complex and engaging, b) the participation structures used to engage with the task where student collaboration is enabled, and c) the social surround framing the task, which should be a safe space for students with affective support for persistence \cite{kapur2012designing}.
By designing computer-supported LbT that prompts PF to occur could help set expectations among university students where they know that failure and imperfect teaching is common in Phase 1. 

{\bf II. Incorporating Micro Learning. }
Once students who are LbT have clearer expectations of potentially failing when teaching unfamiliar content, they can also minimize their own perceived burden of these failures by making mistakes with smaller sections of content. This can be done by integrating micro learning techniques, which is a holistic approach that involves conveying information through short instructional segments known as \textit{micro content}. Micro learning focuses on small content units and narrow topics that can be accessed as part of an informal learning process that fosters active collaboration \cite{hug2007outline}. Research has shown that using shorter content may increase information retention by 20\% compared to longer instructional content (i.e. macro content) \cite{giurgiu2017microlearning}. The length of each instructional segment in a LbT session can affect the quality of the content and knowledge retention. In a system designed by Joshi et al. \cite{joshi2020micromentor}, help sessions between an expert and a novice are time-limited to three minutes per answer. This creates short explanations that are coherent on their own and also easier to link with other snippets to create a more complex series of tutorials. Micro learning has also been integrated into MOOCs (Massive Open Online Courses) to motivate learners to make incremental learning progress in online modules \cite{kamilali2015microlearning} and has been found to make learned subjects more memorable for a longer period \cite{mohammed2018effectiveness}. 

Deconstructing complex tasks into simpler parts that can be mastered individually is a memory-sensitive strategy that improves learning while reducing mental load on learners \cite{wilson2000situated}. Teaching tasks in smaller parts can also give tutors a sense of accomplishment and motivation to continue \cite{wilson2000situated}. Micro teaching has also been shown to reduce teaching anxiety in mathematics teachers \cite{peker2009pre}. By giving teachers the ability to teach in small units, they were able to solicit feedback on the sections to improve and not worry about small mistakes undermining an entire lesson. Since the creation of micro content in collaborative online teaching environments can relieve nervousness, boost confidence levels, and increase knowledge retention in tutors, it would be beneficial to incorporate this method into the LbT web-based scaffolding for university-level students. Furthermore, micro learning could be designed into the PF phases. This has been done by structuring the exploration in PF into shorter iterative cycles of 2-3 minutes and the task involved learning a series of steps \cite{ziegler2021micro}.

{\bf III. Incorporating Preparation Guidance. }
% \subsection{Tools for Teaching Preparation}  
To better understand participants' views on using tools for LbT preparation, which could help establish expectations with using this method, we asked them if they would use such tools and why. Most participants said that they would be comfortable using a web-based support tool to help prepare for their LbT session through a guide or template aimed to help students generate content or lesson plans.
% (P1-8, P10-24). 
Their perceived benefits of using these tools included support for planning the lesson structure (P3, P24), organization (P8, P18, P20, P24), timing (P12), lesson material standardization (P4, P17), expectation setting (P11, P17), improving content (P5) or notes (P19) (e.g., by making them more ``consolidated and cohesive'' (P7)), and template ideation (P24) designed by teaching experts (P14).
P4 also mentioned such tools would make the process of preparing ``easier'' for ``someone who hasn't done it before''. 
Some concerns participants had about this tool included that there could be more ``overhead'' (P2), potentially restraining (P21), and not necessary for their own learning since it might encourage a specific LbT pace (P15).
% Other participants' preference to use these tools depends on a few considerations. P6 would use preparation tools if they ``knew the material'' and were going to teach someone who did not (i.e. regular teaching), but would not use them to teach material that the they themselves do not understand. This is a recurring theme, with P10, P13, and P22 expressing similar opinions that they would not use them if the purpose was to learn the material for themselves through teaching, instead of ensuring the audience learns something.
While many participants commented on the benefits of using preparation tools, the design should avoid solutions that might discourage their use (e.g., lacking flexibility) and carefully consider the balance between the amount of effort and time needed to use these tools.

% Section 6.2 Online peer collaboration for Lbt is moved to 5.4

% \input{7-tools}
\bigskip
\section{Conclusion}\label{sec:conclusion}
% Discussion and Conclusion
%======================================================================

%\subsection{Future Work \& Going Beyond Our Limitations}

\newtext{LbT is a validated classroom technique for learning; however, the process of LbT and the challenges that university students face in the contexts of online teaching and technology usage, are lesser known. In this study, we gave participants an open-ended task of teaching two concepts they wanted to master, but that they were struggling with understanding. 
% The focus of this initial study is not to demonstrate how process affect the success of learning by teaching, but to uncover the unique challenges that students face during the learning by teaching process in the contexts of online teaching and technology usage, in order to derive design suggestions and implications for web-based tools supporting this pedagogical method.
\newtexttwo{The focus of this initial study was not to demonstrate how online teaching affects the success of LbT. Rather, our goal was to uncover the unique challenges that students face during the LbT process in the contexts of online teaching and technology usage, in order to derive design suggestions and implications for web-based tools supporting this pedagogical method.}
These insights can help inform future virtual LbT platform designs to best support students in a peer-to-peer context. Our results highlight three key challenges university-level students face, including psychological barriers relating to self, psychological barriers relating to others, and lack of know-how. \newtexttwo{The online context introduced additional barriers in all of these categories, including the loss of confidence to visualize ideas using digital drawings, being nervous due to being recorded, and being unsure which tools to use to illustrate one's idea when too many options are provided.} We also provide design solutions to overcome these challenges, including ``crash and learn'' and online peer collaboration support for LbT. In addition to this, we outlined students' perceptions of possible features that could be implemented in a LbT platform, the various ways it could be configured, and different configurations' advantages and disadvantages.}

%\newtext{This study investigates the challenges faced by university students in the contexts of online teaching and technology usage. \newtext{Specifically, this paper}presents findings for understanding university students' unguided process and approach to teaching unfamiliar material, the students' individual differences to their approach, their struggles, and provides an outline of design suggestions and implications for web-based tools supporting this pedagogical method. 

\newtext{There are several limitations to our study. First, participants were asked to teach the investigator, instead of the more realistic scenario of teaching a peer (e.g., another student taking the same course). While this more controlled design has less ecological validity, it does allow us to control the personality/interaction style of the tutees, which would be difficult to do if we were to introduce random LbT groups. Furthermore, it allowed for findings of challenges and accompanying design implications for situations where LbT is to be carried out alone, or with strangers, hence expanding our understanding of the various possible LbT configurations and their corresponding challenges. Second,} even though participants knew the intent of the study and met the criteria of teaching two assignments they did not receive full marks on, many were unfamiliar with the idea of teaching to learn, so they may have come into the study with the mindset that they were tutoring, focusing less on how the teaching process makes them learn. \newtext{We cannot control their own goals regardless of the instructions given. Some students may have focused more on learning or more on teaching and this could not have been controlled. Picking a class concept they \newtexttwo{did not} fully grasp but needed to learn for an exam allowed us to investigate through a realistic scenario that LbT can be used for. Moreover, as described in the results, the psychological barriers and lack of know-how can reduce the student’s focus on the actual learning which is why overcoming these challenges is beneficial for LbT. Third, this study was focused on only the university-student age group. For example, since the majority of the students were undergraduates \newtexttwo{it is} possible that their relative youthfulness and/or lack of experience could explain why the perceptions of self and others were prominent themes. While it is likely that at least some of the challenges and design implications discussed in this work are applicable as well to other age groups, further work should be conducted to assess this.}

\newtext{In terms of future work, studies where participants are all given the same content to perform LbT, could be conducted to design specific tools for addressing psychological barriers and to assess the effectiveness of these tools by measuring participants' pre- and post-study knowledge of the materials.} In addition, future studies could explore LbT in a controlled scenario where participants are given unfamiliar material, have the required background to learn it, and can interact with a peer who shares this prerequisite knowledge. This would create an environment where peers perform LbT to each other, and each stage of the LbT method could be observed. Furthermore, implementations of the design suggestions to promote and set expectations around failing and learning from mistakes as well as to support university-level peer collaboration can be investigated. Integration of generalized design principles for PF and micro learning should be carefully considered since they have been studied under specific conditions and settings (e.g. content domain, communication modality, age group, sociocultural factors, etc). 
\bibliographystyle{ACM-Reference-Format}
\bibliography{main}

%%% -*-BibTeX-*-
%%% Do NOT edit. File created by BibTeX with style
%%% ACM-Reference-Format-Journals [18-Jan-2012].

\begin{thebibliography}{119}

%%% ====================================================================
%%% NOTE TO THE USER: you can override these defaults by providing
%%% customized versions of any of these macros before the \bibliography
%%% command.  Each of them MUST provide its own final punctuation,
%%% except for \shownote{}, \showDOI{}, and \showURL{}.  The latter two
%%% do not use final punctuation, in order to avoid confusing it with
%%% the Web address.
%%%
%%% To suppress output of a particular field, define its macro to expand
%%% to an empty string, or better, \unskip, like this:
%%%
%%% \newcommand{\showDOI}[1]{\unskip}   % LaTeX syntax
%%%
%%% \def \showDOI #1{\unskip}           % plain TeX syntax
%%%
%%% ====================================================================

\ifx \showCODEN    \undefined \def \showCODEN     #1{\unskip}     \fi
\ifx \showDOI      \undefined \def \showDOI       #1{#1}\fi
\ifx \showISBNx    \undefined \def \showISBNx     #1{\unskip}     \fi
\ifx \showISBNxiii \undefined \def \showISBNxiii  #1{\unskip}     \fi
\ifx \showISSN     \undefined \def \showISSN      #1{\unskip}     \fi
\ifx \showLCCN     \undefined \def \showLCCN      #1{\unskip}     \fi
\ifx \shownote     \undefined \def \shownote      #1{#1}          \fi
\ifx \showarticletitle \undefined \def \showarticletitle #1{#1}   \fi
\ifx \showURL      \undefined \def \showURL       {\relax}        \fi
% The following commands are used for tagged output and should be
% invisible to TeX
\providecommand\bibfield[2]{#2}
\providecommand\bibinfo[2]{#2}
\providecommand\natexlab[1]{#1}
\providecommand\showeprint[2][]{arXiv:#2}

\bibitem[Akobe et~al\mbox{.}(2019)]%
        {akobe2019web}
\bibfield{author}{\bibinfo{person}{David Akobe}, \bibinfo{person}{Segun~I.
  Popoola}, \bibinfo{person}{Aderemi~A. Atayero},
  \bibinfo{person}{Olasunkanmi~F. Oseni}, {and} \bibinfo{person}{Sanjay
  Misra}.} \bibinfo{year}{2019}\natexlab{}.
\newblock \showarticletitle{A Web Framework for Online Peer Tutoring
  Application in a Smart Campus}.
\newblock In \bibinfo{booktitle}{\emph{Computational Science and Its
  Applications {\textendash} {ICCSA} 2019}}. \bibinfo{publisher}{Springer
  International Publishing}, \bibinfo{address}{Cham, Switzerland},
  \bibinfo{pages}{316--326}.
\newblock
\urldef\tempurl%
\url{https://doi.org/10.1007/978-3-030-24308-1_26}
\showDOI{\tempurl}


\bibitem[Alaimi et~al\mbox{.}(2020)]%
        {mehdi20}
\bibfield{author}{\bibinfo{person}{Mehdi Alaimi}, \bibinfo{person}{Edith Law},
  \bibinfo{person}{Kevin~D. Pantasdo}, \bibinfo{person}{Pierre-Yves Oudeyer},
  {and} \bibinfo{person}{Hélène Sauzeon}.} \bibinfo{year}{2020}\natexlab{}.
\newblock \showarticletitle{Pedagogical Agents for Fostering Question-Asking
  Skills in Children}. In \bibinfo{booktitle}{\emph{Proceedings of the SIGCHI
  Conference on Human Factors in Computing Systems}}
  \emph{(\bibinfo{series}{CHI '20})}. \bibinfo{publisher}{Association for
  Computing Machinery}, \bibinfo{address}{New York, NY, USA},
  \bibinfo{pages}{1--10}.
\newblock
\urldef\tempurl%
\url{https://doi.org/10.1145/3313831.3376776}
\showDOI{\tempurl}


\bibitem[Allais(1953)]%
        {Allais1953}
\bibfield{author}{\bibinfo{person}{M. Allais}.}
  \bibinfo{year}{1953}\natexlab{}.
\newblock \showarticletitle{Le Comportement de l{\textquotesingle}Homme
  Rationnel devant le Risque: Critique des Postulats et Axiomes de
  l{\textquotesingle}Ecole Americaine}.
\newblock \bibinfo{journal}{\emph{Econometrica}} \bibinfo{volume}{21},
  \bibinfo{number}{4} (\bibinfo{date}{Oct.} \bibinfo{year}{1953}),
  \bibinfo{pages}{503}.
\newblock
\urldef\tempurl%
\url{https://doi.org/10.2307/1907921}
\showDOI{\tempurl}


\bibitem[Anderberg et~al\mbox{.}(2013)]%
        {anderberg2013exploring}
\bibfield{author}{\bibinfo{person}{Erik Anderberg}, \bibinfo{person}{Anton
  Axelsson}, \bibinfo{person}{Sanne Bengtsson}, \bibinfo{person}{Maja
  H{\aa}kansson}, {and} \bibinfo{person}{Lisa Lindberg}.}
  \bibinfo{year}{2013}\natexlab{}.
\newblock \showarticletitle{Exploring the use of a teachable agent in a
  mathematical computer game for preschoolers}.
\newblock \bibinfo{journal}{\emph{Intelligent, socially oriented technology}}
  \bibinfo{volume}{154} (\bibinfo{year}{2013}), \bibinfo{pages}{161--171}.
\newblock


\bibitem[Aronson and Patnoe(2011)]%
        {aronson2011cooperation}
\bibfield{author}{\bibinfo{person}{Elliot Aronson} {and}
  \bibinfo{person}{Shelly Patnoe}.} \bibinfo{year}{2011}\natexlab{}.
\newblock \bibinfo{booktitle}{\emph{Cooperation in the classroom}
  (\bibinfo{edition}{3} ed.)}.
\newblock \bibinfo{publisher}{Pinter \& Martin}, \bibinfo{address}{London,
  England}.
\newblock


\bibitem[Bargh and Schul(1980)]%
        {bargh1980cognitive}
\bibfield{author}{\bibinfo{person}{John~A Bargh} {and} \bibinfo{person}{Yaacov
  Schul}.} \bibinfo{year}{1980}\natexlab{}.
\newblock \showarticletitle{On the cognitive benefits of teaching.}
\newblock \bibinfo{journal}{\emph{Journal of Educational Psychology}}
  \bibinfo{volume}{72}, \bibinfo{number}{5} (\bibinfo{year}{1980}),
  \bibinfo{pages}{593}.
\newblock


\bibitem[Benware and Deci(1984)]%
        {benware1984quality}
\bibfield{author}{\bibinfo{person}{Carl~A Benware} {and}
  \bibinfo{person}{Edward~L Deci}.} \bibinfo{year}{1984}\natexlab{}.
\newblock \showarticletitle{Quality of learning with an active versus passive
  motivational set}.
\newblock \bibinfo{journal}{\emph{American educational research journal}}
  \bibinfo{volume}{21}, \bibinfo{number}{4} (\bibinfo{year}{1984}),
  \bibinfo{pages}{755--765}.
\newblock


\bibitem[Berg(1999)]%
        {berg1999}
\bibfield{author}{\bibinfo{person}{E.~C. Berg}.}
  \bibinfo{year}{1999}\natexlab{}.
\newblock \showarticletitle{The effects of trained peer response on ESL
  students’ revision types and writing quality}.
\newblock \bibinfo{journal}{\emph{Journal of Second Language Writing}}
  \bibinfo{volume}{8}, \bibinfo{number}{3} (\bibinfo{year}{1999}),
  \bibinfo{pages}{215--241}.
\newblock


\bibitem[Berthold et~al\mbox{.}(2008)]%
        {Berthold_2008}
\bibfield{author}{\bibinfo{person}{Kirsten Berthold}, \bibinfo{person}{Tessa
  H.~S. Eysink}, {and} \bibinfo{person}{Alexander Renkl}.}
  \bibinfo{year}{2008}\natexlab{}.
\newblock \showarticletitle{Assisting self-explanation prompts are more
  effective than open prompts when learning with multiple representations}.
\newblock \bibinfo{journal}{\emph{Instructional Science}} \bibinfo{volume}{37},
  \bibinfo{number}{4} (\bibinfo{date}{apr} \bibinfo{year}{2008}),
  \bibinfo{pages}{345--363}.
\newblock
\urldef\tempurl%
\url{https://doi.org/10.1007/s11251-008-9051-z}
\showDOI{\tempurl}


\bibitem[Biswas et~al\mbox{.}(2004)]%
        {biswas2004incorporating}
\bibfield{author}{\bibinfo{person}{Gautam Biswas}, \bibinfo{person}{Krittaya
  Leelawong}, \bibinfo{person}{Kadira Belynne}, \bibinfo{person}{Karun
  Viswanath}, \bibinfo{person}{Nancy Vye}, \bibinfo{person}{Daniel Schwartz},
  {and} \bibinfo{person}{Joan Davis}.} \bibinfo{year}{2004}\natexlab{}.
\newblock \showarticletitle{Incorporating self regulated learning techniques
  into learning by teaching environments}. In
  \bibinfo{booktitle}{\emph{Proceedings of the Annual Meeting of the Cognitive
  Science Society}}, Vol.~\bibinfo{volume}{26}.
  \bibinfo{publisher}{eScholarship}, \bibinfo{address}{Oakland, CA, USA},
  \bibinfo{pages}{120--125}.
\newblock


\bibitem[Biswas et~al\mbox{.}(2005)]%
        {biswas2005learning}
\bibfield{author}{\bibinfo{person}{Gautam Biswas}, \bibinfo{person}{Krittaya
  Leelawong}, \bibinfo{person}{Daniel Schwartz}, \bibinfo{person}{Nancy Vye},
  {and} \bibinfo{person}{The Teachable Agents~Group at Vanderbilt}.}
  \bibinfo{year}{2005}\natexlab{}.
\newblock \showarticletitle{Learning by teaching: A new agent paradigm for
  educational software}.
\newblock \bibinfo{journal}{\emph{Applied Artificial Intelligence}}
  \bibinfo{volume}{19}, \bibinfo{number}{3-4} (\bibinfo{year}{2005}),
  \bibinfo{pages}{363--392}.
\newblock


\bibitem[Braun and Clarke(2006)]%
        {braun2006using}
\bibfield{author}{\bibinfo{person}{Virginia Braun} {and}
  \bibinfo{person}{Victoria Clarke}.} \bibinfo{year}{2006}\natexlab{}.
\newblock \showarticletitle{Using thematic analysis in psychology}.
\newblock \bibinfo{journal}{\emph{Qualitative research in psychology}}
  \bibinfo{volume}{3}, \bibinfo{number}{2} (\bibinfo{year}{2006}),
  \bibinfo{pages}{77--101}.
\newblock


\bibitem[Bruffee(1980)]%
        {bruffee1980two}
\bibfield{author}{\bibinfo{person}{Kenneth~A Bruffee}.}
  \bibinfo{year}{1980}\natexlab{}.
\newblock \showarticletitle{Two related issues in peer tutoring: Program
  structure and tutor training}.
\newblock \bibinfo{journal}{\emph{College Composition and Communication}}
  \bibinfo{volume}{31}, \bibinfo{number}{1} (\bibinfo{year}{1980}),
  \bibinfo{pages}{76--80}.
\newblock


\bibitem[Carberry(2008)]%
        {carberry2008learning}
\bibfield{author}{\bibinfo{person}{Adam~R Carberry}.}
  \bibinfo{year}{2008}\natexlab{}.
\newblock \emph{\bibinfo{title}{Learning-by-teaching as a pedagogical approach
  and its implications on engineering education}}.
\newblock \bibinfo{thesistype}{Ph.\,D. Dissertation}. \bibinfo{school}{Tufts
  University}.
\newblock


\bibitem[Ceha et~al\mbox{.}(2021)]%
        {cehaHumour}
\bibfield{author}{\bibinfo{person}{Jessy Ceha}, \bibinfo{person}{Ken~Jen Lee},
  \bibinfo{person}{Elizabeth Nilsen}, \bibinfo{person}{Joslin Goh}, {and}
  \bibinfo{person}{Edith Law}.} \bibinfo{year}{2021}\natexlab{}.
\newblock \bibinfo{booktitle}{\emph{Can a Humorous Conversational Agent Enhance
  Learning Experience and Outcomes?}}
\newblock \bibinfo{publisher}{Association for Computing Machinery},
  \bibinfo{address}{New York, NY, USA}, \bibinfo{pages}{1--14}.
\newblock
\showISBNx{9781450380966}
\urldef\tempurl%
\url{https://doi.org/10.1145/3411764.3445068}
\showURL{%
\tempurl}


\bibitem[Cheng and Yen(1998)]%
        {cheng1998virtual}
\bibfield{author}{\bibinfo{person}{Charles~YY Cheng} {and}
  \bibinfo{person}{Jerome Yen}.} \bibinfo{year}{1998}\natexlab{}.
\newblock \showarticletitle{Virtual learning environment (VLE): a Web-based
  collaborative learning system}. In \bibinfo{booktitle}{\emph{Proceedings of
  the Thirty-First Hawaii International Conference on System Sciences}},
  Vol.~\bibinfo{volume}{1}. IEEE, \bibinfo{publisher}{IEEE Computer Society},
  \bibinfo{address}{Washington, DC, USA}, \bibinfo{pages}{480--491}.
\newblock


\bibitem[Chi et~al\mbox{.}(1989)]%
        {chi1989self}
\bibfield{author}{\bibinfo{person}{Michelene~TH Chi}, \bibinfo{person}{Miriam
  Bassok}, \bibinfo{person}{Matthew~W Lewis}, \bibinfo{person}{Peter Reimann},
  {and} \bibinfo{person}{Robert Glaser}.} \bibinfo{year}{1989}\natexlab{}.
\newblock \showarticletitle{Self-explanations: How students study and use
  examples in learning to solve problems}.
\newblock \bibinfo{journal}{\emph{Cognitive science}} \bibinfo{volume}{13},
  \bibinfo{number}{2} (\bibinfo{year}{1989}), \bibinfo{pages}{145--182}.
\newblock


\bibitem[Chi et~al\mbox{.}(1994)]%
        {chi1994eliciting}
\bibfield{author}{\bibinfo{person}{Michelene~TH Chi}, \bibinfo{person}{Nicholas
  De~Leeuw}, \bibinfo{person}{Mei-Hung Chiu}, {and} \bibinfo{person}{Christian
  LaVancher}.} \bibinfo{year}{1994}\natexlab{}.
\newblock \showarticletitle{Eliciting self-explanations improves
  understanding}.
\newblock \bibinfo{journal}{\emph{Cognitive science}} \bibinfo{volume}{18},
  \bibinfo{number}{3} (\bibinfo{year}{1994}), \bibinfo{pages}{439--477}.
\newblock


\bibitem[Choi and Seong(2011)]%
        {choi2011peer}
\bibfield{author}{\bibinfo{person}{Young~Eun Choi} {and}
  \bibinfo{person}{Guiboke Seong}.} \bibinfo{year}{2011}\natexlab{}.
\newblock \showarticletitle{How Peer Tutoring and Peer Tutor Training Influence
  Korean EFL Students' Writing}.
\newblock \bibinfo{journal}{\emph{English Language \& Literature Teaching}}
  \bibinfo{volume}{17}, \bibinfo{number}{4} (\bibinfo{year}{2011}),
  \bibinfo{pages}{23--47}.
\newblock


\bibitem[Chuang(2015)]%
        {chuang2015sscls}
\bibfield{author}{\bibinfo{person}{Yung-Ting Chuang}.}
  \bibinfo{year}{2015}\natexlab{}.
\newblock \showarticletitle{SSCLS: A smartphone-supported collaborative
  learning system}.
\newblock \bibinfo{journal}{\emph{Telematics and Informatics}}
  \bibinfo{volume}{32}, \bibinfo{number}{3} (\bibinfo{year}{2015}),
  \bibinfo{pages}{463--474}.
\newblock


\bibitem[Damon and Phelps(1989)]%
        {damon1989critical}
\bibfield{author}{\bibinfo{person}{William Damon} {and} \bibinfo{person}{Erin
  Phelps}.} \bibinfo{year}{1989}\natexlab{}.
\newblock \showarticletitle{Critical distinctions among three approaches to
  peer education}.
\newblock \bibinfo{journal}{\emph{International journal of educational
  research}} \bibinfo{volume}{13}, \bibinfo{number}{1} (\bibinfo{year}{1989}),
  \bibinfo{pages}{9--19}.
\newblock


\bibitem[Dhawan(2020)]%
        {dhawan2020online}
\bibfield{author}{\bibinfo{person}{Shivangi Dhawan}.}
  \bibinfo{year}{2020}\natexlab{}.
\newblock \showarticletitle{Online learning: A panacea in the time of COVID-19
  crisis}.
\newblock \bibinfo{journal}{\emph{Journal of educational technology systems}}
  \bibinfo{volume}{49}, \bibinfo{number}{1} (\bibinfo{year}{2020}),
  \bibinfo{pages}{5--22}.
\newblock


\bibitem[Duley et~al\mbox{.}(2005)]%
        {duley2005fear}
\bibfield{author}{\bibinfo{person}{Aaron~R Duley}, \bibinfo{person}{David~E
  Conroy}, \bibinfo{person}{Katherine Morris}, \bibinfo{person}{Jennifer
  Wiley}, {and} \bibinfo{person}{Christopher~M Janelle}.}
  \bibinfo{year}{2005}\natexlab{}.
\newblock \showarticletitle{Fear of failure biases affective and attentional
  responses to lexical and pictorial stimuli}.
\newblock \bibinfo{journal}{\emph{Motivation and Emotion}}
  \bibinfo{volume}{29}, \bibinfo{number}{1} (\bibinfo{year}{2005}),
  \bibinfo{pages}{1--17}.
\newblock


\bibitem[Duran(2016)]%
        {Duran_2016}
\bibfield{author}{\bibinfo{person}{David Duran}.}
  \bibinfo{year}{2016}\natexlab{}.
\newblock \showarticletitle{Learning-by-teaching. Evidence and implications as
  a pedagogical mechanism}.
\newblock \bibinfo{journal}{\emph{Innovations in Education and Teaching
  International}} \bibinfo{volume}{54}, \bibinfo{number}{5}
  (\bibinfo{date}{feb} \bibinfo{year}{2016}), \bibinfo{pages}{476--484}.
\newblock
\urldef\tempurl%
\url{https://doi.org/10.1080/14703297.2016.1156011}
\showDOI{\tempurl}


\bibitem[Duran(2017)]%
        {duran2017learning}
\bibfield{author}{\bibinfo{person}{David Duran}.}
  \bibinfo{year}{2017}\natexlab{}.
\newblock \showarticletitle{Learning-by-teaching. Evidence and implications as
  a pedagogical mechanism}.
\newblock \bibinfo{journal}{\emph{Innovations in Education and Teaching
  International}} \bibinfo{volume}{54}, \bibinfo{number}{5}
  (\bibinfo{year}{2017}), \bibinfo{pages}{476--484}.
\newblock


\bibitem[Engel(2011)]%
        {Engel2011Need}
\bibfield{author}{\bibinfo{person}{Susan Engel}.}
  \bibinfo{year}{2011}\natexlab{}.
\newblock \showarticletitle{{Children's Need to Know: Curiosity in Schools}}.
\newblock \bibinfo{journal}{\emph{Harvard Educational Review}}
  \bibinfo{volume}{81}, \bibinfo{number}{4} (\bibinfo{date}{12}
  \bibinfo{year}{2011}), \bibinfo{pages}{625--645}.
\newblock
\showISSN{0017-8055}
\urldef\tempurl%
\url{https://doi.org/10.17763/haer.81.4.h054131316473115}
\showDOI{\tempurl}
\showeprint{https://meridian.allenpress.com/her/article-pdf/81/4/625/2111551/haer\_81\_4\_h054131316473115.pdf}


\bibitem[Ensergueix and Lafont(2010)]%
        {ensergueix2010reciprocal}
\bibfield{author}{\bibinfo{person}{Pierre~Jean Ensergueix} {and}
  \bibinfo{person}{Lucile Lafont}.} \bibinfo{year}{2010}\natexlab{}.
\newblock \showarticletitle{Reciprocal peer tutoring in a physical education
  setting: influence of peer tutor training and gender on motor performance and
  self-efficacy outcomes}.
\newblock \bibinfo{journal}{\emph{European Journal of Psychology of Education}}
  \bibinfo{volume}{25}, \bibinfo{number}{2} (\bibinfo{year}{2010}),
  \bibinfo{pages}{222--242}.
\newblock


\bibitem[Festinger(1957)]%
        {festinger1957theory}
\bibfield{author}{\bibinfo{person}{Leon Festinger}.}
  \bibinfo{year}{1957}\natexlab{}.
\newblock \bibinfo{booktitle}{\emph{A theory of cognitive dissonance}}.
  Vol.~\bibinfo{volume}{2}.
\newblock \bibinfo{publisher}{Stanford university press},
  \bibinfo{address}{Redwood City, CA, USA}.
\newblock


\bibitem[Fiorella and Kuhlmann(2020)]%
        {Fiorella_2020}
\bibfield{author}{\bibinfo{person}{Logan Fiorella} {and}
  \bibinfo{person}{Shelbi Kuhlmann}.} \bibinfo{year}{2020}\natexlab{}.
\newblock \showarticletitle{Creating drawings enhances learning by teaching.}
\newblock \bibinfo{journal}{\emph{Journal of Educational Psychology}}
  \bibinfo{volume}{112}, \bibinfo{number}{4} (\bibinfo{date}{may}
  \bibinfo{year}{2020}), \bibinfo{pages}{811--822}.
\newblock
\urldef\tempurl%
\url{https://doi.org/10.1037/edu0000392}
\showDOI{\tempurl}


\bibitem[Fiorella and Mayer(2013)]%
        {Fiorella_2013}
\bibfield{author}{\bibinfo{person}{Logan Fiorella} {and}
  \bibinfo{person}{Richard~E. Mayer}.} \bibinfo{year}{2013}\natexlab{}.
\newblock \showarticletitle{The relative benefits of learning by teaching and
  teaching expectancy}.
\newblock \bibinfo{journal}{\emph{Contemporary Educational Psychology}}
  \bibinfo{volume}{38}, \bibinfo{number}{4} (\bibinfo{date}{oct}
  \bibinfo{year}{2013}), \bibinfo{pages}{281--288}.
\newblock
\urldef\tempurl%
\url{https://doi.org/10.1016/j.cedpsych.2013.06.001}
\showDOI{\tempurl}


\bibitem[Fiorella and Mayer(2014)]%
        {Fiorella_2014}
\bibfield{author}{\bibinfo{person}{Logan Fiorella} {and}
  \bibinfo{person}{Richard~E. Mayer}.} \bibinfo{year}{2014}\natexlab{}.
\newblock \showarticletitle{Role of expectations and explanations in learning
  by teaching}.
\newblock \bibinfo{journal}{\emph{Contemporary Educational Psychology}}
  \bibinfo{volume}{39}, \bibinfo{number}{2} (\bibinfo{date}{apr}
  \bibinfo{year}{2014}), \bibinfo{pages}{75--85}.
\newblock
\urldef\tempurl%
\url{https://doi.org/10.1016/j.cedpsych.2014.01.001}
\showDOI{\tempurl}


\bibitem[Flavell and Wellman(1975)]%
        {flavell1975metamemory}
\bibfield{author}{\bibinfo{person}{John~H Flavell} {and}
  \bibinfo{person}{Henry~M Wellman}.} \bibinfo{year}{1975}\natexlab{}.
\newblock \showarticletitle{Metamemory.}. In
  \bibinfo{booktitle}{\emph{Proceedings of the Annual Meeting of the American
  Psychological Association}}. \bibinfo{publisher}{ERIC},
  \bibinfo{address}{Chicago, IL, USA}, \bibinfo{numpages}{66}~pages.
\newblock


\bibitem[Fremouw and Feindler(1978)]%
        {fremouw1978peer}
\bibfield{author}{\bibinfo{person}{William~J Fremouw} {and}
  \bibinfo{person}{Eva~L Feindler}.} \bibinfo{year}{1978}\natexlab{}.
\newblock \showarticletitle{Peer versus professional models for study skills
  training.}
\newblock \bibinfo{journal}{\emph{Journal of Counseling Psychology}}
  \bibinfo{volume}{25}, \bibinfo{number}{6} (\bibinfo{year}{1978}),
  \bibinfo{pages}{576}.
\newblock


\bibitem[Frisch and Baron(1988)]%
        {frisch1988ambiguity}
\bibfield{author}{\bibinfo{person}{Deborah Frisch} {and}
  \bibinfo{person}{Jonathan Baron}.} \bibinfo{year}{1988}\natexlab{}.
\newblock \showarticletitle{Ambiguity and rationality}.
\newblock \bibinfo{journal}{\emph{Journal of Behavioral Decision Making}}
  \bibinfo{volume}{1}, \bibinfo{number}{3} (\bibinfo{year}{1988}),
  \bibinfo{pages}{149--157}.
\newblock
\urldef\tempurl%
\url{https://doi.org/10.1002/bdm.3960010303}
\showDOI{\tempurl}
\showeprint{https://onlinelibrary.wiley.com/doi/pdf/10.1002/bdm.3960010303}


\bibitem[Galbraith and Winterbottom(2011)]%
        {galbraith2011peer}
\bibfield{author}{\bibinfo{person}{Jonathan Galbraith} {and}
  \bibinfo{person}{Mark Winterbottom}.} \bibinfo{year}{2011}\natexlab{}.
\newblock \showarticletitle{Peer-tutoring: what’s in it for the tutor?}
\newblock \bibinfo{journal}{\emph{Educational Studies}} \bibinfo{volume}{37},
  \bibinfo{number}{3} (\bibinfo{year}{2011}), \bibinfo{pages}{321--332}.
\newblock


\bibitem[Garcia-Melgar and Meyers(2020)]%
        {garcia2020stem}
\bibfield{author}{\bibinfo{person}{Ana Garcia-Melgar} {and}
  \bibinfo{person}{Noel Meyers}.} \bibinfo{year}{2020}\natexlab{}.
\newblock \showarticletitle{STEM near peer mentoring for secondary school
  students: a case study of university mentors’ experiences with online
  mentoring}.
\newblock \bibinfo{journal}{\emph{Journal for STEM Education Research}}
  \bibinfo{volume}{3}, \bibinfo{number}{1} (\bibinfo{year}{2020}),
  \bibinfo{pages}{19--42}.
\newblock


\bibitem[Gartner et~al\mbox{.}(1971)]%
        {gartner1971children}
\bibfield{author}{\bibinfo{person}{Alan Gartner} {et~al\mbox{.}}}
  \bibinfo{year}{1971}\natexlab{}.
\newblock \bibinfo{booktitle}{\emph{Children teach children: Learning by
  teaching.}}
\newblock \bibinfo{publisher}{Harper \& Row}, \bibinfo{address}{Manhattan, NY
  10022, USA}.
\newblock


\bibitem[George(1984)]%
        {george1984working}
\bibfield{author}{\bibinfo{person}{Diana George}.}
  \bibinfo{year}{1984}\natexlab{}.
\newblock \showarticletitle{Working with peer groups in the composition
  classroom}.
\newblock \bibinfo{journal}{\emph{College Composition and Communication}}
  \bibinfo{volume}{35}, \bibinfo{number}{3} (\bibinfo{year}{1984}),
  \bibinfo{pages}{320--326}.
\newblock


\bibitem[Giurgiu(2017)]%
        {giurgiu2017microlearning}
\bibfield{author}{\bibinfo{person}{Lumini{\c{t}}a Giurgiu}.}
  \bibinfo{year}{2017}\natexlab{}.
\newblock \showarticletitle{Microlearning an evolving elearning trend}.
\newblock \bibinfo{journal}{\emph{Scientific Bulletin}} \bibinfo{volume}{22},
  \bibinfo{number}{1} (\bibinfo{year}{2017}), \bibinfo{pages}{18--23}.
\newblock


\bibitem[Graesser and Person(1994)]%
        {graesser1994question}
\bibfield{author}{\bibinfo{person}{Arthur~C Graesser} {and}
  \bibinfo{person}{Natalie~K Person}.} \bibinfo{year}{1994}\natexlab{}.
\newblock \showarticletitle{Question asking during tutoring}.
\newblock \bibinfo{journal}{\emph{American educational research journal}}
  \bibinfo{volume}{31}, \bibinfo{number}{1} (\bibinfo{year}{1994}),
  \bibinfo{pages}{104--137}.
\newblock
\urldef\tempurl%
\url{https://doi.org/10.3102/00028312031001104}
\showDOI{\tempurl}


\bibitem[Gurung et~al\mbox{.}(2010)]%
        {gurung2010focusing}
\bibfield{author}{\bibinfo{person}{Regan~AR Gurung}, \bibinfo{person}{Janet
  Weidert}, {and} \bibinfo{person}{Amanda Jeske}.}
  \bibinfo{year}{2010}\natexlab{}.
\newblock \showarticletitle{Focusing on how students study}.
\newblock \bibinfo{journal}{\emph{Journal of the Scholarship of Teaching and
  Learning}} \bibinfo{volume}{10}, \bibinfo{number}{1} (\bibinfo{year}{2010}),
  \bibinfo{pages}{28--35}.
\newblock


\bibitem[Henderson and Harper(2009)]%
        {Henderson_2009}
\bibfield{author}{\bibinfo{person}{Charles Henderson} {and}
  \bibinfo{person}{Kathleen~A. Harper}.} \bibinfo{year}{2009}\natexlab{}.
\newblock \showarticletitle{Quiz Corrections: Improving Learning by Encouraging
  Students to Reflect on Their Mistakes}.
\newblock \bibinfo{journal}{\emph{The Physics Teacher}} \bibinfo{volume}{47},
  \bibinfo{number}{9} (\bibinfo{date}{dec} \bibinfo{year}{2009}),
  \bibinfo{pages}{581--586}.
\newblock
\urldef\tempurl%
\url{https://doi.org/10.1119/1.3264589}
\showDOI{\tempurl}


\bibitem[Hoogerheide et~al\mbox{.}(2016a)]%
        {hoogerheide2016gaining}
\bibfield{author}{\bibinfo{person}{Vincent Hoogerheide}, \bibinfo{person}{Lian
  Deijkers}, \bibinfo{person}{Sofie~MM Loyens}, \bibinfo{person}{Anita
  Heijltjes}, {and} \bibinfo{person}{Tamara van Gog}.}
  \bibinfo{year}{2016}\natexlab{a}.
\newblock \showarticletitle{Gaining from explaining: Learning improves from
  explaining to fictitious others on video, not from writing to them}.
\newblock \bibinfo{journal}{\emph{Contemporary Educational Psychology}}
  \bibinfo{volume}{44} (\bibinfo{year}{2016}), \bibinfo{pages}{95--106}.
\newblock


\bibitem[Hoogerheide et~al\mbox{.}(2016b)]%
        {Hoogerheide_2016}
\bibfield{author}{\bibinfo{person}{Vincent Hoogerheide}, \bibinfo{person}{Lian
  Deijkers}, \bibinfo{person}{Sofie~M.M. Loyens}, \bibinfo{person}{Anita
  Heijltjes}, {and} \bibinfo{person}{Tamara van Gog}.}
  \bibinfo{year}{2016}\natexlab{b}.
\newblock \showarticletitle{Gaining from explaining: Learning improves from
  explaining to fictitious others on video, not from writing to them}.
\newblock \bibinfo{journal}{\emph{Contemporary Educational Psychology}}
  \bibinfo{volume}{44-45} (\bibinfo{date}{jan} \bibinfo{year}{2016}),
  \bibinfo{pages}{95--106}.
\newblock
\urldef\tempurl%
\url{https://doi.org/10.1016/j.cedpsych.2016.02.005}
\showDOI{\tempurl}


\bibitem[Hoogerheide et~al\mbox{.}(2014)]%
        {Hoogerheide_2014}
\bibfield{author}{\bibinfo{person}{Vincent Hoogerheide},
  \bibinfo{person}{Sofie~M.M. Loyens}, {and} \bibinfo{person}{Tamara van Gog}.}
  \bibinfo{year}{2014}\natexlab{}.
\newblock \showarticletitle{Effects of creating video-based modeling examples
  on learning and transfer}.
\newblock \bibinfo{journal}{\emph{Learning and Instruction}}
  \bibinfo{volume}{33} (\bibinfo{date}{oct} \bibinfo{year}{2014}),
  \bibinfo{pages}{108--119}.
\newblock
\urldef\tempurl%
\url{https://doi.org/10.1016/j.learninstruc.2014.04.005}
\showDOI{\tempurl}


\bibitem[Huff and Kline(1987)]%
        {huff1987contemporary}
\bibfield{author}{\bibinfo{person}{Roland Huff} {and}
  \bibinfo{person}{Charles~R Kline}.} \bibinfo{year}{1987}\natexlab{}.
\newblock \bibinfo{booktitle}{\emph{The contemporary writing curriculum:
  Rehearsing, composing, and valuing}}.
\newblock \bibinfo{publisher}{Teachers College Press},
  \bibinfo{address}{Dulles, VA, USA}.
\newblock


\bibitem[Hug and Friesen(2009)]%
        {hug2007outline}
\bibfield{author}{\bibinfo{person}{Theo Hug} {and} \bibinfo{person}{Norm
  Friesen}.} \bibinfo{year}{2009}\natexlab{}.
\newblock \showarticletitle{Outline of a microlearning agenda}. In
  \bibinfo{booktitle}{\emph{eLearning Papers}}.
  \bibinfo{publisher}{elearningeuropa.info}, \bibinfo{address}{Barcelona,
  Spain}, \bibinfo{pages}{1--13}.
\newblock


\bibitem[Jackson(2017)]%
        {jackson2017fear}
\bibfield{author}{\bibinfo{person}{Carolyn Jackson}.}
  \bibinfo{year}{2017}\natexlab{}.
\newblock \showarticletitle{Fear of failure}.
\newblock In \bibinfo{booktitle}{\emph{Understanding Learning and Motivation in
  Youth}}. \bibinfo{publisher}{Routledge}, \bibinfo{address}{Oxfordshire,
  England, UK}, \bibinfo{pages}{30--39}.
\newblock


\bibitem[Joshi et~al\mbox{.}(2020)]%
        {joshi2020micromentor}
\bibfield{author}{\bibinfo{person}{Nikhita Joshi}, \bibinfo{person}{Justin
  Matejka}, \bibinfo{person}{Fraser Anderson}, \bibinfo{person}{Tovi Grossman},
  {and} \bibinfo{person}{George Fitzmaurice}.} \bibinfo{year}{2020}\natexlab{}.
\newblock \showarticletitle{MicroMentor: Peer-to-Peer Software Help Sessions in
  Three Minutes or Less}. In \bibinfo{booktitle}{\emph{Proceedings of the 2020
  CHI Conference on Human Factors in Computing Systems}}.
  \bibinfo{publisher}{ACM}, \bibinfo{address}{New York, NY, USA},
  \bibinfo{pages}{1--13}.
\newblock


\bibitem[Kahneman and Tversky(1979)]%
        {Kahneman1979}
\bibfield{author}{\bibinfo{person}{Daniel Kahneman} {and} \bibinfo{person}{Amos
  Tversky}.} \bibinfo{year}{1979}\natexlab{}.
\newblock \showarticletitle{Prospect Theory: An Analysis of Decision under
  Risk}.
\newblock \bibinfo{journal}{\emph{Econometrica}} \bibinfo{volume}{47},
  \bibinfo{number}{2} (\bibinfo{date}{March} \bibinfo{year}{1979}),
  \bibinfo{pages}{263}.
\newblock
\urldef\tempurl%
\url{https://doi.org/10.2307/1914185}
\showDOI{\tempurl}


\bibitem[Kamilali and Sofianopoulou(2015)]%
        {kamilali2015microlearning}
\bibfield{author}{\bibinfo{person}{Despina Kamilali} {and}
  \bibinfo{person}{Chryssa Sofianopoulou}.} \bibinfo{year}{2015}\natexlab{}.
\newblock \showarticletitle{Microlearning as Innovative Pedagogy for Mobile
  Learning in MOOCs.}. In \bibinfo{booktitle}{\emph{Proceedings of the 11th
  International Conference on Mobile Learning 2015}}.
  \bibinfo{publisher}{International Association for Development of the
  Information Society}, \bibinfo{address}{Portugal},
  \bibinfo{numpages}{5}~pages.
\newblock


\bibitem[Kapur(2008)]%
        {kapur2008productive}
\bibfield{author}{\bibinfo{person}{Manu Kapur}.}
  \bibinfo{year}{2008}\natexlab{}.
\newblock \showarticletitle{Productive failure}.
\newblock \bibinfo{journal}{\emph{Cognition and instruction}}
  \bibinfo{volume}{26}, \bibinfo{number}{3} (\bibinfo{year}{2008}),
  \bibinfo{pages}{379--424}.
\newblock


\bibitem[Kapur(2016)]%
        {kapur2016examining}
\bibfield{author}{\bibinfo{person}{Manu Kapur}.}
  \bibinfo{year}{2016}\natexlab{}.
\newblock \showarticletitle{Examining productive failure, productive success,
  unproductive failure, and unproductive success in learning}.
\newblock \bibinfo{journal}{\emph{Educational Psychologist}}
  \bibinfo{volume}{51}, \bibinfo{number}{2} (\bibinfo{year}{2016}),
  \bibinfo{pages}{289--299}.
\newblock


\bibitem[Kapur and Bielaczyc(2012)]%
        {kapur2012designing}
\bibfield{author}{\bibinfo{person}{Manu Kapur} {and} \bibinfo{person}{Katerine
  Bielaczyc}.} \bibinfo{year}{2012}\natexlab{}.
\newblock \showarticletitle{Designing for productive failure}.
\newblock \bibinfo{journal}{\emph{Journal of the Learning Sciences}}
  \bibinfo{volume}{21}, \bibinfo{number}{1} (\bibinfo{year}{2012}),
  \bibinfo{pages}{45--83}.
\newblock


\bibitem[King et~al\mbox{.}(1998)]%
        {King_1998}
\bibfield{author}{\bibinfo{person}{Alison King}, \bibinfo{person}{Anne
  Staffieri}, {and} \bibinfo{person}{Anne Adelgais}.}
  \bibinfo{year}{1998}\natexlab{}.
\newblock \showarticletitle{Mutual peer tutoring: Effects of structuring
  tutorial interaction to scaffold peer learning.}
\newblock \bibinfo{journal}{\emph{Journal of Educational Psychology}}
  \bibinfo{volume}{90}, \bibinfo{number}{1} (\bibinfo{year}{1998}),
  \bibinfo{pages}{134--152}.
\newblock
\urldef\tempurl%
\url{https://doi.org/10.1037/0022-0663.90.1.134}
\showDOI{\tempurl}


\bibitem[Kuhn(2000)]%
        {kuhn2000metacognitive}
\bibfield{author}{\bibinfo{person}{Deanna Kuhn}.}
  \bibinfo{year}{2000}\natexlab{}.
\newblock \showarticletitle{Metacognitive development}.
\newblock \bibinfo{journal}{\emph{Current directions in psychological science}}
  \bibinfo{volume}{9}, \bibinfo{number}{5} (\bibinfo{year}{2000}),
  \bibinfo{pages}{178--181}.
\newblock


\bibitem[Kuhn and Pease(2006)]%
        {kuhn2006children}
\bibfield{author}{\bibinfo{person}{Deanna Kuhn} {and} \bibinfo{person}{Maria
  Pease}.} \bibinfo{year}{2006}\natexlab{}.
\newblock \showarticletitle{Do children and adults learn differently?}
\newblock \bibinfo{journal}{\emph{Journal of cognition and development}}
  \bibinfo{volume}{7}, \bibinfo{number}{3} (\bibinfo{year}{2006}),
  \bibinfo{pages}{279--293}.
\newblock


\bibitem[Kumar et~al\mbox{.}(1998)]%
        {kumar1998virtual}
\bibfield{author}{\bibinfo{person}{Anup Kumar}, \bibinfo{person}{Raj Pakala},
  \bibinfo{person}{RK Ragade}, {and} \bibinfo{person}{JP Wong}.}
  \bibinfo{year}{1998}\natexlab{}.
\newblock \showarticletitle{The virtual learning environment system}. In
  \bibinfo{booktitle}{\emph{FIE'98. 28th Annual Frontiers in Education
  Conference. Moving from 'Teacher-Centered' to 'Learner-Centered' Education.
  Conference Proceedings (Cat. No. 98CH36214)}}, Vol.~\bibinfo{volume}{2}.
  \bibinfo{publisher}{IEEE}, \bibinfo{address}{Los Alamitos, CA},
  \bibinfo{pages}{711--716}.
\newblock


\bibitem[Kumar(1996)]%
        {kumar1996computer}
\bibfield{author}{\bibinfo{person}{Vivekanandan~Suresh Kumar}.}
  \bibinfo{year}{1996}\natexlab{}.
\newblock \showarticletitle{Computer-supported collaborative learning: issues
  for research}. In \bibinfo{booktitle}{\emph{Eighth annual graduate symposium
  on Computer Science, University of Saskatchewan}}.
  \bibinfo{publisher}{University of Saskatchewan},
  \bibinfo{address}{Saskatchewan, Canada}, \bibinfo{numpages}{22}~pages.
\newblock


\bibitem[Lee et~al\mbox{.}(2021)]%
        {Lee2021}
\bibfield{author}{\bibinfo{person}{Ken~Jen Lee}, \bibinfo{person}{Apoorva
  Chauhan}, \bibinfo{person}{Joslin Goh}, \bibinfo{person}{Elizabeth Nilsen},
  {and} \bibinfo{person}{Edith Law}.} \bibinfo{year}{2021}\natexlab{}.
\newblock \showarticletitle{Curiosity Notebook: The Design of a Research
  Platform for Learning by Teaching}.
\newblock \bibinfo{journal}{\emph{Proc. ACM Hum.-Comput. Interact.}}
  \bibinfo{volume}{5}, \bibinfo{number}{CSCW2}, Article
  \bibinfo{articleno}{394} (\bibinfo{year}{2021}),
  \bibinfo{numpages}{26}~pages.
\newblock


\bibitem[Leelawong and Biswas(2008)]%
        {leelawong2008designing}
\bibfield{author}{\bibinfo{person}{Krittaya Leelawong} {and}
  \bibinfo{person}{Gautam Biswas}.} \bibinfo{year}{2008}\natexlab{}.
\newblock \showarticletitle{Designing learning by teaching agents: The Betty's
  Brain system}.
\newblock \bibinfo{journal}{\emph{International Journal of Artificial
  Intelligence in Education}} \bibinfo{volume}{18}, \bibinfo{number}{3}
  (\bibinfo{year}{2008}), \bibinfo{pages}{181--208}.
\newblock


\bibitem[Leikin(2006)]%
        {leikin2006learning}
\bibfield{author}{\bibinfo{person}{Roza Leikin}.}
  \bibinfo{year}{2006}\natexlab{}.
\newblock \bibinfo{booktitle}{\emph{Learning by teaching: The case of Sieve of
  Eratosthenes and one elementary school teacher}}.
\newblock \bibinfo{publisher}{Routledge}, \bibinfo{address}{Oxfordshire,
  England, UK}, \bibinfo{pages}{115--140}.
\newblock


\bibitem[Loewenstein et~al\mbox{.}(2001)]%
        {loewenstein2001risk}
\bibfield{author}{\bibinfo{person}{George~F Loewenstein},
  \bibinfo{person}{Elke~U Weber}, \bibinfo{person}{Christopher~K Hsee}, {and}
  \bibinfo{person}{Ned Welch}.} \bibinfo{year}{2001}\natexlab{}.
\newblock \showarticletitle{Risk as feelings.}
\newblock \bibinfo{journal}{\emph{Psychological bulletin}}
  \bibinfo{volume}{127}, \bibinfo{number}{2} (\bibinfo{year}{2001}),
  \bibinfo{pages}{267}.
\newblock


\bibitem[Mason and Singh(2010)]%
        {Mason_2010}
\bibfield{author}{\bibinfo{person}{Andrew Mason} {and}
  \bibinfo{person}{Chandralekha Singh}.} \bibinfo{year}{2010}\natexlab{}.
\newblock \showarticletitle{Do advanced physics students learn from their
  mistakes without explicit intervention?}
\newblock \bibinfo{journal}{\emph{American Journal of Physics}}
  \bibinfo{volume}{78}, \bibinfo{number}{7} (\bibinfo{date}{jul}
  \bibinfo{year}{2010}), \bibinfo{pages}{760--767}.
\newblock
\urldef\tempurl%
\url{https://doi.org/10.1119/1.3318805}
\showDOI{\tempurl}


\bibitem[Matsuda et~al\mbox{.}(2011)]%
        {matsuda2011learning}
\bibfield{author}{\bibinfo{person}{Noboru Matsuda}, \bibinfo{person}{Evelyn
  Yarzebinski}, \bibinfo{person}{Victoria Keiser}, \bibinfo{person}{Rohan
  Raizada}, \bibinfo{person}{Gabriel~J Stylianides}, \bibinfo{person}{William~W
  Cohen}, {and} \bibinfo{person}{Kenneth~R Koedinger}.}
  \bibinfo{year}{2011}\natexlab{}.
\newblock \showarticletitle{Learning by teaching SimStudent--An initial
  classroom baseline study comparing with Cognitive Tutor}. In
  \bibinfo{booktitle}{\emph{International Conference on Artificial Intelligence
  in Education}}. \bibinfo{publisher}{Springer}, \bibinfo{address}{New York
  City, NY, USA}, \bibinfo{pages}{213--221}.
\newblock


\bibitem[Matsuda et~al\mbox{.}(2013)]%
        {matsuda2013studying}
\bibfield{author}{\bibinfo{person}{Noboru Matsuda}, \bibinfo{person}{Evelyn
  Yarzebinski}, \bibinfo{person}{Victoria Keiser}, \bibinfo{person}{Rohan
  Raizada}, \bibinfo{person}{Gabriel~J Stylianides}, {and}
  \bibinfo{person}{Kenneth~R Koedinger}.} \bibinfo{year}{2013}\natexlab{}.
\newblock \showarticletitle{Studying the effect of a competitive game show in a
  learning by teaching environment}.
\newblock \bibinfo{journal}{\emph{International Journal of Artificial
  Intelligence in Education}} \bibinfo{volume}{23}, \bibinfo{number}{1-4}
  (\bibinfo{year}{2013}), \bibinfo{pages}{1--21}.
\newblock


\bibitem[McKellar(1986)]%
        {mckellar1986behaviors}
\bibfield{author}{\bibinfo{person}{Nancy~A McKellar}.}
  \bibinfo{year}{1986}\natexlab{}.
\newblock \showarticletitle{Behaviors used in peer tutoring}.
\newblock \bibinfo{journal}{\emph{The Journal of Experimental Education}}
  \bibinfo{volume}{54}, \bibinfo{number}{3} (\bibinfo{year}{1986}),
  \bibinfo{pages}{163--167}.
\newblock


\bibitem[McNamara(2004)]%
        {McNamara_2004}
\bibfield{author}{\bibinfo{person}{Danielle~S. McNamara}.}
  \bibinfo{year}{2004}\natexlab{}.
\newblock \showarticletitle{{SERT}: Self-Explanation Reading Training}.
\newblock \bibinfo{journal}{\emph{Discourse Processes}} \bibinfo{volume}{38},
  \bibinfo{number}{1} (\bibinfo{date}{jul} \bibinfo{year}{2004}),
  \bibinfo{pages}{1--30}.
\newblock
\urldef\tempurl%
\url{https://doi.org/10.1207/s15326950dp3801_1}
\showDOI{\tempurl}


\bibitem[Miller et~al\mbox{.}(1994)]%
        {miller1994group}
\bibfield{author}{\bibinfo{person}{Judith~E Miller} {et~al\mbox{.}}}
  \bibinfo{year}{1994}\natexlab{}.
\newblock \showarticletitle{Group Dynamics: Understanding Group Success and
  Failure in Collaborative Learning.}
\newblock \bibinfo{journal}{\emph{New directions for teaching and learning}}
  \bibinfo{volume}{59} (\bibinfo{year}{1994}), \bibinfo{pages}{33--44}.
\newblock


\bibitem[Modigliani(1971)]%
        {modigliani1971embarrassment}
\bibfield{author}{\bibinfo{person}{Andre Modigliani}.}
  \bibinfo{year}{1971}\natexlab{}.
\newblock \showarticletitle{Embarrassment, facework, and eye contact: Testing a
  theory of embarrassment.}
\newblock \bibinfo{journal}{\emph{Journal of Personality and social
  Psychology}} \bibinfo{volume}{17}, \bibinfo{number}{1}
  (\bibinfo{year}{1971}), \bibinfo{pages}{15}.
\newblock


\bibitem[Mohammed et~al\mbox{.}(2018)]%
        {mohammed2018effectiveness}
\bibfield{author}{\bibinfo{person}{Gona~Sirwan Mohammed},
  \bibinfo{person}{Karzan Wakil}, {and} \bibinfo{person}{Sarkhell~Sirwan
  Nawroly}.} \bibinfo{year}{2018}\natexlab{}.
\newblock \showarticletitle{The effectiveness of microlearning to improve
  students’ learning ability}.
\newblock \bibinfo{journal}{\emph{International Journal of Educational Research
  Review}} \bibinfo{volume}{3}, \bibinfo{number}{3} (\bibinfo{year}{2018}),
  \bibinfo{pages}{32--38}.
\newblock


\bibitem[Moy and Rinke(2012)]%
        {moy2012attitudinal}
\bibfield{author}{\bibinfo{person}{Patricia Moy} {and}
  \bibinfo{person}{Eike~Mark Rinke}.} \bibinfo{year}{2012}\natexlab{}.
\newblock \showarticletitle{Attitudinal and behavioral consequences of
  published opinion polls}.
\newblock In \bibinfo{booktitle}{\emph{Opinion polls and the media}}.
  \bibinfo{publisher}{Springer}, \bibinfo{address}{New York City, NY, USA},
  \bibinfo{pages}{225--245}.
\newblock


\bibitem[Murphy and Stay(2006)]%
        {Murphy2006}
\bibfield{author}{\bibinfo{person}{Christina Murphy} {and}
  \bibinfo{person}{Bryon~L Stay}.} \bibinfo{year}{2006}\natexlab{}.
\newblock \bibinfo{booktitle}{\emph{The writing center director's resource
  book}}.
\newblock \bibinfo{publisher}{Lawrence Erlbaum Associates, Publishers},
  \bibinfo{address}{Mahwah, N.J}.
\newblock


\bibitem[Peker(2009)]%
        {peker2009pre}
\bibfield{author}{\bibinfo{person}{M Peker}.} \bibinfo{year}{2009}\natexlab{}.
\newblock \showarticletitle{Pre-service mathematics teacher perspectives about
  the expanded microteaching experiences}.
\newblock \bibinfo{journal}{\emph{Journal of Turkish Educational Science}}
  \bibinfo{volume}{7}, \bibinfo{number}{2} (\bibinfo{year}{2009}),
  \bibinfo{pages}{353--376}.
\newblock


\bibitem[Person and Graesser(1999)]%
        {graesser2014evolution}
\bibfield{author}{\bibinfo{person}{Natalie~K Person} {and}
  \bibinfo{person}{Arthur~G Graesser}.} \bibinfo{year}{1999}\natexlab{}.
\newblock \bibinfo{booktitle}{\emph{Evolution of Discourse During Cross-Age
  Tutoring}}.
\newblock \bibinfo{publisher}{Lawrence Erlbaum Associates},
  \bibinfo{address}{Mahwah, NJ, USA}, \bibinfo{pages}{69--86}.
\newblock


\bibitem[Phiri et~al\mbox{.}(2017)]%
        {phiri2017peer}
\bibfield{author}{\bibinfo{person}{Lighton Phiri}, \bibinfo{person}{Christoph
  Meinel}, {and} \bibinfo{person}{Hussein Suleman}.}
  \bibinfo{year}{2017}\natexlab{}.
\newblock \showarticletitle{Peer tutoring orchestration: Streamlined
  technology-driven orchestration for peer tutoring}. In
  \bibinfo{booktitle}{\emph{Proceedings of the 9th International Conference on
  Computer Supported Education}}, Vol.~\bibinfo{volume}{1}.
  \bibinfo{publisher}{SciTePress}, \bibinfo{address}{Setúbal, Portugal},
  \bibinfo{pages}{434--441}.
\newblock


\bibitem[Ploetzner et~al\mbox{.}(1999)]%
        {ploetzner1999learning}
\bibfield{author}{\bibinfo{person}{Rolf Ploetzner}, \bibinfo{person}{Pierre
  Dillenbourg}, \bibinfo{person}{Michael Preier}, {and} \bibinfo{person}{David
  Traum}.} \bibinfo{year}{1999}\natexlab{}.
\newblock \showarticletitle{Learning by explaining to oneself and to others}.
\newblock \bibinfo{journal}{\emph{Collaborative learning: Cognitive and
  computational approaches}}  \bibinfo{volume}{1} (\bibinfo{year}{1999}),
  \bibinfo{pages}{103--121}.
\newblock


\bibitem[Post and van~der Molen(2019)]%
        {post2019development}
\bibfield{author}{\bibinfo{person}{Tim Post} {and} \bibinfo{person}{Juliette
  H~Walma van~der Molen}.} \bibinfo{year}{2019}\natexlab{}.
\newblock \showarticletitle{Development and validation of a questionnaire to
  measure primary school children’s images of and attitudes towards curiosity
  (the CIAC questionnaire)}.
\newblock \bibinfo{journal}{\emph{Motivation and emotion}}
  \bibinfo{volume}{43}, \bibinfo{number}{1} (\bibinfo{year}{2019}),
  \bibinfo{pages}{159--178}.
\newblock
\urldef\tempurl%
\url{https://doi.org/10.1007/s11031-018-9728-9}
\showDOI{\tempurl}


\bibitem[Ramadani and Xhaferi(2020)]%
        {ramadani2020teachers}
\bibfield{author}{\bibinfo{person}{Adelina Ramadani} {and}
  \bibinfo{person}{Brikena Xhaferi}.} \bibinfo{year}{2020}\natexlab{}.
\newblock \showarticletitle{Teachers’ experiences with online teaching using
  the zoom platform with efl teachers in high schools in kumanova}.
\newblock \bibinfo{journal}{\emph{Seeu Review}} \bibinfo{volume}{15},
  \bibinfo{number}{1} (\bibinfo{year}{2020}), \bibinfo{pages}{142--155}.
\newblock


\bibitem[Ravari et~al\mbox{.}(2021)]%
        {ravarieffects}
\bibfield{author}{\bibinfo{person}{Parastoo~Baghaei Ravari},
  \bibinfo{person}{Ken~Jen Lee}, \bibinfo{person}{Edith Law}, {and}
  \bibinfo{person}{Dana Kulic}.} \bibinfo{year}{2021}\natexlab{}.
\newblock \showarticletitle{Effects of an Adaptive Robot Encouraging Teamwork
  on Students’ Learning}. In \bibinfo{booktitle}{\emph{2021 30th IEEE
  International Symposium on Robot and Human Interactive Communication
  (RO-MAN)}}. \bibinfo{publisher}{ACM}, \bibinfo{address}{New York, NY, USA},
  \bibinfo{numpages}{8}~pages.
\newblock


\bibitem[Reigstad and McAndrew(1984)]%
        {reigstad1984training}
\bibfield{author}{\bibinfo{person}{Thomas~J Reigstad} {and}
  \bibinfo{person}{Donald~A McAndrew}.} \bibinfo{year}{1984}\natexlab{}.
\newblock \bibinfo{booktitle}{\emph{Training Tutors for Writing Conferences.}}
\newblock \bibinfo{publisher}{ERIC}, \bibinfo{address}{Urbana, IL 61801, USA}.
\newblock


\bibitem[Richardson(1994)]%
        {richardson1994mature}
\bibfield{author}{\bibinfo{person}{John~TE Richardson}.}
  \bibinfo{year}{1994}\natexlab{}.
\newblock \showarticletitle{Mature students in higher education: I. A
  literature survey on approaches to studying}.
\newblock \bibinfo{journal}{\emph{Studies in Higher Education}}
  \bibinfo{volume}{19}, \bibinfo{number}{3} (\bibinfo{year}{1994}),
  \bibinfo{pages}{309--325}.
\newblock


\bibitem[Riemer and Shavitt(2011)]%
        {riemer2011impression}
\bibfield{author}{\bibinfo{person}{Hila Riemer} {and} \bibinfo{person}{Sharon
  Shavitt}.} \bibinfo{year}{2011}\natexlab{}.
\newblock \showarticletitle{Impression management in survey responding: Easier
  for collectivists or individualists?}
\newblock \bibinfo{journal}{\emph{Journal of Consumer Psychology}}
  \bibinfo{volume}{21}, \bibinfo{number}{2} (\bibinfo{year}{2011}),
  \bibinfo{pages}{157--168}.
\newblock


\bibitem[Rienties et~al\mbox{.}(2009)]%
        {rienties2009role}
\bibfield{author}{\bibinfo{person}{Bart Rienties}, \bibinfo{person}{Dirk
  Tempelaar}, \bibinfo{person}{Piet Van~den Bossche}, \bibinfo{person}{Wim
  Gijselaers}, {and} \bibinfo{person}{Mien Segers}.}
  \bibinfo{year}{2009}\natexlab{}.
\newblock \showarticletitle{The role of academic motivation in
  Computer-Supported Collaborative Learning}.
\newblock \bibinfo{journal}{\emph{Computers in Human Behavior}}
  \bibinfo{volume}{25}, \bibinfo{number}{6} (\bibinfo{year}{2009}),
  \bibinfo{pages}{1195--1206}.
\newblock


\bibitem[Rikkers(2002)]%
        {rikkers2002bandwagon}
\bibfield{author}{\bibinfo{person}{Layton~F Rikkers}.}
  \bibinfo{year}{2002}\natexlab{}.
\newblock \bibinfo{title}{The bandwagon effect}.
\newblock
\newblock


\bibitem[Robin and Heselton(1977)]%
        {robin1977proctor}
\bibfield{author}{\bibinfo{person}{Arthur~L Robin} {and}
  \bibinfo{person}{Patricia Heselton}.} \bibinfo{year}{1977}\natexlab{}.
\newblock \showarticletitle{Proctor training: The effects of a manual versus
  direct training.}
\newblock \bibinfo{journal}{\emph{Journal of personalized instruction}}
  \bibinfo{volume}{2}, \bibinfo{number}{1} (\bibinfo{year}{1977}),
  \bibinfo{pages}{19--24}.
\newblock


\bibitem[Roscoe(2014)]%
        {roscoe2014self}
\bibfield{author}{\bibinfo{person}{Rod~D Roscoe}.}
  \bibinfo{year}{2014}\natexlab{}.
\newblock \showarticletitle{Self-monitoring and knowledge-building in learning
  by teaching}.
\newblock \bibinfo{journal}{\emph{Instructional Science}} \bibinfo{volume}{42},
  \bibinfo{number}{3} (\bibinfo{year}{2014}), \bibinfo{pages}{327--351}.
\newblock


\bibitem[Roscoe and Chi(2004)]%
        {roscoe2004influence}
\bibfield{author}{\bibinfo{person}{Rod~D Roscoe} {and}
  \bibinfo{person}{Michelene~TH Chi}.} \bibinfo{year}{2004}\natexlab{}.
\newblock \showarticletitle{The influence of the tutee in learning by peer
  tutoring}. In \bibinfo{booktitle}{\emph{Proceedings of the Annual Meeting of
  the Cognitive Science Society}}, Vol.~\bibinfo{volume}{26}.
  \bibinfo{publisher}{eScholarship}, \bibinfo{address}{Oakland, CA, USA},
  \bibinfo{pages}{1179--1184}.
\newblock


\bibitem[Roscoe and Chi(2007)]%
        {roscoe2007understanding}
\bibfield{author}{\bibinfo{person}{Rod~D Roscoe} {and}
  \bibinfo{person}{Michelene~TH Chi}.} \bibinfo{year}{2007}\natexlab{}.
\newblock \showarticletitle{Understanding tutor learning: Knowledge-building
  and knowledge-telling in peer tutors’ explanations and questions}.
\newblock \bibinfo{journal}{\emph{Review of educational research}}
  \bibinfo{volume}{77}, \bibinfo{number}{4} (\bibinfo{year}{2007}),
  \bibinfo{pages}{534--574}.
\newblock


\bibitem[Roscoe and Chi(2008)]%
        {roscoe2008tutor}
\bibfield{author}{\bibinfo{person}{Rod~D Roscoe} {and}
  \bibinfo{person}{Michelene~TH Chi}.} \bibinfo{year}{2008}\natexlab{}.
\newblock \showarticletitle{Tutor learning: The role of explaining and
  responding to questions}.
\newblock \bibinfo{journal}{\emph{Instructional science}} \bibinfo{volume}{36},
  \bibinfo{number}{4} (\bibinfo{year}{2008}), \bibinfo{pages}{321--350}.
\newblock


\bibitem[S.(2012)]%
        {al2012comparative}
\bibfield{author}{\bibinfo{person}{Ajlan S.}} \bibinfo{year}{2012}\natexlab{}.
\newblock \showarticletitle{A Comparative Study Between E-Learning Features}.
\newblock In \bibinfo{booktitle}{\emph{Methodologies, Tools and New
  Developments for E-Learning}}. \bibinfo{publisher}{{IntechOpen}},
  \bibinfo{address}{London, UK}.
\newblock
\urldef\tempurl%
\url{https://doi.org/10.5772/29854}
\showDOI{\tempurl}


\bibitem[Sayem et~al\mbox{.}(2017)]%
        {sayem2017effective}
\bibfield{author}{\bibinfo{person}{Abu Shadat~Muhammad Sayem},
  \bibinfo{person}{Benjamin Taylor}, \bibinfo{person}{Mitchell McClanachan},
  {and} \bibinfo{person}{Umme Mumtahina}.} \bibinfo{year}{2017}\natexlab{}.
\newblock \showarticletitle{Effective use of zoom technology and instructional
  videos to improve engagement and success of distance students in
  engineering}. In \bibinfo{booktitle}{\emph{Proceedings of the 28th annual
  conference of the Australasian association for engineering education (AAEE
  2017)}}, Vol.~\bibinfo{volume}{926}. \bibinfo{publisher}{Australasian
  Association for Engineering Education}, \bibinfo{address}{Barton, Australia},
  \bibinfo{numpages}{6}~pages.
\newblock


\bibitem[Schein(1995)]%
        {schein1995organizational}
\bibfield{author}{\bibinfo{person}{Edgar~H Schein}.}
  \bibinfo{year}{1995}\natexlab{}.
\newblock \bibinfo{title}{Organizational and managerial culture as a
  facilitator or inhibitor of organizational transformation}.
  (\bibinfo{year}{1995}), \bibinfo{numpages}{28}~pages.
\newblock


\bibitem[Schwartz and Bransford(1998)]%
        {schwartz1998time}
\bibfield{author}{\bibinfo{person}{Daniel~L Schwartz} {and}
  \bibinfo{person}{John~D Bransford}.} \bibinfo{year}{1998}\natexlab{}.
\newblock \showarticletitle{A Time for Telling}.
\newblock \bibinfo{journal}{\emph{Cognition and Instruction}}
  \bibinfo{volume}{16}, \bibinfo{number}{4} (\bibinfo{year}{1998}),
  \bibinfo{pages}{475--5223}.
\newblock


\bibitem[Seaward and Kemp(2000)]%
        {seaward2000optimism}
\bibfield{author}{\bibinfo{person}{Hamish~GW Seaward} {and}
  \bibinfo{person}{Simon Kemp}.} \bibinfo{year}{2000}\natexlab{}.
\newblock \showarticletitle{Optimism bias and student debt}.
\newblock \bibinfo{journal}{\emph{New Zealand Journal of Psychology}}
  \bibinfo{volume}{29}, \bibinfo{number}{1} (\bibinfo{year}{2000}),
  \bibinfo{pages}{17--19}.
\newblock


\bibitem[Shah et~al\mbox{.}(2014)]%
        {shah2014analyzing}
\bibfield{author}{\bibinfo{person}{Niral Shah}, \bibinfo{person}{Colleen
  Lewis}, {and} \bibinfo{person}{Roxane Caires}.}
  \bibinfo{year}{2014}\natexlab{}.
\newblock \showarticletitle{Analyzing equity in collaborative learning
  situations: A comparative case study in elementary computer science}. In
  \bibinfo{booktitle}{\emph{Proceedings of the 11th International Conference of
  the Learning Sciences}}. \bibinfo{publisher}{International Society of the
  Learning Sciences}, \bibinfo{address}{Boulder, CO, USA},
  \bibinfo{pages}{495--502}.
\newblock


\bibitem[Shamoon and Burns(1995)]%
        {shamoon1995critique}
\bibfield{author}{\bibinfo{person}{Linda~K Shamoon} {and}
  \bibinfo{person}{Deborah~H Burns}.} \bibinfo{year}{1995}\natexlab{}.
\newblock \showarticletitle{A critique of pure tutoring}.
\newblock \bibinfo{journal}{\emph{The Writing Center Journal}}
  \bibinfo{volume}{15}, \bibinfo{number}{2} (\bibinfo{year}{1995}),
  \bibinfo{pages}{134--151}.
\newblock


\bibitem[Sharpley and Sharpley(1981)]%
        {sharpley1981}
\bibfield{author}{\bibinfo{person}{A.~M. Sharpley} {and} \bibinfo{person}{C.~E.
  Sharpley}.} \bibinfo{year}{1981}\natexlab{}.
\newblock \showarticletitle{Peer tutoring: A review of the literature}.
\newblock \bibinfo{journal}{\emph{Collected Original Resources in Education}}
  \bibinfo{volume}{5}, \bibinfo{number}{3} (\bibinfo{year}{1981}),
  \bibinfo{pages}{7--11}.
\newblock


\bibitem[Sinha and Kapur(2019)]%
        {sinha2019productive}
\bibfield{author}{\bibinfo{person}{Tanmay Sinha} {and} \bibinfo{person}{Manu
  Kapur}.} \bibinfo{year}{2019}\natexlab{}.
\newblock \showarticletitle{When productive failure fails}.
\newblock \bibinfo{journal}{\emph{Europe (Germany, Switzerland, UK)}}
  \bibinfo{volume}{30} (\bibinfo{year}{2019}), \bibinfo{pages}{31--6}.
\newblock


\bibitem[Slavin(1980)]%
        {slavin1980cooperative}
\bibfield{author}{\bibinfo{person}{Robert~E Slavin}.}
  \bibinfo{year}{1980}\natexlab{}.
\newblock \showarticletitle{Cooperative learning}.
\newblock \bibinfo{journal}{\emph{Review of educational research}}
  \bibinfo{volume}{50}, \bibinfo{number}{2} (\bibinfo{year}{1980}),
  \bibinfo{pages}{315--342}.
\newblock


\bibitem[Stanley(1992)]%
        {stanley1992coaching}
\bibfield{author}{\bibinfo{person}{Jane Stanley}.}
  \bibinfo{year}{1992}\natexlab{}.
\newblock \showarticletitle{Coaching student writers to be effective peer
  evaluators}.
\newblock \bibinfo{journal}{\emph{Journal of second language writing}}
  \bibinfo{volume}{1}, \bibinfo{number}{3} (\bibinfo{year}{1992}),
  \bibinfo{pages}{217--233}.
\newblock


\bibitem[Toney et~al\mbox{.}(2021)]%
        {toney2021fighting}
\bibfield{author}{\bibinfo{person}{Scott Toney}, \bibinfo{person}{Jenn Light},
  {and} \bibinfo{person}{Andrew Urbaczewski}.} \bibinfo{year}{2021}\natexlab{}.
\newblock \showarticletitle{Fighting Zoom fatigue: Keeping the zoombies at
  bay}.
\newblock \bibinfo{journal}{\emph{Communications of the Association for
  Information Systems}} \bibinfo{volume}{48}, \bibinfo{number}{1}
  (\bibinfo{year}{2021}), \bibinfo{pages}{10}.
\newblock


\bibitem[Topping(1988)]%
        {topping1988peer}
\bibfield{author}{\bibinfo{person}{Keith Topping}.}
  \bibinfo{year}{1988}\natexlab{}.
\newblock \bibinfo{booktitle}{\emph{The Peer Tutoring Handbook: Promoting
  Co-Operative Learning.}}
\newblock \bibinfo{publisher}{Brookline Books}, \bibinfo{address}{Cambridge, MA
  02238, USA}.
\newblock


\bibitem[Topping(1996)]%
        {topping1996effectiveness}
\bibfield{author}{\bibinfo{person}{Keith~J Topping}.}
  \bibinfo{year}{1996}\natexlab{}.
\newblock \showarticletitle{The effectiveness of peer tutoring in further and
  higher education: A typology and review of the literature}.
\newblock \bibinfo{journal}{\emph{Higher education}} \bibinfo{volume}{32},
  \bibinfo{number}{3} (\bibinfo{year}{1996}), \bibinfo{pages}{321--345}.
\newblock


\bibitem[Topping(2005)]%
        {Topping2005}
\bibfield{author}{\bibinfo{person}{Keith~J. Topping}.}
  \bibinfo{year}{2005}\natexlab{}.
\newblock \showarticletitle{Trends in Peer Learning}.
\newblock \bibinfo{journal}{\emph{Educational Psychology}}
  \bibinfo{volume}{25}, \bibinfo{number}{6} (\bibinfo{date}{Dec.}
  \bibinfo{year}{2005}), \bibinfo{pages}{631--645}.
\newblock
\urldef\tempurl%
\url{https://doi.org/10.1080/01443410500345172}
\showDOI{\tempurl}


\bibitem[Viscusi et~al\mbox{.}(1987)]%
        {Viscusi1987}
\bibfield{author}{\bibinfo{person}{W.~Kip Viscusi}, \bibinfo{person}{Wesley~A.
  Magat}, {and} \bibinfo{person}{Joel Huber}.} \bibinfo{year}{1987}\natexlab{}.
\newblock \showarticletitle{An Investigation of the Rationality of Consumer
  Valuations of Multiple Health Risks}.
\newblock \bibinfo{journal}{\emph{The {RAND} Journal of Economics}}
  \bibinfo{volume}{18}, \bibinfo{number}{4} (\bibinfo{year}{1987}),
  \bibinfo{pages}{465}.
\newblock
\urldef\tempurl%
\url{https://doi.org/10.2307/2555636}
\showDOI{\tempurl}


\bibitem[Walker et~al\mbox{.}(2008)]%
        {walker2008tutor}
\bibfield{author}{\bibinfo{person}{Erin Walker}, \bibinfo{person}{Nikol
  Rummel}, {and} \bibinfo{person}{Kenneth~R Koedinger}.}
  \bibinfo{year}{2008}\natexlab{}.
\newblock \showarticletitle{To tutor the tutor: Adaptive domain support for
  peer tutoring}. In \bibinfo{booktitle}{\emph{International Conference on
  Intelligent Tutoring Systems}}. \bibinfo{publisher}{Springer},
  \bibinfo{address}{New York City, NY, USA}, \bibinfo{pages}{626--635}.
\newblock


\bibitem[Walker et~al\mbox{.}(2014)]%
        {walker2014adaptive}
\bibfield{author}{\bibinfo{person}{Erin Walker}, \bibinfo{person}{Nikol
  Rummel}, {and} \bibinfo{person}{Kenneth~R Koedinger}.}
  \bibinfo{year}{2014}\natexlab{}.
\newblock \showarticletitle{Adaptive intelligent support to improve peer
  tutoring in algebra}.
\newblock \bibinfo{journal}{\emph{International Journal of Artificial
  Intelligence in Education}} \bibinfo{volume}{24}, \bibinfo{number}{1}
  (\bibinfo{year}{2014}), \bibinfo{pages}{33--61}.
\newblock


\bibitem[Walker et~al\mbox{.}(2007)]%
        {walker2007student}
\bibfield{author}{\bibinfo{person}{Erin Walker}, \bibinfo{person}{Nikol
  Rummel}, \bibinfo{person}{Bruce McLaren}, {et~al\mbox{.}}}
  \bibinfo{year}{2007}\natexlab{}.
\newblock \showarticletitle{The student becomes the master: Integrating peer
  tutoring with cognitive tutoring}. In \bibinfo{booktitle}{\emph{Proceedings
  of the 7th International Conference on Computer Supported Collaborative
  Learning}}. \bibinfo{publisher}{International Society of the Learning
  Sciences, Inc.}, \bibinfo{address}{Bloomington, IN, USA},
  \bibinfo{pages}{750--752}.
\newblock


\bibitem[Webb and Mastergeorge(2003)]%
        {webb2003promoting}
\bibfield{author}{\bibinfo{person}{Noreen~M Webb} {and} \bibinfo{person}{Ann
  Mastergeorge}.} \bibinfo{year}{2003}\natexlab{}.
\newblock \showarticletitle{Promoting effective helping behavior in
  peer-directed groups}.
\newblock \bibinfo{journal}{\emph{International Journal of Educational
  Research}} \bibinfo{volume}{39}, \bibinfo{number}{1-2}
  (\bibinfo{year}{2003}), \bibinfo{pages}{73--97}.
\newblock


\bibitem[Weinberger et~al\mbox{.}(2005)]%
        {weinberger2005epistemic}
\bibfield{author}{\bibinfo{person}{Armin Weinberger}, \bibinfo{person}{Bernhard
  Ertl}, \bibinfo{person}{Frank Fischer}, {and} \bibinfo{person}{Heinz Mandl}.}
  \bibinfo{year}{2005}\natexlab{}.
\newblock \showarticletitle{Epistemic and social scripts in computer--supported
  collaborative learning}.
\newblock \bibinfo{journal}{\emph{Instructional Science}} \bibinfo{volume}{33},
  \bibinfo{number}{1} (\bibinfo{year}{2005}), \bibinfo{pages}{1--30}.
\newblock


\bibitem[Weinstein(1980)]%
        {weinstein1980unrealistic}
\bibfield{author}{\bibinfo{person}{Neil~D Weinstein}.}
  \bibinfo{year}{1980}\natexlab{}.
\newblock \showarticletitle{Unrealistic optimism about future life events.}
\newblock \bibinfo{journal}{\emph{Journal of personality and social
  psychology}} \bibinfo{volume}{39}, \bibinfo{number}{5}
  (\bibinfo{year}{1980}), \bibinfo{pages}{806}.
\newblock


\bibitem[Westera et~al\mbox{.}(2009)]%
        {Westera2009}
\bibfield{author}{\bibinfo{person}{Wim Westera}, \bibinfo{person}{Gijs de
  Bakker}, {and} \bibinfo{person}{Leo Wagemans}.}
  \bibinfo{year}{2009}\natexlab{}.
\newblock \showarticletitle{Self-arrangement of fleeting student pairs: a Web
  2.0 approach for peer tutoring}.
\newblock \bibinfo{journal}{\emph{Interactive Learning Environments}}
  \bibinfo{volume}{17}, \bibinfo{number}{4} (\bibinfo{date}{Dec.}
  \bibinfo{year}{2009}), \bibinfo{pages}{341--349}.
\newblock
\urldef\tempurl%
\url{https://doi.org/10.1080/10494820903195249}
\showDOI{\tempurl}


\bibitem[Wilson and Myers(2000)]%
        {wilson2000situated}
\bibfield{author}{\bibinfo{person}{Brent~G Wilson} {and}
  \bibinfo{person}{Karen~Madsen Myers}.} \bibinfo{year}{2000}\natexlab{}.
\newblock \bibinfo{booktitle}{\emph{Situated cognition in theoretical and
  practical context}}.
\newblock \bibinfo{publisher}{Lawrence Erlbaum Associates},
  \bibinfo{address}{Mahwah, NJ, USA}, \bibinfo{pages}{57--88}.
\newblock


\bibitem[Yerushalmi and Polingher(2006)]%
        {yerushalmi2006guiding}
\bibfield{author}{\bibinfo{person}{Edit Yerushalmi} {and}
  \bibinfo{person}{Corina Polingher}.} \bibinfo{year}{2006}\natexlab{}.
\newblock \showarticletitle{Guiding students to learn from mistakes}.
\newblock \bibinfo{journal}{\emph{Physics Education}} \bibinfo{volume}{41},
  \bibinfo{number}{6} (\bibinfo{year}{2006}), \bibinfo{pages}{532}.
\newblock


\bibitem[Zajonc(1960)]%
        {zajonc1960process}
\bibfield{author}{\bibinfo{person}{Robert~B Zajonc}.}
  \bibinfo{year}{1960}\natexlab{}.
\newblock \showarticletitle{The process of cognitive tuning in communication.}
\newblock \bibinfo{journal}{\emph{The Journal of Abnormal and Social
  Psychology}} \bibinfo{volume}{61}, \bibinfo{number}{2}
  (\bibinfo{year}{1960}), \bibinfo{pages}{159}.
\newblock


\bibitem[Zajonc(1966)]%
        {zajonc1966social}
\bibfield{author}{\bibinfo{person}{Robert~Boleslaw Zajonc}.}
  \bibinfo{year}{1966}\natexlab{}.
\newblock \bibinfo{booktitle}{\emph{Social psychology: An experimental
  approach}}.
\newblock \bibinfo{publisher}{Wadsworth Publishing Company},
  \bibinfo{address}{Belmont, CA, USA}.
\newblock


\bibitem[Zhang et~al\mbox{.}(2012)]%
        {zhang2012human}
\bibfield{author}{\bibinfo{person}{Haoqi Zhang}, \bibinfo{person}{Edith Law},
  \bibinfo{person}{Rob Miller}, \bibinfo{person}{Krzysztof Gajos},
  \bibinfo{person}{David Parkes}, {and} \bibinfo{person}{Eric Horvitz}.}
  \bibinfo{year}{2012}\natexlab{}.
\newblock \showarticletitle{Human computation tasks with global constraints}.
  In \bibinfo{booktitle}{\emph{Proceedings of the SIGCHI Conference on Human
  Factors in Computing Systems}}. \bibinfo{publisher}{ACM},
  \bibinfo{address}{New York, NY, USA}, \bibinfo{pages}{217--226}.
\newblock


\bibitem[Ziegler et~al\mbox{.}(2021)]%
        {ziegler2021micro}
\bibfield{author}{\bibinfo{person}{Esther Ziegler}, \bibinfo{person}{Dragan
  Trninic}, {and} \bibinfo{person}{Manu Kapur}.}
  \bibinfo{year}{2021}\natexlab{}.
\newblock \showarticletitle{Micro productive failure and the acquisition of
  algebraic procedural knowledge}.
\newblock \bibinfo{journal}{\emph{Instructional Science}}  \bibinfo{volume}{49}
  (\bibinfo{year}{2021}), \bibinfo{pages}{313--336}.
\newblock


\end{thebibliography}
% \newpage
%%
%% If your work has an appendix, this is the place to put it.
\appendix

\section{Thematic Codes \& Descriptions}
\label{AppendixA}

\subsection{Codes for LbT Session Analysis}\label{sec:app}
\begin{figure*} [ht]
  \centering
  \includegraphics[width=1\columnwidth]{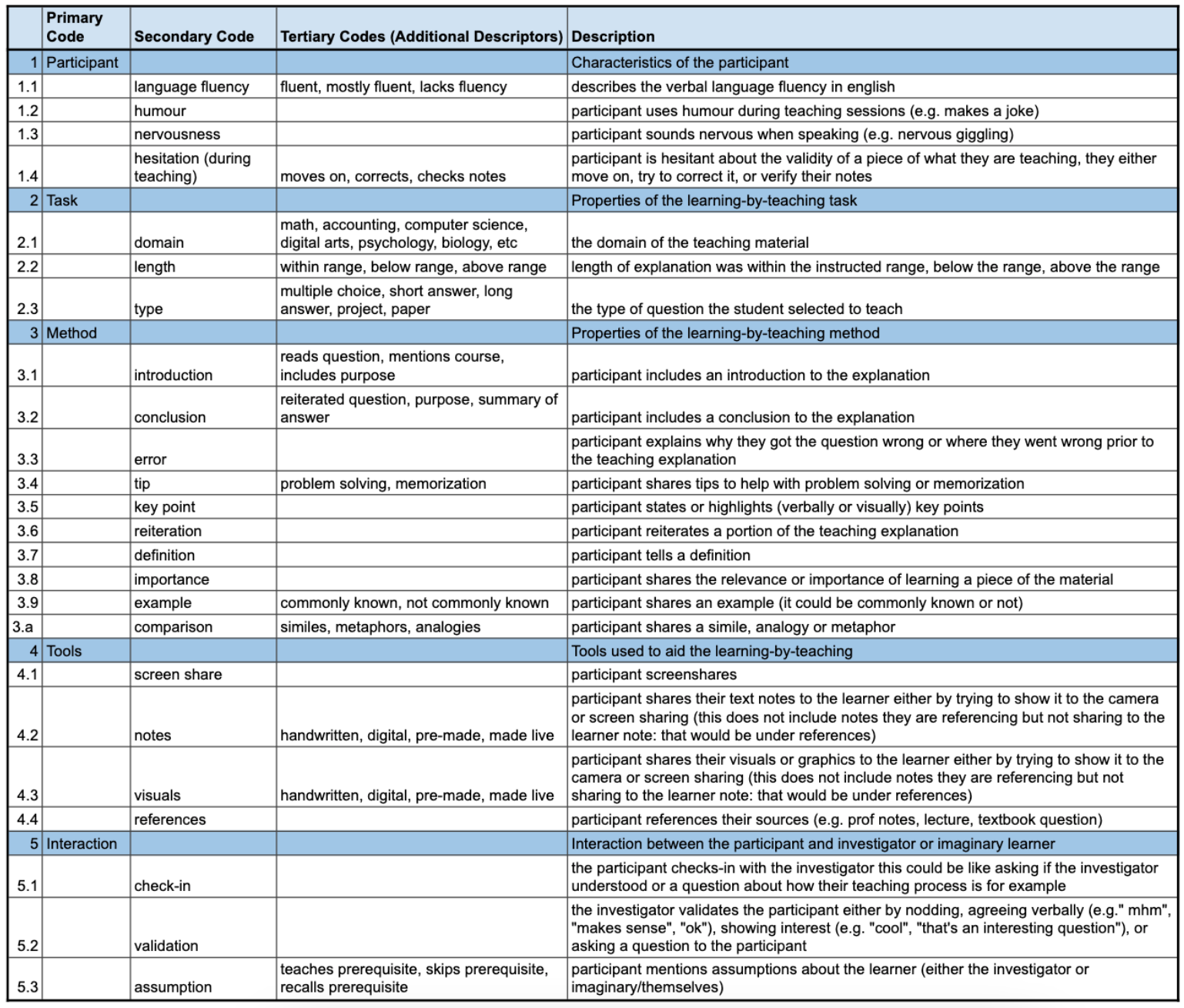}
%   \caption[Short version for LoF]{Long version to appear next to the figure}
  \caption[Codes for LbT Session Analysis]{The codes above were used to analyse the participants' LbT sessions. If data from certain codes did not provide enough information to answer our research questions or were not relevant they were omitted from the analysis.}
%   \label{test}
\end{figure*}

% \cleardoublepage
\clearpage

\subsection{Codes for Semi-Structured Interview Analysis}
\begin{figure*} [ht]
  \centering
  \includegraphics[width=1\columnwidth]{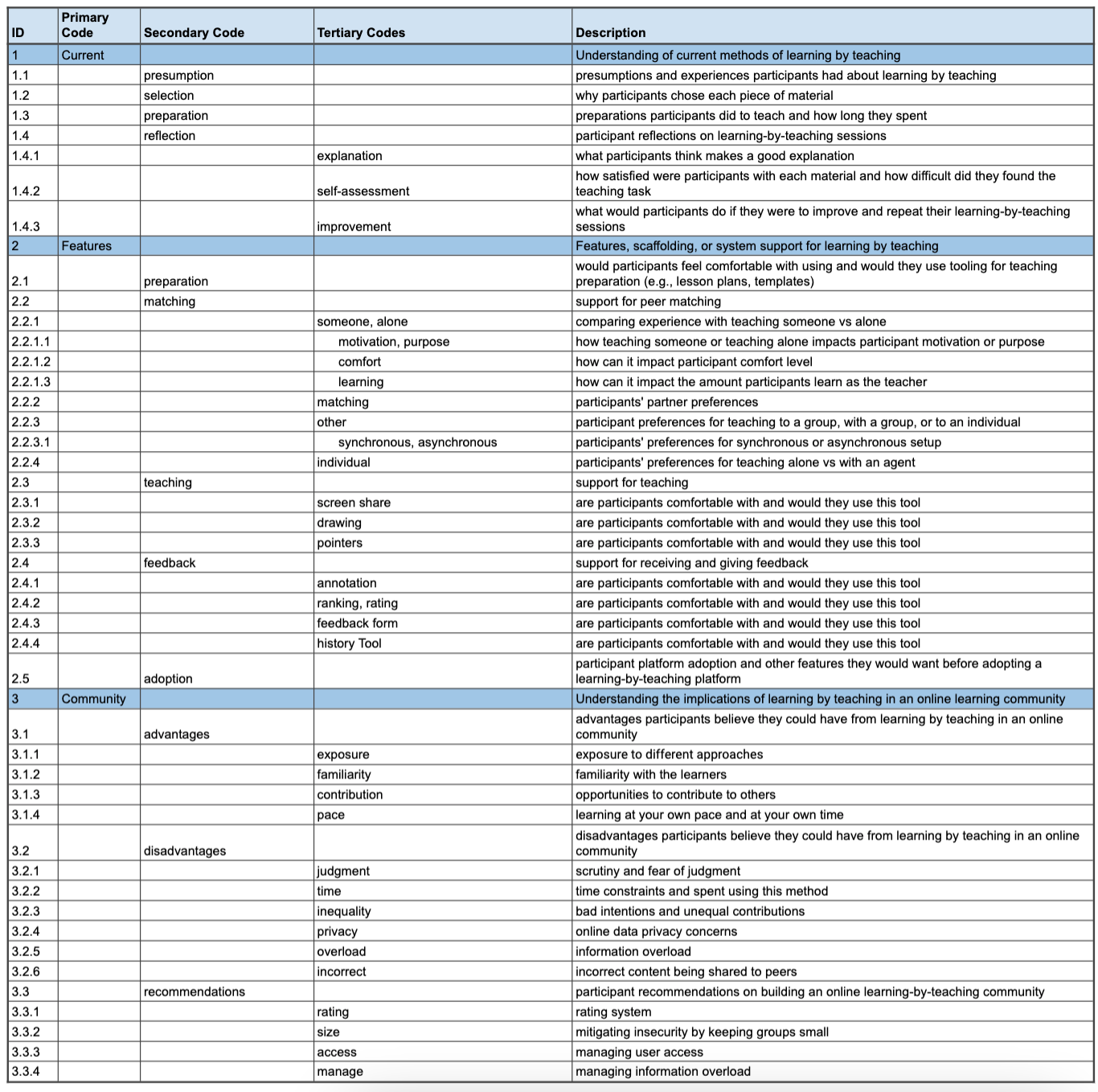}
  \caption[Codes for Semi-Structured Interview Analysis]{The codes above were used to analyse the participants' semi-structured interviews. If data from certain codes did not provide enough information to answer our research questions or were not relevant they were omitted from the analysis.}
%   \label{ppt}
\end{figure*}

% from thesis:
% \chapter[Participant Recruitment Information]{Participant Recruitment Information}
% \label{AppendixA}

% \begin{figure*} [h]
%   \centering
%   \includegraphics[width=1\columnwidth]{Study Information.png}
%   \caption[Participant Recruitment Information]{The Study Information was published on the University of Waterloo's SONA system for Human-Computer Interaction}
% %   \label{ppt}
% \end{figure*}

% \chapter[Thematic Codes \& Descriptions]{Thematic Codes \& Descriptions}
% \label{AppendixB}

% \section{Codes for LbT Session Analysis}
% \begin{figure*} [h]
%   \centering
%   \includegraphics[width=1\columnwidth]{Codes Table 1 .png}
% %   \caption[Short version for LoF]{Long version to appear next to the figure}
%   \caption[Codes for LbT Session Analysis]{The codes above were used to analyse the participants' LbT sessions. If data from certain codes did not provide enough information to answer our research questions or were not relevant they were omitted from the analysis.}
% %   \label{ppt}
% \end{figure*}

% \cleardoublepage

% \section{Codes for Semi-Structured Interview Analysis}
% \begin{figure*} [h]
%   \centering
%   \includegraphics[width=1\columnwidth]{Codes Table 2.png}
%   \caption[Codes for Semi-Structured Interview Analysis]{The codes above were used to analyse the participants' semi-structured interviews. If data from certain codes did not provide enough information to answer our research questions or were not relevant they were omitted from the analysis.}
% %   \label{ppt}
% \end{figure*}

\received{January 2022}
\received[revised]{July 2022}
\received[accepted]{November 2022}

\end{document}